\documentclass[11pt,a4paper]{article}
\pdfoutput=1

\usepackage[utf8]{inputenc}
\usepackage{a4wide}
\usepackage{amsmath}
\usepackage{cite}
\usepackage{xcolor}
\usepackage{amssymb}
\usepackage{amsfonts}
\usepackage{booktabs}
\usepackage{makecell}
\usepackage{multirow}
\usepackage{pdflscape}
\usepackage{textcomp}
\usepackage{gensymb}
\usepackage{bigstrut}
\usepackage{array}
\usepackage{enumerate}
\usepackage[small,bf]{caption}
\usepackage{graphicx}
\usepackage{hyperref}
\usepackage{mathbbol}
\usepackage{amssymb}
\usepackage{appendix}
\usepackage{verbatim}
\usepackage{geometry}
\usepackage{makecell}
\usepackage{listings}
\usepackage{mathtools}
\usepackage{dsfont}

\DeclareMathOperator{\diag}{diag}
\DeclareMathOperator{\im}{Im}
\DeclareMathOperator{\re}{Re}

\newcommand{\id}{\mathds{1}}

\DeclarePairedDelimiter{\vevdel}{\langle}{\rangle}
\newcommand{\vev}{\vevdel}
\allowdisplaybreaks
\numberwithin{equation}{section}

\begin{document}

\begin{titlepage}

\vspace*{-15mm}
\begin{flushright}
SISSA 12/2020/FISI \\
IPMU20-0065 \\
CFTP/20-006
\end{flushright}
\vspace*{5mm}

\begin{center}
{\bf\LARGE Double Cover of Modular $S_4$
\\[3mm] for Flavour Model Building} \\[8mm]
P.~P.~Novichkov$^{\,a,}$\footnote{E-mail: \texttt{pavel.novichkov@sissa.it}},
J.~T.~Penedo$^{\,b,}$\footnote{E-mail: \texttt{joao.t.n.penedo@tecnico.ulisboa.pt}},
S.~T.~Petcov$^{\,a,c,}$\footnote{Also at
Institute of Nuclear Research and Nuclear Energy,
Bulgarian Academy of Sciences, 1784 Sofia, Bulgaria.}\\
 \vspace{5mm}
$^{a}$\,{\it SISSA/INFN, Via Bonomea 265, 34136 Trieste, Italy} \\
\vspace{2mm}
$^{b}$\,{\it CFTP, Departamento de Física, Instituto Superior Técnico, Universidade de Lisboa,\\
Avenida Rovisco Pais 1, 1049-001 Lisboa, Portugal} \\
\vspace{2mm}
$^{c}$\,{\it Kavli IPMU (WPI), University of Tokyo, 5-1-5 Kashiwanoha, 277-8583 Kashiwa, Japan}
\end{center}
\vspace{2mm}

\begin{abstract}
We develop the formalism of the finite modular group $\Gamma'_4 \equiv S'_4$,
a double cover of the modular permutation group $\Gamma_4 \simeq S_4$,
for theories of flavour.
The integer weight $k>0$ of the level 4 modular forms indispensable for the formalism can be even or odd.
We explicitly construct the lowest-weight ($k=1$) modular forms in terms of 
two Jacobi theta constants, denoted as $\varepsilon(\tau)$ and $\theta(\tau)$, $\tau$ being the modulus.
We show that these forms furnish a 3D representation of $S'_4$ not present for $S_4$.
Having derived the $S'_4$ multiplication rules and Clebsch-Gordan coefficients,
we construct multiplets of modular forms of weights up to $k=10$.
These are expressed as polynomials in $\varepsilon$ and $\theta$,
bypassing the need to search for non-linear constraints.
We further show that within $S'_4$ there are two options to define 
the (generalised) CP transformation and we discuss the possible residual symmetries 
in theories based on modular and CP invariance.
Finally, we provide two examples of application of our results,
constructing phenomenologically viable lepton flavour models. 
\end{abstract}

\end{titlepage}
\setcounter{footnote}{0}
%

\section{Introduction}
\label{sec:intro}

The origin of the flavour structures of quarks and leptons remains a fundamental mystery in particle physics. In the lepton sector in particular, data from neutrino oscillation experiments~\cite{PDG2019} has revealed a mixing pattern of two large and one small mixing angles, which suggests a non-Abelian discrete flavour symmetry may be at play~\cite{Altarelli:2010gt,Ishimori:2010au,King:2013eh,Tanimoto:2015nfa,Petcov:2017ggy}. Future observations are expected to put such symmetry-based scenarios to the test
via, e.g.,~high precision measurements of the neutrino mixing angles and 
of the amount of leptonic Dirac CP Violation (CPV). Of paramount 
importance are also the measurement of the absolute scale of neutrino masses 
and the determination of the neutrino mass ordering.
Recent global data analyses (see, e.g.,~\cite{Esteban:2018azc,Capozzi:2020qhw})
show that data favour values 
of the leptonic Dirac CPV phase $\delta$ close to $3\pi/2$,%
\footnote{The best fit value of $\delta$ obtained by the T2K 
Collaboration in the latest data analysis is also close to 
$3\pi/2$, 
while the CP conserving values $\delta = 0$ and $\pi$ 
are disfavoured by the T2K data respectively 
at $3\sigma$ and $2\sigma$~\cite{Abe:2019vii}.
}
and a neutrino mass spectrum with normal 
ordering (NO) over the one with inverted ordering (IO),
the IO spectrum being disfavoured at $\sim 3\sigma$ confidence level.
Upper bounds on the sum of neutrino masses 
in the range of
$\Sigma < 0.12 - 0.69$ eV 
(at the $2\sigma$ level) are also found
in the most recent analysis~\cite{Capozzi:2020qhw},
where the quoted largest upper limit corresponds to  
the cosmological data set used as input which leads to the most 
conservative result.

Within the approach of postulating a discrete symmetry and its breaking 
pattern, one can generically predict correlations between some
of the three neutrino mixing angles and/or between some 
of, or all, these angles and $\delta$ (see, e.g. \cite{Petcov:2017ggy}).
Majorana CPV phases remain instead unconstrained, unless one combines 
the discrete symmetry with a generalised CP (gCP)
symmetry~\cite{Feruglio:2012cw,Holthausen:2012dk}. While the latter scenarios 
are more predictive, one is still required to construct specific models 
to obtain predictions for neutrino masses. These models typically rely on 
the introduction of a plethora of so-called flavon scalar fields, 
acquiring specifically
aligned vacuum expectation values (VEVs),
which require a rather elaborate potential and 
additional large shaping symmetries.

The modular invariance approach to the flavour problem
put forward in Ref.~\cite{Feruglio:2017spp} has opened up
a new direction in flavour model building. Modular symmetry is introduced
into the supersymmetric (SUSY) flavour picture, with quotients
$\Gamma_N$ of the modular group ($N=2,3,\dots$) playing the role of
non-Abelian discrete symmetry groups.
For $N\leq 5$, these finite
modular groups are isomorphic to the permutation groups
$S_3$, $A_4$, $S_4$ and $A_5$,
widely used in flavour model building. The traditional approach to
flavour is thus generalised, since fields can carry non-trivial
modular weights $k$, further constraining their couplings in the
superpotential. Furthermore, no flavons need to be introduced in the model. 
In such a case, Yukawa couplings and fermion mass
matrices in the Lagrangian of the theory are obtained from
combinations of modular forms, which are 
holomorphic functions of a
single complex number -- the VEV of the modulus $\tau$ --
and have specific transformation properties under the 
action of the modular symmetry group.
 Models of
flavour based on modular invariance have then an increased predictive
power, constraining fermion masses, mixing and CPV phases.%
\footnote{Possible non-minimal additions to the Kähler potential, compatible with the modular symmetry, may jeopardise the predictive power of the framework~\cite{Chen:2019ewa}.
This problem is the subject of ongoing research.
}

Bottom-up modular invariance approaches to the lepton flavour problem
have been exploited using the groups
$\Gamma_2 \simeq S_3$~\cite{Kobayashi:2018vbk, Okada:2019xqk},
$\Gamma_3 \simeq A_4$~\cite{Feruglio:2017spp, Kobayashi:2018vbk, Criado:2018thu, Kobayashi:2018scp, Novichkov:2018yse, Nomura:2019jxj, Nomura:2019yft, Ding:2019zxk, Okada:2019mjf, Nomura:2019lnr, Asaka:2019vev, Gui-JunDing:2019wap, Zhang:2019ngf, Nomura:2019xsb, Kobayashi:2019gtp, Wang:2019xbo, Okada:2020dmb, Ding:2020yen},
$\Gamma_4 \simeq S_4$~\cite{Penedo:2018nmg, Novichkov:2018ovf, Kobayashi:2019mna, Okada:2019lzv, Kobayashi:2019xvz, Gui-JunDing:2019wap, Wang:2019ovr},
$\Gamma_5 \simeq A_5$~\cite{Novichkov:2018nkm, Ding:2019xna},
and $\Gamma_7 \simeq PSL(2,\mathbb{Z}_7)$~\cite{Ding:2020msi}.
Similarly, attempts have been made to construct viable models of 
quark flavour~\cite{Okada:2018yrn} and of quark-lepton unification~\cite{Kobayashi:2018wkl, Okada:2019uoy, Kobayashi:2019rzp, Abbas:2020qzc, Okada:2020rjb}.
The interplay of modular and gCP symmetries
has also been investigated~\cite{Novichkov:2019sqv, Kobayashi:2019uyt}, 
as were the problem of fermion mass hierarchies~\cite{Criado:2019tzk, King:2020qaj} and the possibility of coexistence of multiple
 moduli~\cite{deMedeirosVarzielas:2019cyj, King:2019vhv}, 
 considered first phenomenologically in~\cite{Novichkov:2018ovf, Novichkov:2018yse}.
Such bottom-up analyses are expected to eventually connect with top-down results~\cite{Kobayashi:2018rad, Kobayashi:2018bff, deAnda:2018ecu, Baur:2019kwi, Kariyazono:2019ehj, Baur:2019iai, Nilles:2020nnc, Kobayashi:2020hoc, Abe:2020vmv, Ohki:2020bpo,Nilles:2020kgo,Kikuchi:2020frp, Kobayashi:2020uaj}
based on ultraviolet-complete theories. 
While the aforementioned finite quotients~$\Gamma_N$ of the modular
group -- also known as inhomogeneous finite modular groups -- have
been widely used in the literature to construct modular-invariant
models of flavour from the bottom-up perspective, top-down
constructions typically lead to their double covers~$\Gamma'_N$ (see,
e.g.,~\cite{Ferrara:1989qb, Baur:2019kwi, Baur:2019iai, Nilles:2020nnc}).
The formalism of such double covers has been explored in
Ref.~\cite{Liu:2019khw}, where the case of $\Gamma'_3 \simeq T'$ was
considered (see also~\cite{Lu:2019vgm}).

In the present work, we analyse the double cover of the $\Gamma_4
\simeq S_4$ finite modular group, namely $\Gamma'_4 \simeq S'_4 \equiv
SL(2,\mathbb{Z}_4)$,
which, in view of the above discussion, appears theoretically more motivated than $\Gamma_4$.
We start by briefly reviewing the modular symmetry
approach to flavour
in Section~\ref{sec:framework} (see Ref.~\cite{Feruglio:2019ktm} for a recent review),
considering also its generalisation to the case of double covers of
finite modular groups.
In Section~\ref{sec:modforms}, we compute the fundamental object
required for flavour model building: the modular multiplet of lowest
modular weight ($k=1$). In particular, we write the $k=1$ modular
forms in terms of two ``weight 1/2'' functions $\theta(\tau)$ and
$\varepsilon(\tau)$, which are obtained from the Dedekind eta function
and present interesting features. The $k=1$ forms are then found to
arrange themselves into a triplet $\mathbf{\hat{3}}$ of $S_4'$.
This fundamental triplet is used to derive higher-weight
multiplets, up to $k=10$, via tensor products. We thus obtain new,
odd-weight modular multiplets specific to $\Gamma_4' \simeq S_4'$, and recover the
even-weight modular multiplets of $\Gamma_4 \simeq S_4$~\cite{Penedo:2018nmg,
Novichkov:2018ovf, Novichkov:2019sqv}. Given our construction in terms of $\theta$
and $\varepsilon$, the derivation of $k>1$ modular multiplets
automatically bypasses a typical need to search for non-linear
constraints, which would relate dependent multiplets coming from
tensor products.
In Section~\ref{sec:gCP} we discuss the problem of combining modular and 
CP invariance in theories based on  $S_4'$, 
while in Section~\ref{sec:residual}
we analyse the possible residual symmetries in such theories.
In Section~\ref{sec:pheno}, we illustrate phenomenological applications
of our results, by building and analysing
two viable models of lepton masses and mixing based on $S'_4$ modular
symmetry.
We finally summarise our results
and conclude in Section~\ref{sec:conclusions}.

\section{Framework}
\label{sec:framework}
\subsection{The Modular Group and Transformation of Fields}
We introduce a complex scalar field~$\tau$, called the \textit{modulus}, 
whose VEV is restricted to the upper half-plane $\mathcal{H} \equiv \left\{ \tau \in \mathbb{C} : \im \tau > 0 \right\}$.%
\footnote{We use $\tau$ to denote both the modulus and its VEV.}
The modulus~$\tau$ plays the role of a spurion and transforms non-trivially
under the \textit{modular group}~$\Gamma$, which is the special linear group of $2 \times 2$ integer matrices with unit determinant, i.e.
\begin{equation}
  \label{eq:full_mod_group}
  \Gamma \equiv SL(2, \mathbb{Z}) \equiv
  \left\{
    \begin{pmatrix}
      a & b \\ c & d
    \end{pmatrix}
    \middle\vert \,
    a, b, c, d \in \mathbb{Z}, \,
    ad - bc = 1
  \right\} \,.
\end{equation}
%
The group~$\Gamma$ is generated by three matrices
\begin{equation}
  \label{eq:STR_def}
  S =
  \begin{pmatrix}
    0 & 1 \\ -1 & 0
  \end{pmatrix}
  \,, \quad
  T =
  \begin{pmatrix}
    1 & 1 \\ 0 & 1
  \end{pmatrix}
  \,, \quad
  R =
  \begin{pmatrix}
    -1 & 0 \\ 0 & -1
  \end{pmatrix}\,,
\end{equation}
%
subject to the following relations:
\begin{equation}
  \label{eq:STR_rel}
  S^2 = R, \quad
  (ST)^3 = \id, \quad
  R^2 = \id, \quad
  RT = TR \,,
\end{equation}
%
where $\id$ denotes the identity element of a group.

The modular group~$\Gamma$ acts on the modulus with fractional
linear transformations:
\begin{equation}
  \label{eq:tau_mod_trans}
  \gamma =
  \begin{pmatrix}
    a & b \\ c & d
  \end{pmatrix}
  \in \Gamma : \,
  \tau \to \gamma \tau = \frac{a\tau + b}{c\tau + d} \,.
\end{equation}
%
The matter superfields transform under~$\Gamma$
as ``weighted'' multiplets~\cite{Ferrara:1989bc,Ferrara:1989qb,Feruglio:2017spp}:
\begin{equation}
  \label{eq:psi_mod_trans}
  \psi_i \to (c\tau + d)^{-k} \, \rho_{ij}(\gamma) \, \psi_j \,,
\end{equation}
%
where $(c\tau + d)^{-k}$ is the \textit{automorphy factor}, $k \in \mathbb{Z}$ is the \textit{modular weight}%
\footnote{While we restrict ourselves to integer modular weights, it is also possible to have fractional weights $k$ \cite{Dixon:1989fj,Ibanez:1992hc,Olguin-Trejo:2017zav,Nilles:2020nnc}.}
and $\rho$ is a unitary representation of~$\Gamma$.

Note that the group action~\eqref{eq:tau_mod_trans} has a non-trivial kernel~$\mathbb{Z}_2^R = \{\id, R\}$, i.e.~the modulus~$\tau$ does not transform under the action of $R$.
For this reason one typically defines the {\it (inhomogeneous) modular group} as the quotient~$\overline{\Gamma} \equiv PSL(2, \mathbb{Z}) \equiv SL(2, \mathbb{Z}) \, / \, \mathbb{Z}_2^R$, which is the projective version of~$SL(2, \mathbb{Z})$ with matrices $\gamma$ and $-\gamma$ being identified.
However, matter fields of a modular-invariant theory are in general allowed to transform under $R$, as can be seen from~\eqref{eq:psi_mod_trans}.
Therefore the symmetry group of such theory is $\Gamma$ rather than $\overline{\Gamma}$, as was stressed recently in~\cite{Nilles:2020nnc}.
The inclusion of the $R$ generator is crucial in extending
finite modular groups to their double covers, as we will see shortly. 

We assume that representations of matter fields are trivial when restricted to the so-called \textit{principal congruence subgroup},
\begin{equation}
  \label{eq:congr_subgr}
  \Gamma(N) \equiv
  \left\{
    \begin{pmatrix}
      a & b \\ c & d
    \end{pmatrix}
    \in SL(2, \mathbb{Z}), \,
    \begin{pmatrix}
      a & b \\ c & d
    \end{pmatrix}
    \equiv
    \begin{pmatrix}
      1 & 0 \\ 0 & 1
    \end{pmatrix}
    (\text{mod } N)
  \right\}\,,
\end{equation}
%
with a fixed integer $N \geq 2$ called the \textit{level}.
In other words, $\rho(\gamma)$ of eq.~\eqref{eq:psi_mod_trans} is the identity matrix whenever $\gamma \in \Gamma(N)$, so that $\rho$ is effectively a representation of the quotient group
\begin{equation}
  \label{eq:hom_fin_mod_group}
  \Gamma_N' \equiv \Gamma \, \big/ \, \Gamma(N) \simeq SL(2, \mathbb{Z}_N)\,,
\end{equation}
%
called the \textit{homogeneous finite modular group}.
Unlike $\Gamma$, $\Gamma_N'$ is finite as the name suggests.
For $N\leq 5$, this group admits the presentations%
\footnote{For $N>5$, additional relations are needed in order to render the group finite~\cite{deAdelhartToorop:2011re}.}
\begin{equation}
  \label{eq:hom_fin_mod_group_pres}
  \begin{aligned}
    \Gamma'_N &= \left\langle S, \, T, \, R \mid S^2 = R, \, (ST)^3 = \id, \, R^2 = \id, \, RT = TR, \, T^N = \id \right\rangle \\
    &= \left\langle S, \, T \mid S^4 = \id, \, (ST)^3 = \id, \, S^2 T = T S^2, \, T^N = \id \right\rangle \,,
  \end{aligned}
\end{equation}
%
where with a slight abuse of notation we denote by $S$, $T$, $R$ the
equivalence classes of the corresponding generators~\eqref{eq:STR_def}
of the full modular group.

In the special case when $\rho$ does not distinguish between $\gamma$
and $-\gamma$, i.e.~$\rho(R)$ is identity, we see that $\rho$ is a
representation of a smaller quotient group
\begin{equation}
  \label{eq:inhom_fin_mod_group}
    \Gamma_N \equiv \Gamma \, \big/ \left\langle \, \Gamma(N) \cup \mathbb{Z}_2^R \, \right\rangle \simeq SL(2, \mathbb{Z}_N) \, \big/ \left\langle R \right\rangle\,,
\end{equation}
%
called the \textit{(inhomogeneous) finite modular group}.
For $N\leq 5$, $\Gamma_N$ has the following presentation:
\begin{equation}
  \label{eq:inhom_fin_mod_group_pres}
    \Gamma_N = \left\langle S, \, T \mid S^2 = \id, \, (ST)^3 = \id, \, T^N = \id \right\rangle \,.
\end{equation}
%
Note that $R \in \Gamma(2)$, hence $\Gamma^{\phantom{\prime}}_2 = \Gamma'_2$.
In contrast, for $N \geq 3$ one has $R \notin \Gamma(N)$, 
and $\Gamma_N'$ is a double cover of $\Gamma_N$.
For small values of $N$, the groups $\Gamma^{\phantom{\prime}}_N$ and
$\Gamma'_N$ are isomorphic to permutation groups and their double covers,
see Table~\ref{tab:fin_mod_group}.
\begin{table}
  \centering
  \begin{tabular}{ccccc}
    \toprule
    $N$ & 2 & 3 & 4 & 5 \\
    \midrule
    $\Gamma_N$ & $S_3$ & $A_4$ & $S_4$ & $A_5$ \\
    $\Gamma'_N$ & $S_3$ & $A'_4 \equiv T'$ & $S'_4 \equiv SL(2, \mathbb{Z}_4)$ & $A'_5 \equiv SL(2, \mathbb{Z}_5)$ \\
    \midrule
    $\dim \mathcal{M}_k(\Gamma(N))$ & $k/2 + 1$ & $k+1$ & $2k + 1$ & $5k + 1$ \\
    \bottomrule
  \end{tabular}
  \caption{Finite modular groups and dimensionality of the corresponding spaces of modular forms, for $N \leq 5$.
  Note that for $N = 2$ only even-weighted modular forms exist.}
  \label{tab:fin_mod_group}
\end{table}

As a final remark, let us stress that the level~$N$ defining the finite
modular group is common to all matter fields~$\psi_I$, which may however carry different modular weights~$k_I$.

%
\subsection{Modular Forms and Modular-Invariant Actions}
%

The Lagrangian of a $\mathcal{N} = 1$ global supersymmetric theory is given by
\begin{equation}
  \label{eq:susy_lagrangian}
  \mathcal{L} = \int \text{d}^2 \theta \, \text{d}^2 \bar{\theta} \, K(\Phi, \bar{\Phi})
  + \left[ \, \int \text{d}^2 \theta \, W(\Phi) + \text{h.c.} \right] \,,
\end{equation}
%
where $K$ is the Kähler potential, $W$ is the superpotential, $\theta$
and $\bar{\theta}$ are Graßmann variables, and $\Phi$ collectively
denotes chiral superfields of the theory.
In modular-invariant supersymmetric theories, $\tau$ is the scalar component
of a chiral superfield, and the superpotential has to be modular-invariant, $W(\Phi) \xrightarrow{\gamma} W(\Phi)$~\cite{Ferrara:1989bc}.
In theories of supergravity, the superpotential is instead coupled to the Kähler potential and
has to transform with a
certain weight $-h$ under modular transformations (up to a field-independent
phase)~\cite{Ferrara:1989bc,Ferrara:1989qb}:
\begin{equation}
  \label{eq:W_mod_trans}
  \gamma =
  \begin{pmatrix}
    a & b \\ c & d
  \end{pmatrix}
  \in \Gamma : \,\,
  W(\Phi) \to e^{i \alpha(\gamma)} \, (c\tau + d)^{-h} \, W(\Phi) \,.
\end{equation}
%
The superpotential can be expanded in powers of matter superfields $\psi_I$ as:
\begin{equation}
  \label{eq:W_series}
  W(\tau, \psi_I) = \sum \left( \vphantom{\sum} Y_{I_1 \ldots I_n}(\tau) \, \psi_{I_1} \ldots \psi_{I_n} \right)_{\mathbf{1}} \,,
\end{equation}
%
where the sum is taken over all possible combinations of fields
$\{ I_1, \ldots, I_n\}$ and all independent singlets of
$\Gamma'_N$, denoted by $(\ldots)_{\mathbf{1}}$.%
\footnote{
Since the field-independent phase factor in~\eqref{eq:W_mod_trans} does not affect the supergravity scalar potential, these singlets need not be trivial. 
All terms in~\eqref{eq:W_series} should nevertheless transform in the same way under the modular group.
}

In order to satisfy~\eqref{eq:W_mod_trans} given the field transformation
rules~\eqref{eq:psi_mod_trans}, the field couplings
$Y_{I_1 \ldots I_n}(\tau)$ have to be \textit{modular forms} of level $N$ and
weight $k_Y = k_{I_1} + \ldots + k_{I_n} - h$, i.e., transform under $\Gamma$ as
\begin{equation}
  \label{eq:Y_mod_trans}
  Y_{I_1 \ldots I_n}(\tau) \,\to\, Y_{I_1 \ldots I_n}(\gamma \tau) = (c\tau + d)^{k_Y} \rho(\gamma) \,Y_{I_1 \ldots I_n}(\tau) \,,
\end{equation}
%
where $\rho$ is a unitary representation of the homogeneous finite modular
group~$\Gamma'_N$ such that
$\rho \,\otimes\, \rho_{I_1} \,\otimes \ldots \otimes\, \rho_{I_n} \supset \mathbf{1}$.
Apart from that, due to holomorphicity of the superpotential, modular forms have to be holomorphic functions of~$\tau$.
Together with the transformation property~\eqref{eq:Y_mod_trans}, this significantly constrains the space of modular forms.
In fact, non-trivial modular forms of a given level~$N$ exist only for positive integer weights~$k \in \mathbb{N}$ and form finite-dimensional linear spaces~$\mathcal{M}_k(\Gamma(N))$ which decompose into multiplets of~$\Gamma'_N$.
As can be seen from Table~\ref{tab:fin_mod_group}, the spaces~$\mathcal{M}_k(\Gamma(N))$ have low dimensionalities for small values of~$k$ and $N$. Therefore it is possible to form only a few independent Yukawa couplings, which yields predictive models of flavour.

By analysing eq.~\eqref{eq:Y_mod_trans}, one notes that odd-weighted modular forms necessarily have~$\rho(R) = -\id$ in order to compensate the minus sign arising from the automorphy factor, while for even-weighted modular forms one has~$\rho(R) = \id$.
Therefore, in modular-invariant theories based on inhomogeneous modular groups~$\Gamma_N$ only even-weighted modular forms appear.

%
\section{Modular Forms of Level 4}
\label{sec:modforms}
%

%
\subsection{``Weight 1/2'' Modular Forms}
%

Modular forms of level 4 and weight $k$ form a linear space of
dimension $2k+1$ given by~\cite{Schultz}:
\begin{equation}
  \begin{aligned}
    \mathcal{M}_k(\Gamma(4)) &=
    \bigoplus_{\substack{m+n=2k,\\m,n \geq 0}} \mathbb{C} \, \frac{\eta^{2n-2m}(4\tau) \, \eta^{5m-n}(2\tau)}{\eta^{2m}(\tau)} \\
    &= \bigoplus_{\substack{m+n=2k,\\m,n \geq 0}} \mathbb{C} \left( \frac{\eta^5(2\tau)}{\eta^2(\tau) \eta^2(4\tau)} \right)^m \left( \frac{\eta^2(4\tau)}{\eta(2\tau)} \right)^n \,,
  \end{aligned}
  \label{eq:mkg4}
\end{equation}
%
where $m$ and $n$ are non-negative integers, and $\eta(\tau)$ is the Dedekind
eta function (we collect all the necessary definitions and properties of
special functions in Appendix~\ref{app:eta_theta}).
In other words, $\mathcal{M}_k(\Gamma(4))$ is spanned by polynomials of even
degree $2k$ in two functions $\theta(\tau)$ and $\varepsilon(\tau)$ defined as
\begin{equation}
  \theta(\tau)\, \equiv\,
  \frac{\eta^5(2\tau)}
{\eta^2(\tau) \eta^2(4\tau)} \,=\, 
\Theta_3(2\tau) \,, \quad\,\,
 \varepsilon(\tau) \,\equiv\, \frac{2\,\eta^2(4\tau)}{\eta(2\tau)} \,
=\, \Theta_2(2\tau) \,.
  \label{eq:theta_eps_def}
\end{equation}
%
Here $\Theta_2(\tau)$ and $\Theta_3(\tau)$ are the Jacobi theta constants
related to the Dedekind eta by eq.~\eqref{eq:theta_eta_rel}.
In particular, we conclude from eq.~\eqref{eq:mkg4} that the space of
weight 1 modular forms of level 4 is formed by the homogeneous quadratic
polynomials in $\theta$ and $\varepsilon$, or equivalently, in the theta
constants $\Theta_2$ and $\Theta_3$ of double argument (for more details on
the correspondence between modular forms of level 4 and the theta constants,
see Appendix~\ref{app:mod4_theta}).

From eqs.~\eqref{eq:theta_eps_def} and \eqref{eq:theta_def} we find
immediately that $\theta(\tau)$ and $\varepsilon(\tau)$ admit the following
$q$-expansions, i.e.~power series expansions in $q_4 \equiv \exp(i\pi \tau/2)$:
\begin{equation}
  \begin{aligned}
    \theta(\tau) \,&=\, 1
    + 2\sum_{k=1}^{\infty} q_4^{(2k)^2}
    \,=\, 1 + 2 \,q_4^4 + 2 \,q_4^{16} + \ldots \,, \\
    \varepsilon(\tau) \,&=\, 2\sum_{k=1}^{\infty} q_4^{(2k-1)^2} \,=\, 2 \,q_4^{\phantom{1}} + 2 \,q_4^9 + 2 \,q_4^{25} + \ldots \,,
  \end{aligned}
  \label{eq:theta_eps_qexp}
\end{equation}
%
so that $\theta \to 1$, $\varepsilon \to 0$ in the ``large volume''
limit $\im\tau \to \infty$. In fact, $\varepsilon \sim 2\,q_4$ in this limit and
it can be used as an expansion parameter instead of $q_4$, which justifies
the notation. Note that, due to quadratic dependence in the exponents of
$q_4$, the series~\eqref{eq:theta_eps_qexp} converge rapidly in the
fundamental domain of the modular group, where one has
$|q_4| \leq \exp(-\pi \sqrt{3}/4) \simeq 0.26.$
We give below the values of  $\theta(\tau)$
and $\varepsilon(\tau)$ at values of $\tau$,
namely $\tau_C$, $\tau_L$, and $\tau_T$,
at which there exist residual symmetries (see Section~\ref{sec:residual} for details):
\begin{equation}
\begin{aligned}
\theta(\tau_C) &= 1 + 2\,e^{-2\pi} + O(10^{-11}) \,\simeq\, 1.00373\,,\\
\varepsilon(\tau_C) &= 2\,e^{-\pi/2} + O(10^{-6})  \,\simeq\, 0.415761\,; \\[3mm]
\theta(\tau_L) &=  1 - 2\,e^{-\sqrt{3}\, \pi}+ O(10^{-9}) \,\simeq\, 0.991333\,,\\
\varepsilon(\tau_L) &= 2\,e^{-i\,\pi/4}\left[e^{-\sqrt{3}\,\pi/4} + O(10^{-5})\right] \,\simeq\, 0.512152\;e^{-i\,\pi/4}\,;\\[3mm]
\theta(\tau_T) &= 1\,,\\\varepsilon(\tau_T) &= 0\,,
\end{aligned}
\label{eq:thepsres}
\end{equation}
where $\tau_C \equiv i$, $\tau_L \equiv -\,1/2 + i\sqrt{3}/2$ 
and  $\tau_T \equiv i\,\infty$.
We further find the exact relations at symmetric points:
\begin{equation}
    \frac{\varepsilon(\tau_C)}{\theta(\tau_C)} \,=\, \frac{1}{1+\sqrt{2}}\,,
    \qquad
    \frac{\varepsilon(\tau_L)}{\theta(\tau_L)} \,=\, \frac{1-i}{1+\sqrt{3}}\,.
\end{equation}

The action of the $T$ generator on $\theta$ and $\varepsilon$ follows from
the corresponding transformation of the theta
constants~\eqref{eq:theta_mod_trans}:
\begin{equation}
  \theta(\tau)\, \xrightarrow{T}\, \theta(\tau) \,, \quad
  \varepsilon(\tau)\, \xrightarrow{T}\, i \, \varepsilon(\tau) \,.
  \label{eq:theta_eps_T}
\end{equation}
%
Similarly, one can obtain the action of the $S$ generator on
$\theta$ from eq.~\eqref{eq:theta_mod_trans} with the help of
identity~\eqref{eq:theta_double_arg}:
\begin{equation}
  \begin{aligned}
    \theta(\tau)
    &=\Theta_3(2\tau)
    = \frac{1}{2} \left[ \Theta_3 \left(\frac{\tau}{2}\right) + \Theta_4 \left(\frac{\tau}{2}\right) \right]
    \,\,\xrightarrow{S}\,\, \frac{1}{2} \left[ \Theta_3 \left(-\frac{1}{2\tau}\right) + \Theta_4 \left(-\frac{1}{2\tau}\right) \right] \\[2mm]
    &= \frac{1}{2} \sqrt{-i 2\tau} \left[ \Theta_3(2\tau) + \Theta_2(2\tau) \right]
    = \sqrt{-i \tau} \, \frac{\theta(\tau) + \varepsilon(\tau)}{\sqrt{2}} \,.
  \end{aligned}
\end{equation}
%
By requiring that the second action of $S$ should transform the result back
to $\theta(\tau)$, we find the corresponding action on
$\varepsilon(\tau)$, and conclude that
\begin{equation}
  \theta(\tau) \,\xrightarrow{S}\, \sqrt{-i \tau} \; \frac{\theta(\tau) + \varepsilon(\tau)}{\sqrt{2}} \,, \qquad
  \varepsilon(\tau) \,\xrightarrow{S}\, \sqrt{-i \tau} \; \frac{\theta(\tau) - \varepsilon(\tau)}{\sqrt{2}} \,.
  \label{eq:theta_eps_S}
\end{equation}
%
From the transformation properties \eqref{eq:theta_eps_T} and
\eqref{eq:theta_eps_S}, one sees that $\theta$ and $\varepsilon$ work
as ``weight 1/2'' modular forms. Their even powers produce integer
weight modular forms, which we consider in the following subsection.

%
\subsection{Weight 1 Modular Forms}
%

We have seen that the linear space of weight 1 modular forms of level 4
is spanned by three quadratic monomials in $\theta(\tau)$ and
$\varepsilon(\tau)$, namely:
\begin{equation}
  \theta(\tau)^2\,, \quad
  \theta(\tau) \varepsilon(\tau)\,, \quad
  \varepsilon(\tau)^2\,,
\end{equation}
%
such that the linear space of weight $k=1$ has the correct dimension, $2k+1=3$.

These three functions can be arranged into a triplet furnishing a
representation of $S_4' \equiv SL(2, \mathbb{Z}_4)$, which is a double cover%
\footnote{Strictly speaking, the term ``double cover of symmetric group''
is used for a special kind of a double cover called the Schur cover. There are
two double covers of $S_4$ of this kind: the binary octahedral group
(group ID \texttt{[48,28]} in GAP~\cite{GAP4,SmallGroups}) and
$GL(2, 3)$ (group ID \texttt{[48,29]}). Our double cover
$SL(2, \mathbb{Z}_4)$ is not a Schur cover of $S_4$.
It has group ID \texttt{[48,30]}, hence it is a double cover of $S_4$ in
a broader sense.}
of the permutation group $S_4$~\cite{Liu:2019khw}.
We summarise the group theory of $S_4'$ in Appendix~\ref{app:S4p}.

In the group representation basis of Table~\ref{tab:basis}, the relevant
triplet has the form
\begin{equation}
  Y_{\mathbf{\hat{3}}}^{(1)}(\tau) =
  \begin{pmatrix}
    \sqrt{2} \, \varepsilon \, \theta \\[1mm]
    \varepsilon^2 \\[1mm]
    -\theta^2
  \end{pmatrix}
  \label{eq:k1triplet}
\end{equation}
%
and furnishes an irreducible representation $\mathbf{\hat{3}}$.
Indeed, using the transformation rules~\eqref{eq:theta_eps_T},
\eqref{eq:theta_eps_S} it is easy to check that the triplet~
\eqref{eq:k1triplet} transforms under the generators of the modular group as
expected:
\begin{equation}
\renewcommand{\arraystretch}{1.5}
\begin{array}{rclclr}
    Y_{\mathbf{\hat{3}}}^{(1)}(\tau) \quad& \xrightarrow{T} \quad& Y_{\mathbf{\hat{3}}}^{(1)}(\tau + 1) & \!= & & \rho_{\mathbf{\hat{3}}}(T) \;Y_{\mathbf{\hat{3}}}^{(1)}(\tau) \,, \\
    Y_{\mathbf{\hat{3}}}^{(1)}(\tau) & \xrightarrow{S} & Y_{\mathbf{\hat{3}}}^{(1)}(-1 / \tau) & \!= & \!\!(-\tau) \!\!\!\!\!& \rho_{\mathbf{\hat{3}}}(S) \; Y_{\mathbf{\hat{3}}}^{(1)}(\tau) \,, \\
    Y_{\mathbf{\hat{3}}}^{(1)}(\tau) & \xrightarrow{R} & Y_{\mathbf{\hat{3}}}^{(1)}(\tau) & \!= & \!\!(-1)\!\!\!\!\! & \rho_{\mathbf{\hat{3}}}(R) \; Y_{\mathbf{\hat{3}}}^{(1)}(\tau) \,.
  \end{array}
\end{equation}
%
The $\mathbf{\hat{3}}$ modular triplet of eq.~\eqref{eq:k1triplet}
is the base result of our construction. It can be used to generate
all modular forms entering and determining the fermion Yukawa couplings
and mass matrices, as we will see in what follows.

%
\subsection{Modular Forms of Higher Weights}
%

Modular multiplets of higher weights $Y^{(k>1)}_{\mathbf{r},\,\mu}$
may be obtained from those of lower weight via tensor products.
Here, the index $\mu$ labels linearly independent multiplets (in case more
than one is present) for a given weight $k$ and irreducible $S_4'$
representation $\mathbf{r}$.
The lowest weight multiplet in eq.~\eqref{eq:k1triplet} works then as
a `seed' multiplet, since all higher weight modular multiplets can be
recovered from a sufficient number of tensor products of
$Y_{\mathbf{\hat{3}}}^{(1)}(\tau)$ with itself.
Note that the latter has been written in terms of a minimal set of functions
of $\tau$ from the start, namely $\theta(\tau)$ and $\varepsilon(\tau)$.
By doing so, tensor products directly provide spaces of modular forms
with the correct dimensions, bypassing the typical need to look for
constraints relating redundant higher weight multiplets. In other words,
these constraints are manifestly verified given the explicit forms of the
multiplet components.

First of all, we recover the known~\cite{Penedo:2018nmg} modular $S_4$
lowest-weight multiplets, a doublet and a triplet($^\prime$), which are now
expressed in terms of $\theta(\tau)$ and $\varepsilon(\tau)$ and read
\begin{equation}
  Y_{\mathbf{2}}^{(2)}(\tau) =
  \begin{pmatrix}
    \frac{1}{\sqrt{2}}\left(\theta^4 + \varepsilon^4\right) \\[1mm]
    -\sqrt{6}\, \varepsilon^2\, \theta^2
  \end{pmatrix}\,,
  \qquad
    Y_{\mathbf{3'}}^{(2)}(\tau) =
  \begin{pmatrix}
    \frac{1}{\sqrt{2}}\left(\theta^4 - \varepsilon^4\right) \\[1mm]
    -2 \,\varepsilon \, \theta^3\\[1mm]
    -2 \,\varepsilon^3 \, \theta
  \end{pmatrix}\,.
  \label{eq:k2}
\end{equation}
%
Our construction reduces to that of modular $\Gamma_4 \simeq S_4$ for even
weights (see also Appendix~\ref{app:propirreps}).
In order to compare the results in eq.~\eqref{eq:k2} with those of
Ref.~\cite{Penedo:2018nmg}, one needs to work in compatible group
representation bases, i.e.~bases in which the representation matrices
$\rho_\mathbf{r}(S)$ and $\rho_\mathbf{r}(T)$ coincide, for irreducible representations
$\mathbf{r}$ common to $S_4$ and $S_4'$ (those without hats). The basis
for $S_4$ compatible with the one for $S_4'$ we here consider, together with
the expressions for modular multiplets in that basis, can be found in
Ref.~\cite{Novichkov:2019sqv} (see Appendices B and C therein).
Then, by looking at the $q$-expansions,
\begin{equation}
\begin{aligned}
  Y_{\mathbf{2}}^{(2)}(\tau) &=
  \begin{pmatrix*}[l]
  	\:\:\:\:\:\:\frac{1}{\sqrt{2}}\left(1+24\,q_4^{4}+24\,q_4^{8}+96\,q_4^{12}+24\,q_4^{16}+144\,q_4^{20}+96\,q_4^{24} + \ldots\right)\\[1mm]
  	-4\sqrt{6}\left(
q_4^{2}+4\,q_4^{6}+6\,q_4^{10}+8\,q_4^{14}+13\,q_4^{18}+12\,q_4^{22} + \ldots\right)
  \end{pmatrix*}\,,
\\[2mm]
    Y_{\mathbf{3'}}^{(2)}(\tau) &=
  \begin{pmatrix*}[l]
\:\:\,\frac{1}{\sqrt{2}}\left(1-8\,q_4^{4}+24\,q_4^{8}-32\,q_4^{12}+24\,q_4^{16}-48\,q_4^{20}+96\,q_4^{24} +\ldots \right) \\[1mm]
\:\:\,-4\left(q_4+6\,q_4^{5}+13\,q_4^{9}+14\,q_4^{13}+18\,q_4^{17}+32\,q_4^{21}+31\,q_4^{25}+\ldots \right) \\[1mm]
\,-16\left(q_4^{3}+2\,q_4^{7}+3\,q_4^{11}+6\,q_4^{15}+5\,q_4^{19}+6\,q_4^{23}+\ldots\right)
  \end{pmatrix*}\,,
\end{aligned}
\label{eq:k2q}
\end{equation}
%
one can see that the modular multiplets in question indeed match,
up to normalisation.

Further tensor products with $Y_{\mathbf{\hat{3}}}^{(1)}$ produce new modular
multiplets of odd weight. At weight $k=3$, a non-trivial singlet and two
triplets exclusive to $S_4'$ arise:
\begin{equation}
\begin{aligned}
  Y_{\mathbf{\hat{1}'}}^{(3)}(\tau) &= \sqrt{3} \left(
  \varepsilon \,\theta^5-\varepsilon^5\, \theta
  \right)\,,
  \\[2mm]
  Y_{\mathbf{\hat{3}}}^{(3)}(\tau) &=
  \begin{pmatrix}
 \varepsilon ^5\, \theta +\varepsilon \, \theta ^5\\[1mm]
 \frac{1}{2\sqrt{2}}\left(5 \,\varepsilon ^2 \, \theta ^4-\varepsilon ^6 \right)\\[1mm]
 \frac{1}{2\sqrt{2}}\left(\theta ^6-5\, \varepsilon ^4 \,\theta ^2\right)
  \end{pmatrix}\,,\quad
    Y_{\mathbf{\hat{3}'}}^{(3)}(\tau) = \frac{1}{2}
  \begin{pmatrix}
 -4 \sqrt{2}\, \varepsilon ^3 \,\theta ^3 \\[1mm]
 \theta ^6 + 3 \,\varepsilon ^4\, \theta ^2 \\[1mm]
 - 3\, \varepsilon ^2\, \theta ^4 -\varepsilon ^6
  \end{pmatrix}\,.
\end{aligned}
\label{eq:k3}
\end{equation}
%

Finally, at weight $k=4$ one again recovers the $S_4$ result. We obtain:
\begin{equation}
\begin{aligned}
  Y_{\mathbf{1}}^{(4)}(\tau) &=
  \frac{1}{2 \sqrt{3}} \left(
  \theta^8 + 14\, \varepsilon^4\, \theta^4 + \varepsilon^8
  \right)\,,\quad
  Y_{\mathbf{2}}^{(4)}(\tau) =
  \begin{pmatrix}
  \frac{1}{4} \left(\theta^8 - 10 \,\varepsilon^4\, \theta^4 + \varepsilon^8\right) \\[1mm]
  \sqrt{3}\left(\varepsilon ^2\,\theta^6 + \varepsilon ^6\, \theta^2\right)
  \end{pmatrix}\,,
  \\[2mm]
  Y_{\mathbf{3}}^{(4)}(\tau) &=
  \frac{3}{2\sqrt{2}}
  \begin{pmatrix}
 \sqrt{2}\left(\varepsilon ^2\, \theta^6 -\varepsilon^6 \, \theta ^2\right)\\[1mm]
 \varepsilon ^3 \,\theta ^5 - \varepsilon ^7 \,\theta \\[1mm]
 - \varepsilon  \,\theta ^7 + \varepsilon ^5 \,\theta ^3
  \end{pmatrix}\,,\quad
  Y_{\mathbf{3'}}^{(4)}(\tau) =
  \begin{pmatrix}
  \frac{1}{4}\left(\theta ^8-\varepsilon ^8\right)\\[1mm]
  \frac{1}{2\sqrt{2}}\left(\varepsilon \, \theta ^7 + 7 \,\varepsilon ^5\, \theta ^3\right) \\[1mm]
  \frac{1}{2\sqrt{2}}\left(7 \,\varepsilon ^3\, \theta ^5 + \varepsilon ^7 \,\theta\right)
  \end{pmatrix}\,,
\end{aligned}
\label{eq:k4}
\end{equation}
%
which can be seen to match known multiplets (up to normalisation) by
comparing $q$-expansions.
We collect the explicit expressions of $S_4'$ modular multiplets with
higher weights, up to $k=10$ and written in terms of $\theta(\tau)$ and
$\varepsilon(\tau)$, in Appendix~\ref{app:multiplets}.
Note that odd(even)-weighted modular forms always furnish (un)hatted representations,
since in our notation hatted representations are exactly the ones for which~$\rho(R) = -\id$.

%
\section{Combining gCP and Modular Symmetries}
\label{sec:gCP}
%

In models possessing a flavour symmetry, one can define a \textit{generalised CP (gCP) transformation} acting on the matter fields as
\begin{equation}
  \psi_i (x) \xrightarrow{\text{CP}} X_{ij} \, \overline{\psi}_j (x_P) \,,
  \label{eq:psi_CP}
\end{equation}
with a bar denoting the conjugate field, and where $x = (t, \mathbf{x})$, $x_P = (t, -\mathbf{x})$ and $X$ is a unitary matrix acting on flavour space.
Modular symmetry, which plays the role of a flavour symmetry, can be consistently combined with a generalised CP symmetry.
This has been done from a bottom-up perspective in~\cite{Novichkov:2019sqv} for the inhomogeneous modular group $\overline \Gamma$.
The result of~\cite{Novichkov:2019sqv} can be generalised to the case of the full modular group $\Gamma$ as follows.

Starting with eq.~\eqref{eq:psi_CP} one can show that the modulus $\tau$ should transform under CP as
\begin{equation}
    \tau \xrightarrow{\text{CP}} -\tau^* \,,
    \label{eq:tau_CP}
\end{equation}
%
without loss of generality (cf.~Ref.~\cite{Novichkov:2019sqv}).
The corresponding action on the modular group $\Gamma$ is given by an outer automorphism $u(\gamma) \equiv \text{CP} \circ \gamma \circ \text{CP}^{-1}$.
The form of $u(\gamma)$ is determined by eq.~\eqref{eq:tau_CP}: for a transformation $\gamma = \begin{psmallmatrix}
    a & b \\
    c & d
  \end{psmallmatrix} \in \Gamma$ one has the chain
\begin{equation}
  \tau \,\xrightarrow{\text{CP}}\, -\tau^{*}
  \,\xrightarrow{\gamma}\, - \frac{a\tau^{*} + b}{c\tau^{*} + d}
  \,\xrightarrow{\text{CP}^{-1}}\, \frac{a\tau - b}{-c\tau + d} = u(\gamma) \tau \,,
\end{equation}
%
which implies
\begin{equation}
  \label{eq:mod_CP}
  u(\gamma) \,=\, \sigma(\gamma)
  \begin{pmatrix}
    a & -b \\
    -c & d
  \end{pmatrix} \in \Gamma \,,
\end{equation}
where $\sigma(\gamma) = \pm 1$.
Note that the signs $\sigma(\gamma)$ are irrelevant in the case of the inhomogeneous modular group $\overline{\Gamma}$ since $\gamma$ is identified with $-\gamma$, and therefore eq.~\eqref{eq:mod_CP} uniquely determines the automorphism $u(\gamma)$.
This is no longer the case for the full modular group $\Gamma$, and one has to treat the signs carefully.

Since $u$ is an automorphism, it is sufficient to define its action on the group generators.
From eq.~\eqref{eq:mod_CP} one has:
\begin{equation}
  u(S) = \sigma(S)\, S^{-1} \,, \qquad
  u(T) = \sigma(T)\, T^{-1} \,, \qquad
  u(R) = \sigma(R)\, R \,.
\end{equation}
The fact that $u(\gamma)$ is an automorphism implies $u(R) \neq \id = -R$, and so $\sigma(R)=+1$ and $u(R) = + R$. Furthermore, the signs $\sigma(\gamma)$ must be chosen in a way consistent with the group relations in~\eqref{eq:STR_rel}. In particular, one finds:
\begin{equation}
  (ST)^3=
  \id \,\,\xrightarrow{u} \,\, (\sigma(S)\sigma(T))^3 \, (TS)^{-3} = \id\,,
\end{equation}
implying that $\sigma(S) = \sigma(T)$, since $(TS)^3 = \id$.
Thus, from the outset, two different outer automorphisms may be realised (see also~\cite{Hua:1951}):
\begin{align}
  \label{eq:mod_CP_gens_1}
  (\text{CP}_1) \,\,u\hphantom{'}&: &u(S) = S^{-1},    &  &u(T) = T^{-1},    &  &u(R) = R \,, \\
  \label{eq:mod_CP_gens_2}
  (\text{CP}_2) \,\, u'&:  &u'(S) = -S^{-1},  &  &u'(T) = -T^{-1}, &  &u'(R) = R \,.
\end{align}
We note that $S^{-1} = -\,S$.

\subsection{\texorpdfstring{CP$_1$}{CP1}}

The first option~\eqref{eq:mod_CP_gens_1}, which we call CP$_1$, corresponds to a trivial sign choice $\sigma(\gamma) = +1$ and therefore admits an explicit formula for generic $\gamma$:
\begin{equation}
  \label{eq:mod_CP_1}
  u :
  \begin{pmatrix}
    a & b \\ c & d
  \end{pmatrix} \to
  \begin{pmatrix}
    a & -b \\
    -c & d
  \end{pmatrix} \,.
\end{equation}
This automorphism can be realised as a similarity transformation within $GL(2,\mathbb{Z})$:
\begin{equation}
  u(\gamma) = \text{CP}_1 \, \gamma \, \text{CP}_1^{-1}
  \quad \text{with} \quad \text{CP}_1 =
  \begin{pmatrix}
    1 & 0 \\ 0 & -1
  \end{pmatrix}
  \notin \Gamma \,.
\end{equation}
Applying the chain $\text{CP}_1 \to  \gamma  \to \text{CP}_1^{-1}$ to the matter field~$\psi$, which transforms under $\Gamma$ and CP as in eqs.~\eqref{eq:psi_mod_trans} and~\eqref{eq:psi_CP}, one arrives at the \textit{gCP consistency condition} on the matrix~$X$:
\begin{equation}
  \label{eq:cons_CP_1}
  X \, \rho^{*}(\gamma) \, X^{-1} = \rho(u(\gamma)) \quad \forall \, \gamma \in \Gamma \,,
\end{equation}
or, equivalently,
\begin{equation}
  \label{eq:cons_CP_1_gens}
  X \, \rho^{*}(S) \, X^{-1} = \rho^{-1}(S) \,, \quad
  X \, \rho^{*}(T) \, X^{-1} = \rho^{-1}(T) \,
\end{equation}
(see also~\cite{Nilles:2020nnc}),
which coincide with the corresponding expressions in the case of $\overline{\Gamma}$~\cite{Novichkov:2019sqv}.

In a basis where $S$ and $T$ are represented by symmetric matrices, eq.~\eqref{eq:cons_CP_1_gens} is satisfied by the canonical CP transformation $X = \id$~\cite{Novichkov:2019sqv}.
Such a basis exists for all irreducible representations of the inhomogeneous finite modular groups $\Gamma_N$ with $N = 2, 3, 4, 5$ (see~\cite{Novichkov:2019sqv} and references therein) and $N = 7$~\cite{Ding:2020msi}, as well as for all irreps of the homogeneous modular groups $\Gamma'_3$ and $\Gamma'_4$ (see Appendix~\ref{app:sym_basis}).%
\footnote{
One can obtain a symmetric basis for $\Gamma'_3$ starting from the one typically considered in the literature~\cite{Liu:2019khw}
and performing a change of basis for all 2-dimensional irreps via the matrix $\diag(e^{-7i\pi/12}, 1)$.}
This means that CP$_1$ allows to define a CP transformation consistently and uniquely for all irreps of the aforementioned finite modular groups, hence $u$ acts as a class-inverting automorphism on these groups~\cite{Chen:2014tpa}.%
\footnote{Note however that, at the level of the full modular group, $u$ is not class-inverting. Taking for instance $\gamma = \begin{psmallmatrix} 11 & 9 \\ 17 & 14\end{psmallmatrix}$, one can show that $u(\gamma)$ and $\gamma^{-1}$ are not in the same $SL(2,\mathbb{Z})$ conjugacy class, via e.g.~the LLS invariant of Ref.~\cite{Karpenkov2013}.}

The action of CP$_1$ on fields (and $\tau$) obeys $\text{CP}_1^2 = \id$, since 
$\psi_i(x) \xrightarrow{\text{CP}^2} (XX^*)_{ij} \, \psi_j(x)$ and
$X = \id \Rightarrow XX^*=\id$ in the symmetric basis. It further follows that $X$ is symmetric in any representation basis~\cite{Novichkov:2019sqv}.
The modular group $\Gamma = SL(2,\mathbb{Z})$ is then extended to
\begin{equation}
\begin{aligned}
  \label{eq:mod_CP_1_presentation}
GL(2,\mathbb{Z}) \,&\simeq\, 
SL(2,\mathbb{Z}) \rtimes \mathbb{Z}_2^{\text{CP}_1}\\
\,&=\, \big\langle S,\, T,\, R,\, \text{CP}_1\, \big|\, 
  S^2 = R, \,
  (ST)^3 = R^2 = \text{CP}_1^2 = \id, \,
  RT = TR,\\
  &\qquad\qquad\qquad\qquad\,\,
  \text{CP}_1 \,S \,\text{CP}_1^{-1} = S^{-1},\,
  \text{CP}_1 \,T \,\text{CP}_1^{-1} = T^{-1}
  \big\rangle\,.
\end{aligned}
\end{equation}

Finally, in a basis where $S$ and $T$ are symmetric, 
where Clebsch-Gordan coefficients are real
and with modular multiplets normalised to satisfy $Y(-\tau^*)=Y^*(\tau)$,%
\footnote{
It is possible to meet these conditions for the aforementioned homogeneous and inhomogeneous finite modular groups.
In Section 3.4 of Ref.~\cite{Novichkov:2019sqv} it is shown that the choice $Y(-\tau^*)= Y^*(\tau)$ is possible if Clebsch-Gordan coefficients are real and one has at most one copy of each irrep at lowest weight.
While for $\Gamma_7$ this last condition is not met (cf.~Ref.~\cite{Ding:2020msi}), one can check that the modular multiplets also satisfy $Y(-\tau^*)= Y^*(\tau)$ in the appropriate basis.
}
the requirement of CP$_1$ invariance reduces to reality of the couplings~\cite{Novichkov:2019sqv}, i.e.~of the numerical coefficients in front of the independent singlets in eq.~\eqref{eq:W_series}.
In such theories, CP symmetry is broken spontaneously by the VEV of the modulus~$\tau$, thus providing a common origin of CP and flavour symmetry violation.
We will make use of CP$_1$ in the upcoming phenomenological examples of Section~\ref{sec:pheno}.

\subsection{\texorpdfstring{CP$_2$}{CP2}}

Let us now discuss the second possibility~\eqref{eq:mod_CP_gens_2} for the modular group outer automorphism, $u'$.
This choice, which we call CP$_2$, is formally defined by
\begin{equation}
  u'(\gamma) = \text{CP}_2 \, \gamma \, \text{CP}_2^{-1}\,,
\end{equation}
but cannot be realised as a similarity transformation within $GL(2,\mathbb{Z})$.
It leads to a different consistency condition on the matrix~$X$, namely:
\begin{equation}
  \label{eq:cons_CP_2}
  X \, \rho^{*}(\gamma) \, X^{-1} = \sigma(\gamma)^k \,\rho(u'(\gamma)) \quad \forall \, \gamma \in \Gamma \,,
\end{equation}
or, in terms of the generators $S$ and $T$,
\begin{equation}
  \label{eq:cons_CP_2_gens}
  X \, \rho^{*}(S) \, X^{-1} = (-1)^k \, \rho(R) \, \rho^{-1}(S) \,, \quad
  X \, \rho^{*}(T) \, X^{-1} = (-1)^k \, \rho(R) \, \rho^{-1}(T) \,,
\end{equation}
which are equivalent to \eqref{eq:cons_CP_2}, since $\sigma(\gamma_1)\sigma(\gamma_2)=\sigma(\gamma_1\gamma_2)$.

In practice, the consistency condition~\eqref{eq:cons_CP_2_gens} differs from that of eq.~\eqref{eq:cons_CP_1_gens}
and CP$_2$ differs from CP$_1$ only when
$(-1)^k \, \rho(R) \neq \id$, i.e.~whenever the matter field~$\psi$ transforms non-trivially under~$R$.
For these $R$-odd fields, however,
it is only possible to satisfy the consistency condition 
if
\begin{itemize}
    \item[i)] both the characters of $T$ and $S$ vanish, $\chi(S)=\chi(T)=0$, which follows from eq.~\eqref{eq:cons_CP_2_gens} after taking traces,
    \item[ii)] the dimension of the representation of $\psi$ is even, which follows from eq.~\eqref{eq:cons_CP_2_gens} after taking determinants, and
    \item[iii)] the level $N$ of the finite group is even, which follows from taking the $N$-th power of the second relation in eq.~\eqref{eq:cons_CP_2_gens}.%
    \footnote{An associated fact is that $\Gamma(N)$ with $N\geq 2$ is only stable under $u'$ for even $N$.}
\end{itemize}
This means that, given a finite modular group of level $N$, CP$_2$ is incompatible with certain combinations of modular weights and irreps.

In particular, combining the groups~$\Gamma_N$ with $N = 3, 5, 7$ and $\Gamma'_3$ with CP$_2$ means that any matter field must be $R$-even, i.e.~satisfy $(-1)^k \, \rho(R) = \id$, and transform canonically under CP, $X_{\text{CP}_2} = \id$, in the symmetric basis.
In the case of $\Gamma_2$, $\Gamma_4$ and $\Gamma'_4$ there is the additional option to have $R$-odd fields, $(-1)^k \, \rho(R) = -\id$, but only for the doublet representations, all of which verify $\chi(S)=\chi(T)=0$. These fields are constrained to transform under CP with
\begin{equation}
X_{\text{CP}_2} = \begin{pmatrix} 0 & 1 \\ -1 & 0 \end{pmatrix}
\end{equation}
in the symmetric basis. Notice that $\text{CP}_2^2 \neq \id$. Instead, the action of $\text{CP}_2^2$ on fields, forms and $\tau$ coincides with that of $R$ for these finite groups.
Equating these two actions, the modular group is in this context minimally extended to the semidirect product%
\footnote{The non-trivial automorphism defining this outer semidirect product is $\gamma \mapsto \text{CP}_2\,S\,\gamma\,S^{-1}\,\text{CP}_2^{-1}$.}
\begin{equation}
  \label{eq:mod_CP_2_presentation}
\begin{aligned}
SL(2,\mathbb{Z})\rtimes \mathbb{Z}_2^{\text{CP}_2\,S}
= \big\langle S,\, T,\, R,\, \text{CP}_2\, \big|\, 
  &S^2 = \text{CP}_2^2 = R, \,
  (ST)^3 = R^2 = \id, \,
  RT = TR,\\
  &\text{CP}_2 \,S = S\,\text{CP}_2,\,
  \text{CP}_2 \,T \,\text{CP}_2^{-1} = R\,T^{-1}
  \big\rangle\,.
\end{aligned}
\end{equation}

Keeping our focus on $\Gamma_2$, $\Gamma_4$, and $\Gamma_4'$,
with their respective symmetric bases and Clebsch-Gordan coefficients
given in Ref.~\cite{Novichkov:2019sqv} and Appendix~\ref{app:S4p},
let us briefly comment on the consequences of implementing CP$_2$
for the couplings in the superpotential $W$. We start by writing the latter as
a sum of independent singlets,
\begin{equation}
 W \supset \sum_s g_s \left( Y_s(\tau) \,\psi_1 \ldots \psi_n \right)_{\mathbf{1},s} \, ,
  \label{eq:Wterm}
\end{equation}
where $Y_s(\tau)$ are modular multiplets of a certain weight and irrep,
and $g_s$ are complex coupling constants.
To be non-vanishing, each term must contain an even number of $R$-odd fields $\psi_i$, if any,
which are in doublet representations of the finite groups at hand. 
Taking $\psi_i$ to be $R$-odd for $i\leq 2m$ and $R$-even for $i > 2m$, with $m$ being a non-negative integer such that $2 m\leq n$,
one can explicitly check that
\begin{equation}
\begin{aligned}
  &g_s \left( Y_s(\tau)\, \psi_1 \ldots \psi_{2m}\,\psi_{2m+1} \ldots \psi_n \right)_{\mathbf{1},s} \\
  \xrightarrow{\text{CP}_2}\,\,
  &g_s \left( Y_s(-\tau^{*}) 
  \left(X_{\text{CP}_2}\overline\psi_1\right)  \ldots 
  \left(X_{\text{CP}_2} \overline\psi_{2k}\right) \overline\psi_{2m+1} \ldots  \overline\psi_n \right)_{\mathbf{1},s} 
  \\
  =\,&g_s \,\overline{\left( Y_s(\tau) \left(X_{\text{CP}_2}\psi_1\right)  \ldots 
  \left(X_{\text{CP}_2} \psi_{2m} \right) \psi_{2m+1} \ldots  \psi_n\right)_{\mathbf{1},s}} 
  \\
  =&\pm \,g_s \,\overline{\left( Y_s(\tau)\, \psi_1  \ldots \psi_n\right)_{\mathbf{1},s}} \, ,
\end{aligned}
\end{equation}
where we have used the reality and symmetry properties of the Clebsch-Gordan coefficients.%
\footnote{
For each pair of $R$-odd doublets,
$(X_{\text{CP}_2} \psi_i \otimes X_{\text{CP}_2} \psi_j)_{\mathbf{r}} = \pm
(\psi_i \otimes \psi_j)_{\mathbf{r}}$,
where the sign depends on $\mathbf{r}$.
}
Under CP$_2$, a term in eq.~\eqref{eq:Wterm} transforms into the conjugate of
\begin{equation}
\pm\,  g_s^{*} \left( Y_s(\tau) \psi_1 \ldots \psi_n \right)_{\mathbf{1},s} \, ,
\end{equation}
which should coincide with the original term.
The independence of singlets then implies the constraint $g_s = \pm\,g_s^{*}$, meaning that all coupling constants $g_s$ have to be real or purely imaginary (depending on the sign) to conserve CP.

It should be noted that it is difficult to build phenomenologically viable models of fermion masses and mixing exploiting the novelty of CP$_2$ with $R$-odd fields, as i) the choice of irreps for such fields is quite limited and ii) the $R$-odd and $R$-even sectors are segregated by the $\mathbb{Z}_2^R$ symmetry. Taken together, these facts imply the vanishing of some mixing angles or masses in simple models based on the combination of the novel CP$_2$ with $\Gamma_2$, $\Gamma_4$, or $\Gamma_4'$. We will not pursue this model-building avenue in what follows.

%
\section{Residual Symmetries}
\label{sec:residual}
%
%

Modular symmetry is spontaneously broken by the VEV of the modulus $\tau$: in fact, there is no value of $\tau$ which is left invariant by the modular group action~\eqref{eq:tau_mod_trans}.
However, certain values of $\tau$ (called \textit{symmetric} or \textit{fixed points}) break the modular group $\Gamma$ only partially, with the unbroken generators giving rise to \textit{residual symmetries}~\cite{Novichkov:2018ovf}.
These unbroken symmetries can play an important role in flavour model building~\cite{Novichkov:2018yse, Gui-JunDing:2019wap}.

\begin{figure}[ht!]
  \centering
  \includegraphics[width=0.5\textwidth]{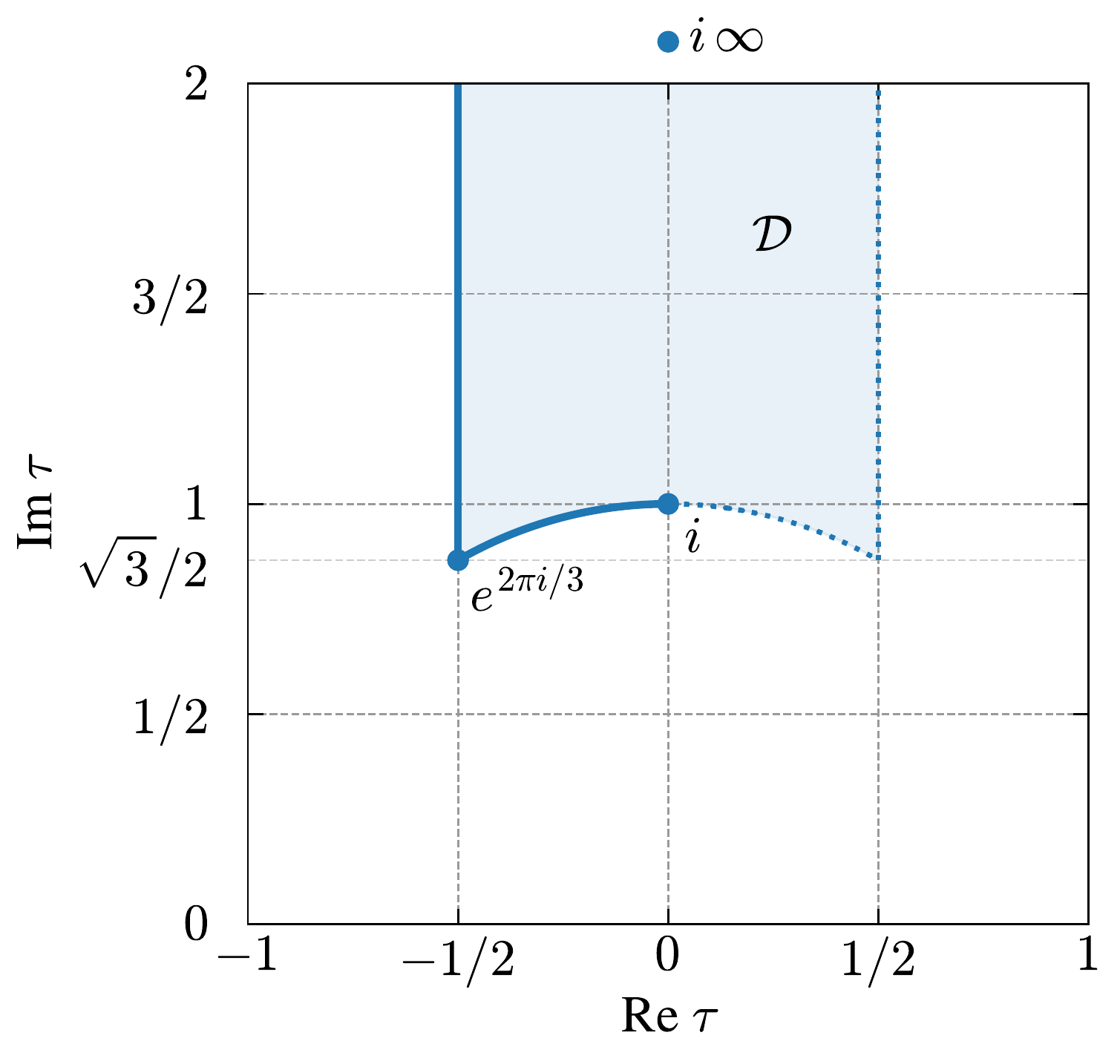}
  \caption{The fundamental domain~$\mathcal{D}$ of the modular group~$\Gamma$, and its three symmetric points $\tau_C = i$, $\tau_L = e^{2\pi i / 3}$ and $\tau_T = i\, \infty$.}
  \label{fig:fund_domain}
\end{figure}

To classify the possible residual symmetries, one first notices that with a proper ``gauge choice'' $\tau$ can always be restricted to the \textit{fundamental domain}~$\mathcal{D}$ of the modular group~$\Gamma$:
\begin{equation}
  \label{eq:fund_domain}
  \mathcal{D} \equiv \left\{ \tau \in \mathcal{H} : -\frac{1}{2} \leq \re \tau < \frac{1}{2},\, |\tau| > 1 \right\} \cup \left\{ \tau \in \mathcal{H} : -\frac{1}{2} < \re \tau \leq 0,\, |\tau| = 1 \right\} \,,
\end{equation}
%
which describes all possible values of $\tau$ up to a modular transformation (see Fig.~\ref{fig:fund_domain}).
Note that, by convention, the right half of the boundary~$\partial \mathcal{D}$ is not included into~$\mathcal{D}$, since it is related to the left half by suitable modular transformations.

In the fundamental domain~$\mathcal{D}$, there exist only three symmetric points, namely~\cite{Novichkov:2018ovf}:
\begin{itemize}
\item[i)] $\tau_C \equiv i$ invariant under $S$;
\item[ii)] $\tau_L \equiv -\,1/2 +i\,\sqrt{3}/2$ (``the left cusp'') invariant under $ST$;
\item[iii)] $\tau_T \equiv i \infty$ invariant under $T$.
\end{itemize}
In addition, the $R$ generator is unbroken for any value of $\tau$.
Finally, if a theory is also CP-invariant (i.e.~its couplings satisfy the constraints discussed in Section~\ref{sec:gCP}), then the CP symmetry is spontaneously broken by any $\tau \in \mathcal{D}$ except for the values lying on the fundamental domain boundary or the imaginary axis~\cite{Novichkov:2019sqv}:
\begin{itemize}
\item[i)] $\re \tau = 0$ (the imaginary axis) is invariant under $\text{CP}$;
\item[ii)] $\re \tau = -1/2$ (the left vertical boundary) is invariant under $\text{CP}\, T$;
\item[iii)] $|\tau| = 1$ (the boundary arc) is invariant under $\text{CP}\, S$.
\end{itemize}
Recall that CP always acts on $\tau$ as in eq.~\eqref{eq:tau_CP}, meaning the above statement does not depend on the choice of CP automorphism (CP$_1$ vs.~CP$_2$).

For a given value of~$\tau$, the residual symmetry group is simply a group generated by the unbroken transformations subject to relations which can be deduced from eqs.~\eqref{eq:mod_CP_1_presentation},~\eqref{eq:mod_CP_2_presentation}.
For instance, the symmetric point $\tau = i$ is invariant under $S$, $R$ and $\text{CP}_1$ in the case of the full modular group~$\Gamma$ enhanced by $\text{CP}_1$.
The corresponding symmetry group is
\begin{equation}
\left\langle S,\, R,\, \text{CP}_1 \right\rangle = \left\langle S,\, \text{CP}_1 \, \middle\vert \, S^4 = \id, \, \text{CP}_1^2 = \id, \, \text{CP}_1 \, S \, \text{CP}_1^{-1} = S^{-1} \right\rangle \simeq D_4 \,,
\end{equation}
where $D_4$ is the dihedral group of order 8 (the symmetry group of a square).
One can find the residual symmetry groups for other values of~$\tau$ in a similar fashion; we collect the results in Table~\ref{tab:residual}.

When considering finite modular versions $\Gamma_N^{(\prime)}$ of the modular group, the residual symmetry groups may be reduced, due to the extra relation $T^N = \id$ (recall that for $N > 5$ further constraints are present). For $N \leq 5$, the instances of $\mathbb{Z}^T$ in Table~\ref{tab:residual} should be replaced by $\mathbb{Z}_N^T$. 

Since every symmetric point outside the fundamental domain~$\mathcal{D}$ is physically equivalent to a symmetric point inside~$\mathcal{D}$, its residual symmetry group is isomorphic to one of the groups listed in Table~\ref{tab:residual}.
For instance, ``the right cusp'' $\tau_R \equiv 1/2 + i\,\sqrt{3}/2$ is related to the left cusp as $\tau_R = T \, \tau_L$, so the residual symmetry group at~$\tau_R$ is isomorphic to that at~$\tau_L$, and the isomorphism is given by a conjugation with~$T^{-1}$.
In particular, the unbroken generators are mapped as $ST \to T (ST) T^{-1} = TS$, $R \to T R T^{-1} = R$ and $\text{CP} \, T \to T (\text{CP} \, T) T^{-1} = T\, \text{CP}$.

\begin{table}
  \footnotesize
  \renewcommand{\arraystretch}{1.8}
  \centering
  \begin{tabular}{rccccc}
    \toprule
    & $\Gamma$ & $\overline{\Gamma}$ & $\Gamma \rtimes \text{CP}_1$ & $\Gamma \rtimes \text{CP}_2$ & $\overline{\Gamma} \rtimes \text{CP}$\\
    \midrule
    $\tau = i$ & $\mathbb{Z}_4^S$ & $\mathbb{Z}_2^S$ & $\mathbb{Z}_4^S \rtimes \mathbb{Z}_2^{\text{CP}_1} \simeq D_4$ & $\mathbb{Z}_4^S \times \mathbb{Z}_2^{\text{CP}_2  S}$ & $\mathbb{Z}_2^S \times \mathbb{Z}_2^{\text{CP}}$\\[3mm]
    $\tau = e^{2\pi i /3}$ & $\mathbb{Z}_3^{ST} \times \mathbb{Z}_2^R$ & $\mathbb{Z}_3^{ST}$ &
    \makecell{$\left(\mathbb{Z}_3^{ST} \rtimes \mathbb{Z}_2^{\text{CP}_1  T}\right) \times \mathbb{Z}_2^R$ \\ $\simeq S_3 \times \mathbb{Z}_2 \simeq D_6$}
    & 
    \makecell{$\left(\mathbb{Z}_3^{ST} \rtimes \mathbb{Z}_2^{\text{CP}_2  T}\right) \times \mathbb{Z}_2^R$ \\ $\simeq S_3 \times \mathbb{Z}_2 \simeq D_6$}
    &
    \makecell{$\mathbb{Z}_3^{ST} \rtimes \mathbb{Z}_2^{\text{CP} \, T}$ \\ $\simeq S_3$}
    \\[3mm]
    $\tau = i \infty$ & $\mathbb{Z}^T \times \mathbb{Z}_2^R$ & $\mathbb{Z}^T$ & $\left( \mathbb{Z}^T \rtimes \mathbb{Z}_2^{\text{CP}_1} \right) \times \mathbb{Z}_2^R$ & $\left( \mathbb{Z}^T \times \mathbb{Z}_2^R \right) \rtimes \mathbb{Z}_2^{\text{CP}_2  T}$ & $\mathbb{Z}^T \rtimes \mathbb{Z}_2^{\text{CP}}$ \\[3mm]
    $\re \tau = 0$ & $\mathbb{Z}_2^R$ & $1$ & $\mathbb{Z}_2^{\text{CP}_1} \times \mathbb{Z}_2^R$ & $\mathbb{Z}_4^{\text{CP}_2}$ & $\mathbb{Z}_2^{\text{CP}}$ \\[3mm]
    $|\tau| = 1$ & $\mathbb{Z}_2^R$ & $1$ & $\mathbb{Z}_2^{\text{CP}_1  S} \times \mathbb{Z}_2^R$ & $\mathbb{Z}_2^{\text{CP}_2  S} \times \mathbb{Z}_2^R$ & $\mathbb{Z}_2^{\text{CP}\, S}$ \\[3mm]
    $\re \tau = -\frac{1}{2}$ & $\mathbb{Z}_2^R$ & $1$ & $\mathbb{Z}_2^{\text{CP}_1  T} \times \mathbb{Z}_2^R$ & $\mathbb{Z}_2^{\text{CP}_2  T} \times \mathbb{Z}_2^R$ & $\mathbb{Z}_2^{\text{CP}\, T}$ \\[3mm]
    generic $\tau$ & $\mathbb{Z}_2^R$ & $1$ & $\mathbb{Z}_2^R$ & $\mathbb{Z}_2^R$ & $1$ \\[1mm]
    \bottomrule
\end{tabular}
  \caption{Residual symmetry groups for different values of $\tau$ and different choices of the full symmetry group.} 
  \label{tab:residual}
\end{table}

%
\section{Phenomenology}
\label{sec:pheno}
%
%
To illustrate how the results of the previous sections can be applied to
model building, we now consider examples of $S_4'$ modular-invariant models of
lepton flavour. 
As in previous bottom-up works,
the Kähler potential is taken to be
\begin{equation}
K(\tau, \overline{\tau}, \psi, \overline{\psi})
= - \Lambda_0^2 \log(-i\tau + i \overline{\tau})
  + \sum_I \frac{|\psi_I|^2}{(-i\tau + i \overline{\tau})^{k_I}} \,, 
\label{eq:Kahler}
\end{equation}
%
with $\Lambda_0$ having mass dimension one.

\subsection{Weinberg Operator Model}
\label{sec:ModelWeinberg}
We first assume that neutrino masses are generated from the Weinberg
operator, and assign both lepton doublets $L$ and charged lepton
singlets $E^c$ to full triplets of the discrete flavour group.
Such an assignment provides a justification for three lepton
generations and contrasts with most previous bottom-up modular approaches to
flavour.
The relevant superpotential is
\begin{align}
W =
\sum_s \alpha_s \left(Y_{\mathbf{r}_s}^{(k_Y)}(\tau)\, E^c \,L\, H_d \right)_{\mathbf{1},s}
+
\frac{1}{\Lambda} \sum_s g_s \left(Y_{\mathbf{r}_s}^{(k_W)}(\tau) \,L^2\, H_u^2\right)_{\mathbf{1},s}
\,,
\label{eq:wl}
\end{align}
%
where one has summed over independent singlets $s$.

In particular,
we take $L \sim \mathbf{3}$ with weight $k_L = 2$,
and $E^c \sim \mathbf{\hat{3}}$ with weight $k_{E^c} = 1$. Higgs doublets
$H_u$ and $H_d$ are assumed to be $S_4'$ trivial singlets of zero modular
weight.
To compensate the modular weights of field monomials, the modular forms entering the Weinberg term need to have weight $k_W = 4$, while those in the Yukawa term need instead $k_Y = 3$.
Note that $E^c$ transforms with an odd modular weight
and in an irrep which is absent from the usual $\Gamma_4 \simeq S_4$
modular construction. Aiming at a minimal and predictive example, we further
impose a gCP symmetry (CP$_1$, see Section~\ref{sec:gCP}) on the model.
Then, eq.~\eqref{eq:wl} explicitly reads
\begin{equation}
\begin{aligned}
W &=
   \alpha_1 \left(Y_{\mathbf{\hat{1}'}}^{(3)}\, E^c \,L \right)_{\mathbf{1}} H_d
+  \alpha_2 \left(Y_{\mathbf{\hat{3}'}}^{(3)}\, E^c \,L \right)_{\mathbf{1}} H_d
+  \alpha_3 \left(Y_{\mathbf{\hat{3}}}^{(3)}\, E^c \,L \right)_{\mathbf{1}} H_d
\\[2mm]
&\,\,
+\frac{g_1}{\Lambda}  \left(Y_{\mathbf{1}}^{(4)}\,L^2\right)_{\mathbf{1}} H_u^2
+\frac{g_2}{\Lambda}  \left(Y_{\mathbf{2}}^{(4)}\,L^2\right)_{\mathbf{1}} H_u^2
+\frac{g_3}{\Lambda}  \left(Y_{\mathbf{3}}^{(4)}\,L^2\right)_{\mathbf{1}} H_u^2
\,,
\end{aligned}
\label{eq:w}
\end{equation}
%
where the $g_s$ and the $\alpha_s$ ($s=1,2,3$) are real as a result of imposing gCP in the working symmetric basis for the $S_4'$ group generators (see Appendix~\ref{app:sym_basis}).
This superpotential results in the following Lagrangian, containing the
mass matrices of neutrinos and charged leptons,
\begin{align}
\mathcal{L} \,\supset\,
-\frac{1}{2}\,\big(M_\nu\big)_{ij}\,\overline{\nu_{iR}^c}\,\nu_{jL}
-\big(M_e\big)_{ij}\,\overline{e_{iL}}\,e_{jR} + \text{h.c.}\,,
\label{eq:4spinors}
\end{align}
%
which is written in terms of four-spinors,
with $\langle H_u\rangle = (0,v_u)^T$, $\langle H_d\rangle = (v_d,0)^T$,
and $\nu_{iR}^c \equiv C \,\overline{\nu_{iL}}^T$,
$C$ being the charge conjugation matrix.
The matrices $M_\nu$ and $M_e$ can be obtained from eq.~\eqref{eq:w} and read:%
\footnote{We have kept in these expressions the canonical Clebsch-Gordan
normalisations, included in Appendix~\ref{app:CGCs}.}
\begin{equation}
\begin{aligned}
    \frac{1}{2\,v_u^2}\, M_\nu &\,=\,
    \frac{1}{\sqrt{3}}\frac{g_1}{\Lambda}\begin{pmatrix}
    Y_1 & 0 & 0 \\ 0 & 0 & Y_1 \\ 0 & Y_1 & 0 \end{pmatrix}_{Y_{\mathbf{1}}^{(4)}}
- \frac{1}{2\sqrt{3}}\frac{g_2}{\Lambda}\begin{pmatrix}
    2\,Y_1 & 0 & 0 \\ 0 & \sqrt{3}\,Y_2 & -Y_1 \\ 0 & -Y_1 & \sqrt{3}\,Y_2
\end{pmatrix}_{Y_{\mathbf{2}}^{(4)}} \\[2mm]
    & \,\,
    + \frac{1}{\sqrt{6}}\frac{g_3}{\Lambda}\begin{pmatrix}
    0 & -Y_2 & Y_3 \\ -Y_2 & -Y_1 & 0 \\ Y_3 & 0 & Y_1
\end{pmatrix}_{Y_{\mathbf{3}}^{(4)}}\,,
\end{aligned}
\label{eq:MnuYs}
\end{equation}
%
and
\begin{equation}
\!\!\!\!\!\!\!\!\!\!\!\!\!\!\!\!\!\!\!\!
\!\!\!\!\!\!\!\!\!\!\!\!\!\!\!\!\!\!\!
\begin{aligned}
    \frac{1}{v_d}\,M_e^\dagger &\,=\,
    \frac{\alpha_1}{\sqrt{3}}
    \begin{pmatrix}
    Y_1 & 0 & 0 \\ 0 & 0 & Y_1 \\ 0 & Y_1 & 0 \end{pmatrix}_{Y_{\mathbf{\hat{1}'}}^{(3)}}
    +\frac{\alpha_2}{\sqrt{6}}
    \begin{pmatrix}
    0 & -Y_2 & Y_3 \\ -Y_2 & -Y_1 & 0 \\ Y_3 & 0 & Y_1
    \end{pmatrix}_{Y_{\mathbf{\hat{3}'}}^{(3)}}\\[2mm]
    & \,\,
    +\frac{\alpha_3}{\sqrt{6}}
    \begin{pmatrix}
    0 & Y_3 & -Y_2 \\ -Y_3 & 0 & Y_1 \\ Y_2 & -Y_1 & 0
\end{pmatrix}_{Y_{\mathbf{\hat{3}}}^{(3)}}\,.
\end{aligned}
\label{eq:MeYs}
\end{equation}
%
In the above, the $Y_{\mathbf{r}}^{(k)}$ subscript attached to each matrix
denotes the modular form multiplet $Y$ to be used within that matrix.
The explicit expressions for these mass matrices in terms of the $\theta$ and
$\varepsilon$ functions are given in Appendix~\ref{app:full}.

 Notice that the 13 independent $Y_i$ are all determined once the value of
the complex modulus $\tau$ is specified. Hence, this model contains 8 real parameters (6 real couplings and $\tau$) while aiming to explain 12 observables (3 charged-lepton masses, 3 neutrino masses, 3 mixing angles, and 3 CPV phases). Since 8 of these observables are rather well-determined, one expects to predict within the model the lightest neutrino mass and 
the Dirac CPV phase $\delta$,
as well as the Majorana phases $\alpha_{21}$ and $\alpha_{31}$,
and hence the effective Majorana mass $|\vev{m}|$ entering the expression for the rate of neutrinoless double beta ($(\beta\beta)_{0\nu}$-)decay~\cite{PDG2019}.

The functions $\theta(\tau)$ and $\varepsilon(\tau)$ are particularly well suited to analyse models in the ``vicinity'' of the symmetric point $\tau_T = i\infty$, i.e.~for models where $\im \tau$ is large. 
In this case, one can use $\varepsilon(\tau)$ as an expansion parameter and obtain the approximate forms of the neutrino and charged-lepton mass matrices given above:%
\footnote{We use stars to denote repeated elements of a symmetric matrix.}
\begin{align}
M_\nu &\,\simeq\,
\frac{v_u^2\,g_1\,\theta^8}{3 \,\Lambda}
\begin{pmatrix}
  1- \dfrac{\sqrt{3}}{2}\,\tilde g_2 
& - \dfrac{3\sqrt{3}}{2}\,\tilde g_3\,\left (\dfrac{\varepsilon}{\theta}  \right )^3
& - \dfrac{3\sqrt{3}}{2}\,\tilde g_3\,\dfrac{\varepsilon}{\theta}
\\[4mm]
  *
& - {3\sqrt{3}}\left (\tilde g_2+\dfrac{\tilde g_3}{\sqrt{2}} \right )\left (\dfrac{\varepsilon}{\theta}  \right )^2
& 1 + \dfrac{\sqrt{3}}{4}\,\tilde g_2 
\\[4mm]
  *
& *
& - {3\sqrt{3}}\left (\tilde g_2-\dfrac{\tilde g_3}{\sqrt{2}} \right )\left (\dfrac{\varepsilon}{\theta}  \right )^2
\end{pmatrix} \,,
\label{eq:mnuapprox}
\\
M_e^\dagger &\,\simeq\,
v_d \, \alpha_1\,\theta^6\,
\begin{pmatrix}
   \dfrac{\varepsilon}{\theta}
& -\dfrac{1}{2\sqrt{6}}
\left(\tilde\alpha_2 - \dfrac{\tilde\alpha_3}{\sqrt{2}}\right)  
& 
-\dfrac{1}{2}\sqrt{\dfrac{3}{2}}\left(\tilde\alpha_2
+\dfrac{5\,\tilde\alpha_3}{3\sqrt{2}}\right) 
\left (\dfrac{\varepsilon}{\theta}  \right )^2
\\[6mm]
-\dfrac{1}{2\sqrt{6}}
\left(\tilde\alpha_2 + \dfrac{\tilde\alpha_3}{\sqrt{2}}\right)  
& 
\dfrac{2}{\sqrt{3}}\,\tilde\alpha_2\, \left (\dfrac{\varepsilon}{\theta}  \right )^3
&  \left(1 + \dfrac{\tilde\alpha_3}{\sqrt{6}}\right)\, \dfrac{\varepsilon}{\theta}
\\[6mm]
-\dfrac{1}{2}\sqrt{\dfrac{3}{2}}\left(\tilde\alpha_2
-\dfrac{5\,\tilde\alpha_3}{3\sqrt{2}}\right) 
\left (\dfrac{\varepsilon}{\theta}  \right )^2
& 
\left(1 - \dfrac{\tilde\alpha_3}{\sqrt{6}}\right)\, \dfrac{\varepsilon}{\theta} 
&
-\dfrac{2}{\sqrt{3}}\,\tilde\alpha_2\, \left (\dfrac{\varepsilon}{\theta}  \right )^3
\end{pmatrix}\,,
\label{eq:meapprox}
\end{align}
\\
where we have omitted $O(\varepsilon^4/\theta^4)$ corrections,
included in full in Appendix~\ref{app:full}. In the above expressions,
we have further defined $\tilde{g}_{2(3)} = g_{2(3)}/g_1$ and
$\tilde \alpha_{2(3)} \equiv \alpha_{2(3)} / \alpha_1$.

The statistical analysis, the details and results of which will be reported in
subsection~\ref{sec:numerical},
shows that a successful description of the neutrino oscillation 
data and of charged-lepton masses can be achieved for a value 
of $\tau$ close to $\tau_C=i$ for NO, and close to $1.6\,i$ for IO. 
In both cases,
one cannot rely on the approximations used in eqs.~\eqref{eq:mnuapprox} and \eqref{eq:meapprox},
and the full expressions given in Appendix~\ref{app:full} are required.

\subsection{Type I Seesaw Model}
\label{sec:ModelSeesaw}
We now assume instead that neutrino masses are generated from interactions with gauge singlets $N^c$ in a type I seesaw, taking 
$L \sim \mathbf{\hat{3}}$ with weight $k_L = 4$, 
$E^c \sim \mathbf{3}$ with weight $k_L = -1$, and
$N^c \sim \mathbf{2}$ with weight $k_L = 1$.
Once more, Higgs doublets
$H_u$ and $H_d$ are assumed to be trivial $S_4'$ singlets of zero modular
weight.
The modular forms entering the Majorana mass term need to have weights $k_M = 2$, while those in the Yukawa terms of charged leptons and neutrinos need $k_{Y_E} = 3$ and $k_{Y_N} = 5$, respectively.
Note that here both $E^c$ and $N^c$ transform with odd modular weights, while 
$L$ transforms in an irrep which is absent from $\Gamma_4 \simeq S_4$.
We further impose a gCP symmetry (CP$_1$) on the model, 
whose superpotential reads:
\begin{equation}    
\begin{aligned}
W &=
   \alpha_1 \left(Y_{\mathbf{\hat{1}'}}^{(3)}\, E^c \,L \right)_{\mathbf{1}} H_d
+  \alpha_2 \left(Y_{\mathbf{\hat{3}'}}^{(3)}\, E^c \,L \right)_{\mathbf{1}} H_d
+  \alpha_3 \left(Y_{\mathbf{\hat{3}}}^{(3)}\, E^c \,L \right)_{\mathbf{1}} H_d
\\[2mm]
&\,\,
+  g_1 \left(Y_{\mathbf{\hat{3}'}}^{(5)}\,N^c\,L\right)_{\mathbf{1}} H_u
+  g_2 \left(Y_{\mathbf{\hat{3}},1}^{(5)}\,N^c\,L\right)_{\mathbf{1}} H_u
+  g_3 \left(Y_{\mathbf{\hat{3}},2}^{(5)}\,N^c\,L\right)_{\mathbf{1}} H_u
\\[2mm]
&\,\,
+ \Lambda_1 \left(Y_{\mathbf{2}}^{(2)}\,{N^c}\,{N^c}\right)_{\mathbf{1}}
\,,
\end{aligned}
\label{eq:wseesaw}
\end{equation}
%
where $\Lambda_1$, the $g_s$ and the $\alpha_s$ ($s=1,2,3$) are real, given the working symmetric basis.
This superpotential can be cast in the form
\begin{align}
W =   \lambda_{ij} \,E^c_i\,L_j\,H_d 
    + \mathcal{Y}_{ij} \,N^c_i\,L_j
    + \frac{1}{2}\,M_{ij} \,N^c_i\, N^c_j\,,
\end{align}
%
with
\begin{equation}
\begin{aligned}
    M &\,=\,  \Lambda_1
    \begin{pmatrix}
    - Y_1 & Y_2 \\ Y_2 & Y_1
    \end{pmatrix}_{Y_{\mathbf{2}}^{(2)}} \,,
\end{aligned}
\end{equation}
%
and
\begin{equation}
\begin{aligned}
\mathcal{Y} \,=\,
    \frac{g_1}{2\sqrt{3}}
    &
    \begin{pmatrix}
     2\,Y_1 & -Y_3          & -Y_2          \\ 
     0      & \sqrt{3}\,Y_2 & \sqrt{3}\,Y_3 
    \end{pmatrix}_{Y_{\mathbf{\hat{3}}'}^{(5)}}
+   \frac{g_2}{2\sqrt{3}}
    \begin{pmatrix}
     0        & \sqrt{3}\,Y_2 & \sqrt{3}\,Y_3 \\
     - 2\,Y_1 & Y_3           & Y_2  
    \end{pmatrix}_{Y_{\mathbf{\hat{3}},1}^{(5)}}
    \\[2mm]
+ \,  \frac{g_3}{2\sqrt{3}}
    &
    \begin{pmatrix}
     0        & \sqrt{3}\,Y_2 & \sqrt{3}\,Y_3 \\
     - 2\,Y_1 & Y_3           & Y_2  
    \end{pmatrix}_{Y_{\mathbf{\hat{3}},2}^{(5)}}
\,.
\end{aligned}
\end{equation}
%
In the conventions of eq.~\eqref{eq:4spinors},
the light neutrino mass matrix $M_\nu$ is then obtained from the seesaw relation,
\begin{align}
M_\nu = - v_u^2\, \mathcal{Y}^T\, M^{-1}\, \mathcal{Y}\,,
\label{eq:seesaw}
\end{align}
%
while the charged-lepton mass matrix $M_e = v_d\,\lambda^\dagger$
is given by eq.~\eqref{eq:MeYs} with $\alpha_3 \to -\alpha_3$.
Note that, due to the seesaw relation~\eqref{eq:seesaw}, changes in the scale of the $g_s$ can be compensated by adjusting the scale of $\Lambda_1$. Hence, this model is effectively described by 8 real parameters at low energy (6 real combinations of couplings and $\tau$).

\subsection{Numerical Analysis and Results}
\label{sec:numerical}

Our models are constrained by the observed ratios of charged-lepton masses, neutrino mass-squared differences, and leptonic mixing angles. The experimental best fit values and $1\sigma$ ranges considered for these observables are collected in Table~\ref{tab:globalFit}.
We do not take into account the $1\sigma$ range of the Dirac CPV phase $\delta$ in our fit.
As a measure of goodness of fit, we use $N\sigma \equiv \sqrt{\Delta \chi^2}$, where $\Delta \chi^2$ is approximated as a sum of one-dimensional chi-squared projections.
The reader is referred to Ref.~\cite{Novichkov:2018ovf} for further details on the numerical procedure.

\begin{table}[t]
\centering
\renewcommand{\arraystretch}{1.2}
\begin{tabular}{l|cc} 
\toprule
Observable & \multicolumn{2}{c}{Best fit value and $1\sigma$ range} \\ 
\midrule
$m_e / m_\mu$ & \multicolumn{2}{c}{$0.0048 \pm 0.0002$} \\
$m_\mu / m_\tau$ & \multicolumn{2}{c}{$0.0565 \pm 0.0045$} \\ 
\midrule
& NO & IO \\
$\delta m^2/(10^{-5}\text{ eV}^2)$ & \multicolumn{2}{c}{$7.34^{+0.17}_{-0.14}$} \\
$|\Delta m^2|/(10^{-3}\text{ eV}^2)$ & $2.485^{+0.029}_{-0.032}$ & $2.465^{+0.030}_{-0.031}$ \\
$r \equiv \delta m^2/|\Delta m^2|$ & $0.0295\pm0.0008$ & $0.0298\pm0.0008$\\
$\sin^2\theta_{12}$ & $0.305^{+0.014}_{-0.013}$ & $0.303^{+0.014}_{-0.013}$ \\
$\sin^2\theta_{13}$ & $0.0222^{+0.0006}_{-0.0008}$ & $0.0223^{+0.0007}_{-0.0006}$ \\
$\sin^2\theta_{23}$ & $0.545^{+0.020}_{-0.047}$ & $0.551^{+0.016}_{-0.034}$ \\
$\delta/\pi$ & $1.28^{+0.38}_{-0.18}$  & $1.52^{+0.13}_{-0.15}$ \\
\bottomrule
\end{tabular}
\caption{Best fit values and 1$\sigma$ ranges for 
neutrino oscillation parameters, obtained from the global analysis
of Ref.~\cite{Capozzi:2020qhw}, and for charged-lepton mass ratios,
given at the scale $2\times 10^{16}$ GeV with the $\tan \beta$ averaging
described in~\cite{Feruglio:2017spp}, obtained from Ref.~\cite{Ross:2007az}.
The parameters entering the definition of $r$ are $\delta m^2 \equiv m_2^2-m_1^2$
and $\Delta m^2 \equiv m_3^2 - (m_1^2+m_2^2)/2$.
The best fit value and $1\sigma$ range of $\delta$
did not drive the numerical searches here reported,
i.e.~the value of $\delta$ does not affect
the value of $N\sigma$ defined in the text.}
\label{tab:globalFit}
\end{table}
%

\begin{figure}[t]
 \makebox[\textwidth][c]{  \includegraphics[width=1.2\textwidth]{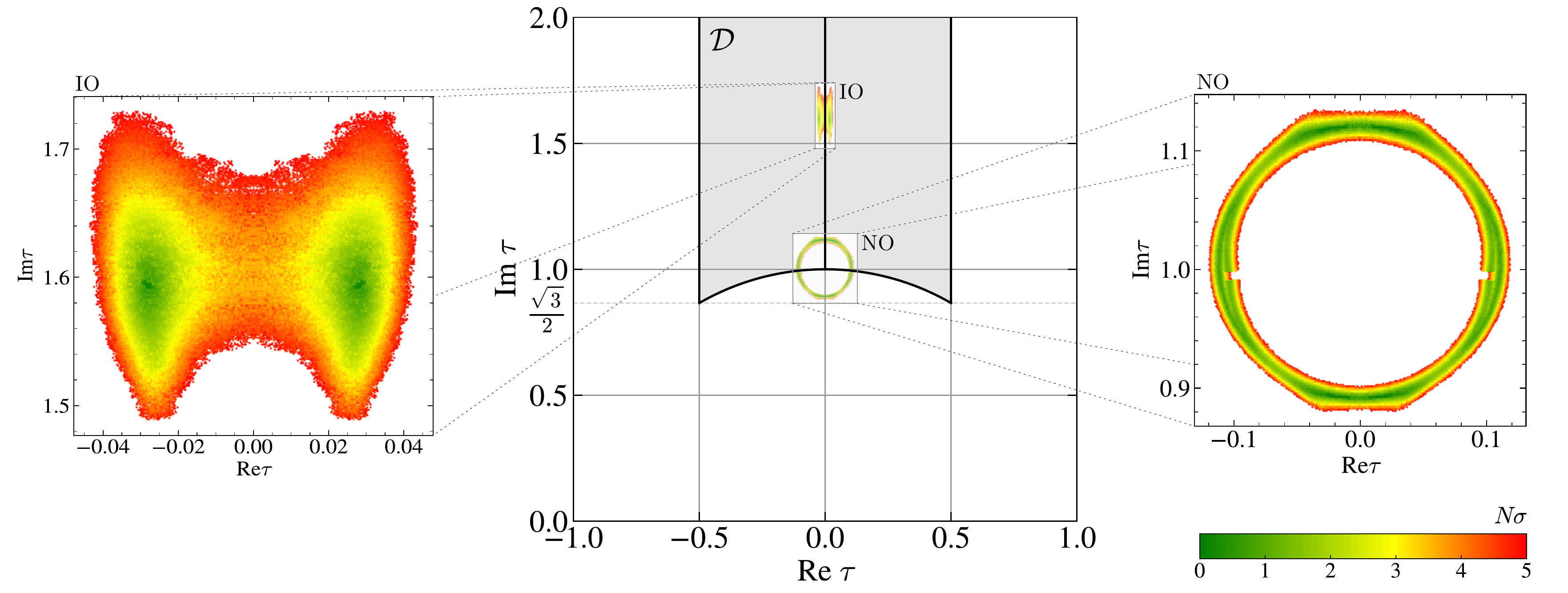} \hphantom{--}}
  \caption{
  Allowed regions in the $\tau$ plane for the fit of the $S_4'$ Weinberg operator model of subsection~\ref{sec:ModelWeinberg}, for IO (left inset) and NO (right inset).
  Here and in what follows, the green, yellow and red colours correspond to different confidence levels,
  as indicated in the legend.
  Points in the NO case which are outside the fundamental domain $\mathcal{D}$ are redundant, since they are equivalent to points inside $\mathcal{D}$ via the action of $S$. We nevertheless keep them for illustrative purposes.
  }
  \label{fig:Wtau}
\end{figure}

Through numerical search, we find that the model of subsection~\ref{sec:ModelWeinberg} can lead to acceptable fits of the leptonic sector ($N\sigma \simeq 0.07$), with the values of $\tau$, $\alpha_i$ and $g_i$ indicated in Tables~\ref{tab:WNO} and~\ref{tab:WIO} for NO and IO, respectively.
The phenomenologically viable region in the $\tau$ plane is shown, for both orderings, in Figure~\ref{fig:Wtau}. While for IO the fit is possible with $\tau \simeq 1.6\,i$, for NO an annular region close to $\tau_C = i$ is selected, with $|\tau-i| \simeq 0.12$.
As one can see from the tables,
independent singlets in the superpotential of eq.~\eqref{eq:w}
can provide comparable contributions to the mass matrices.
There is however some fine-tuning present in the coupling constants $\alpha_i$
in order to accommodate charged-lepton mass hierarchies.

\begin{figure}[t!]
  \centering
  \includegraphics[width=\textwidth]{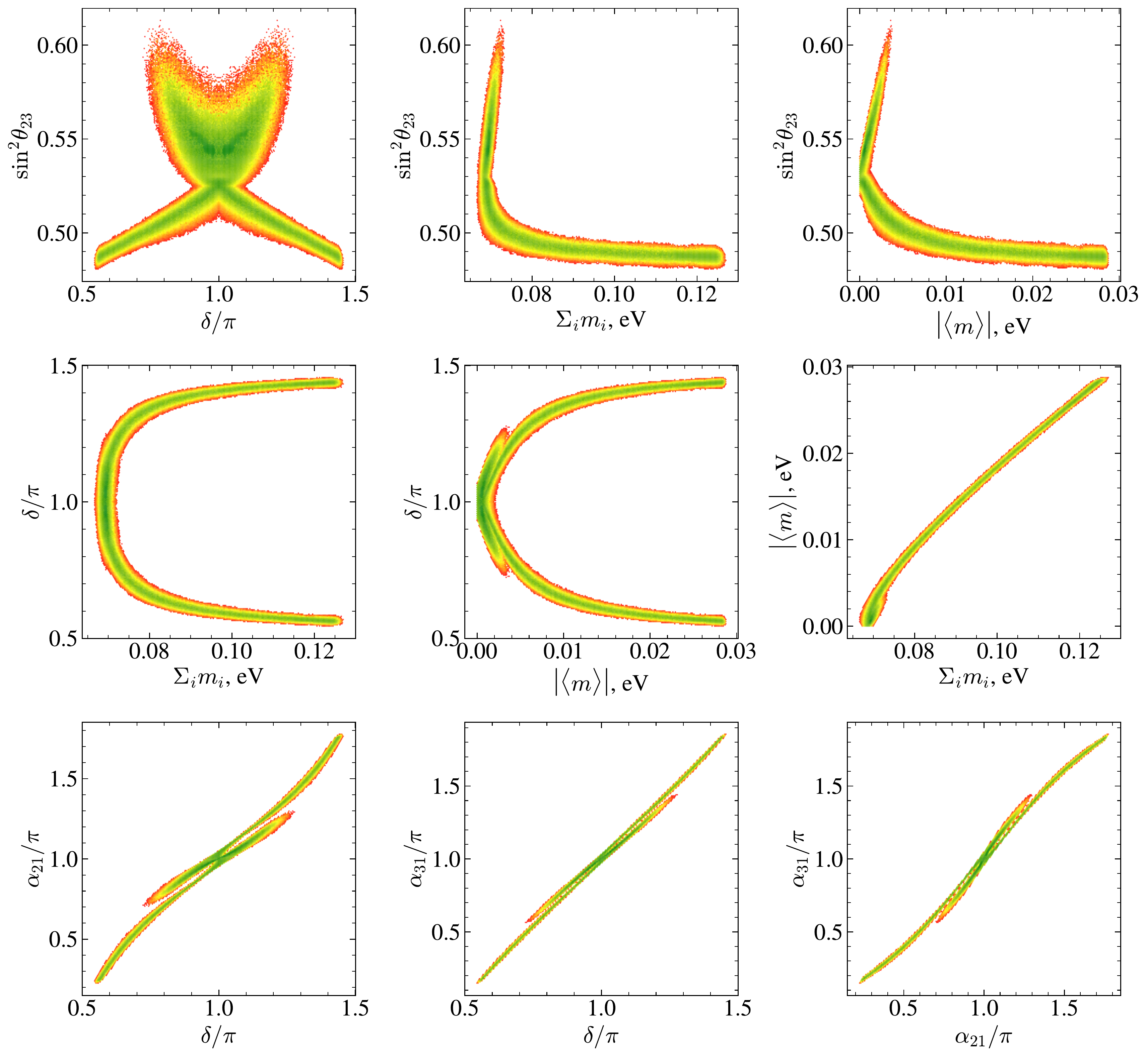}
  \caption{
  Correlations between pairs of observables for the NO fit of the $S_4'$ Weinberg operator model of subsection~\ref{sec:ModelWeinberg}.
  }
  \label{fig:WNO}
\end{figure}

\begin{figure}[ht!]
  \centering
  \includegraphics[width=\textwidth]{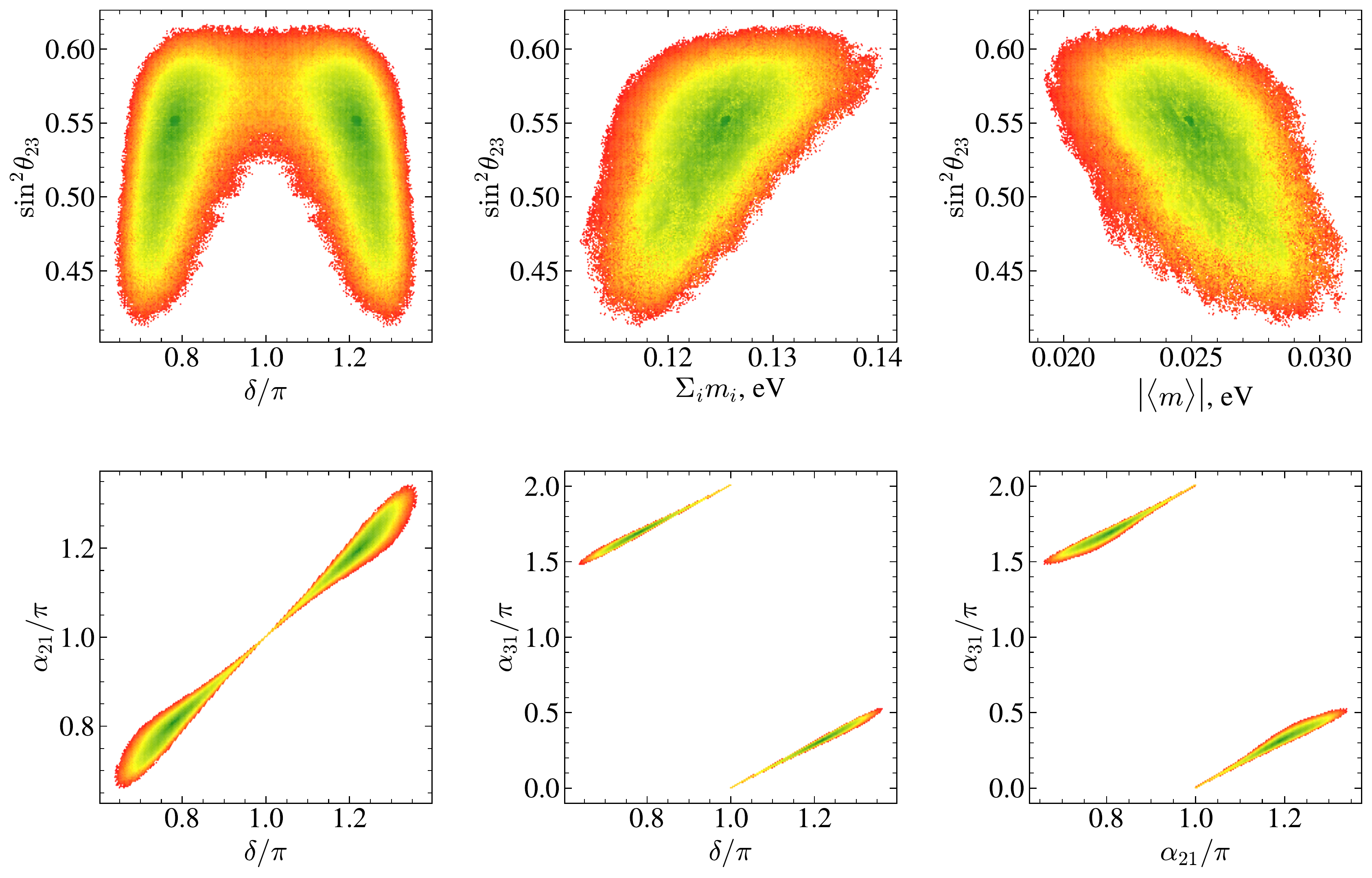}
  \caption{
  Correlations between pairs of observables for the IO fit of the $S_4'$ Weinberg operator model of subsection~\ref{sec:ModelWeinberg}.
  }
  \label{fig:WIO}
\end{figure}

This model additionally predicts peculiar correlations between observables, which are shown in Figures~\ref{fig:WNO} and~\ref{fig:WIO}, for NO and IO, respectively. 
One can see that, in the NO case,
a Dirac phase $\delta$ deviated from $\pi$
is tied to smaller values of the atmospheric angle,
which are in turn associated with larger values of  
the effective Majorana mass $|\vev{m}|$ and of the sum
of neutrino masses $\Sigma_i m_i$, i.e.~with a larger absolute neutrino mass scale.
In the IO case, a deviation of $\delta$ from $\pi$
also favours smaller values of $\sin^2 \theta_{23}$.
In both cases, the values of all three CPV phases are highly correlated.

\begin{figure}[ht!]
  \centering
    \includegraphics[width=0.4\textwidth]{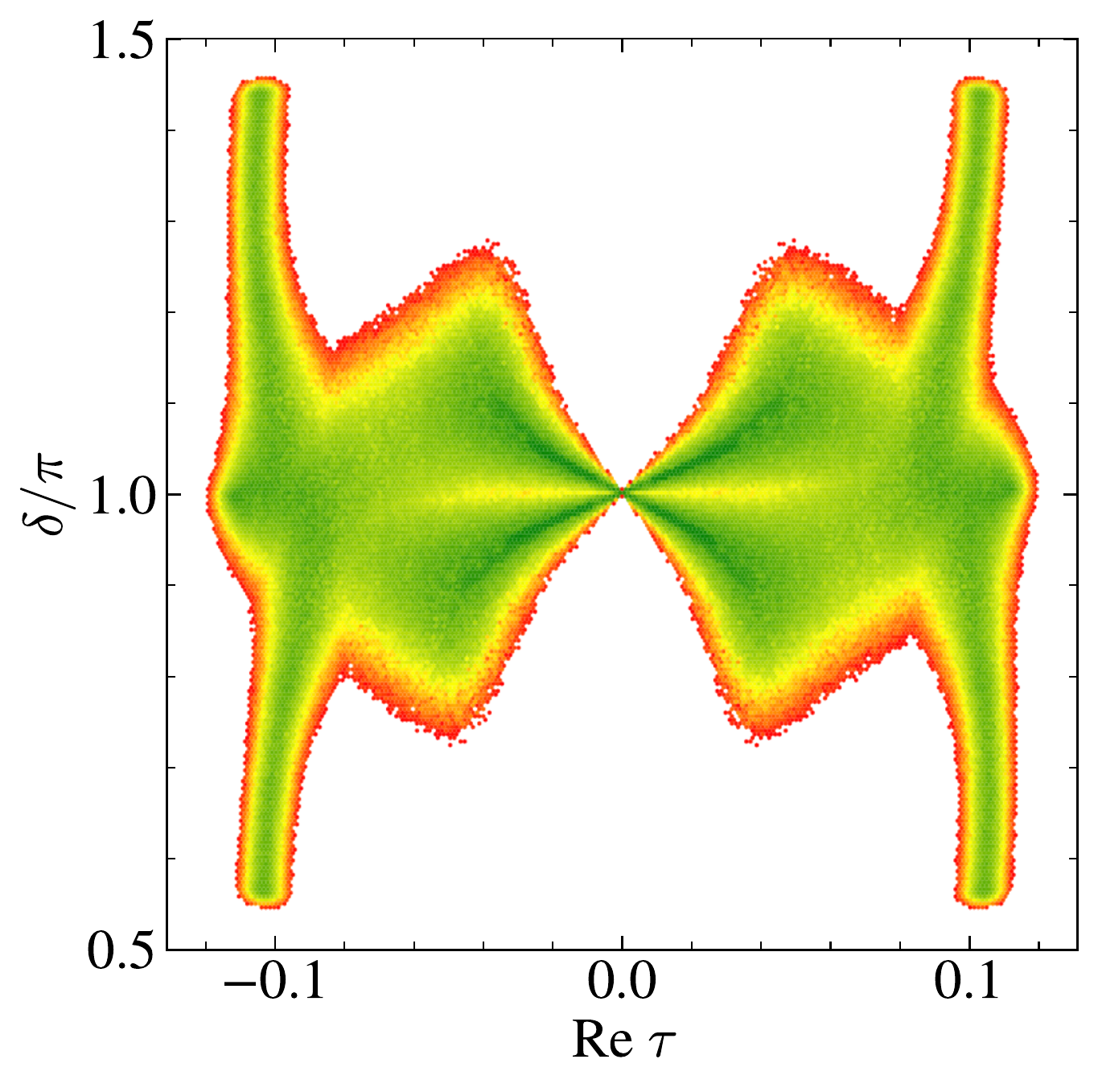}
    \qquad
    \includegraphics[width=0.4\linewidth]{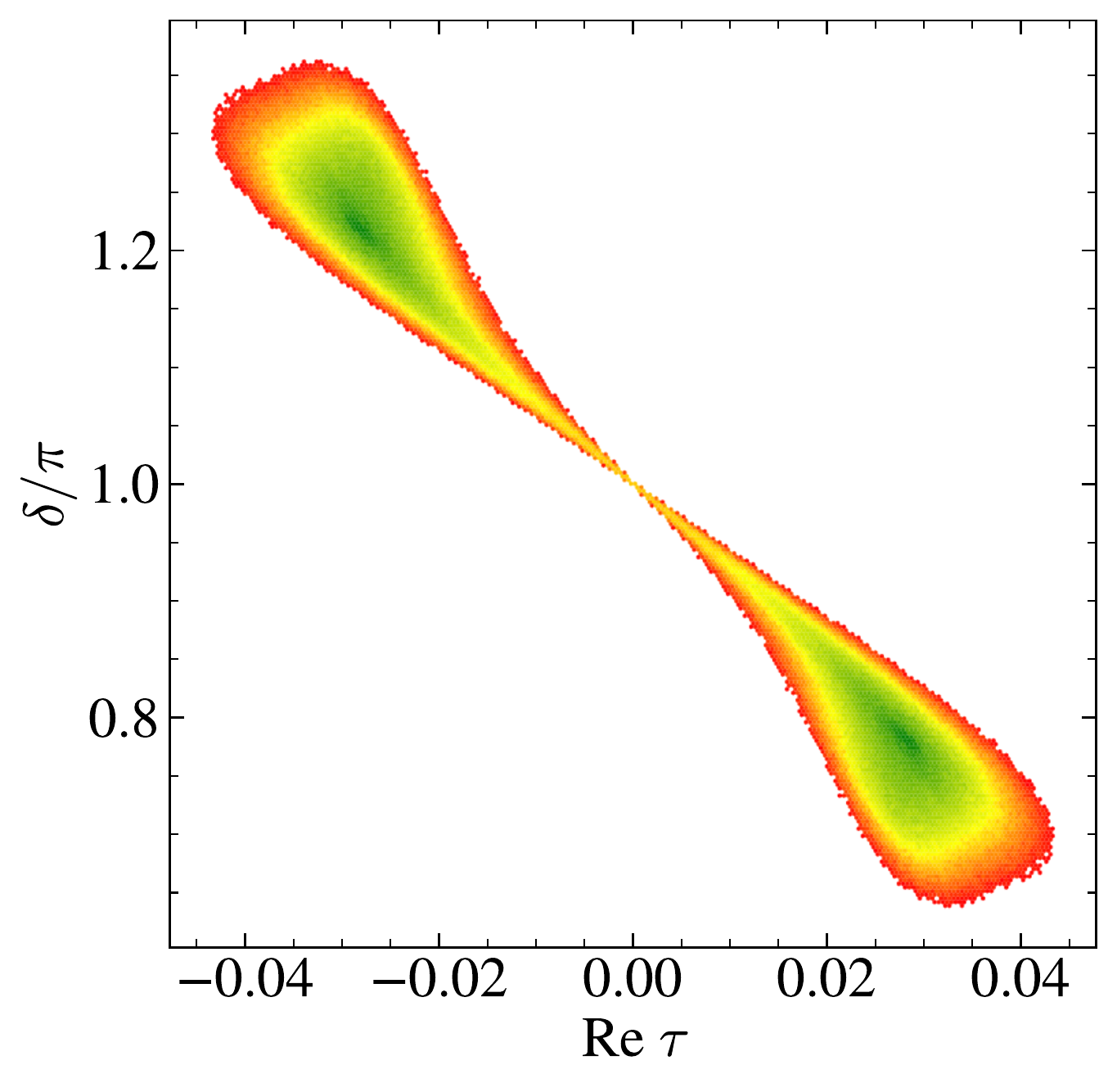}
    \caption{
    Correlation between the Dirac CPV phase $\delta$ and $\re \tau$ for the NO (left) and IO (right) fits of the $S_4'$ Weinberg operator model of subsection~\ref{sec:ModelWeinberg}.
    }
  \label{fig:WCP}
\end{figure}

\begin{figure}[ht!]
  \centering
  \includegraphics[width=\textwidth]{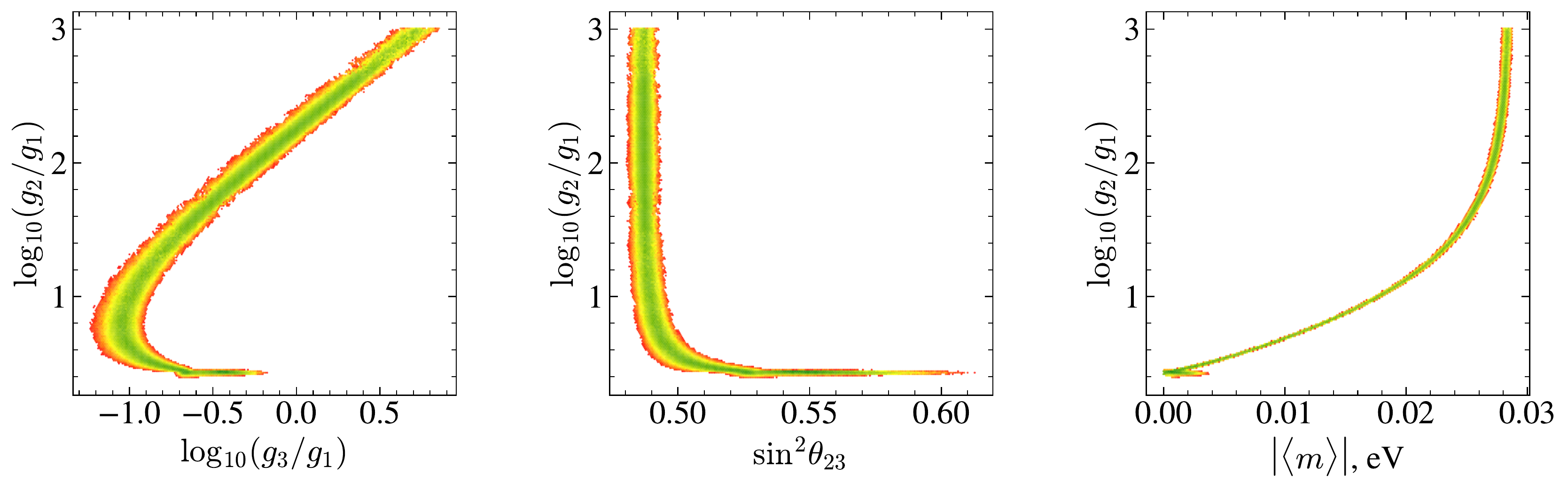}
  \caption{
    Correlations of $\log_{10}(g_2/g_1)$ with $\log_{10}(g_3/g_1)$ and with the observables $\sin^2 \theta_{23}$ and $|\vev{m}|$ for the NO fit of the $S_4'$ Weinberg operator model of subsection~\ref{sec:ModelWeinberg}.
    }
  \label{fig:WNOg1}
\end{figure}

Analysing instead correlations between observables and model parameters,
one can verify that CP is conserved for $\re \tau = 0$, as anticipated in Section~\ref{sec:residual}.
Recall that, in this model, CP symmetry is spontaneously broken by the VEV of $\tau$.
The correlation between the Dirac CP phase and the value of $\re \tau$ is shown in Figure~\ref{fig:WCP} for both orderings, with $\delta$ taking a CP conserving value for purely imaginary $\tau$, as expected. 
We note also that, in the case of NO, the viable fit region for the ratios $g_2/g_1$ 
and $g_3/g_1$
seems to be unbounded, see Figure~\ref{fig:WNOg1}. However, correlations between $g_2/g_1$ and observables suggest that the limit $g_1 \to 0$ is phenomenologically viable, with larger values of the ratio not affecting the values of observables (cf.~$\sin^2 \theta_{23}$ and $\vev{m}$ in the figure). We are then free to limit the range of the ratio $g_2/g_1$ to $10^3$ in our numerical exploration.

\begin{figure}[ht!]
  \centering
  \includegraphics[width=0.9\textwidth]{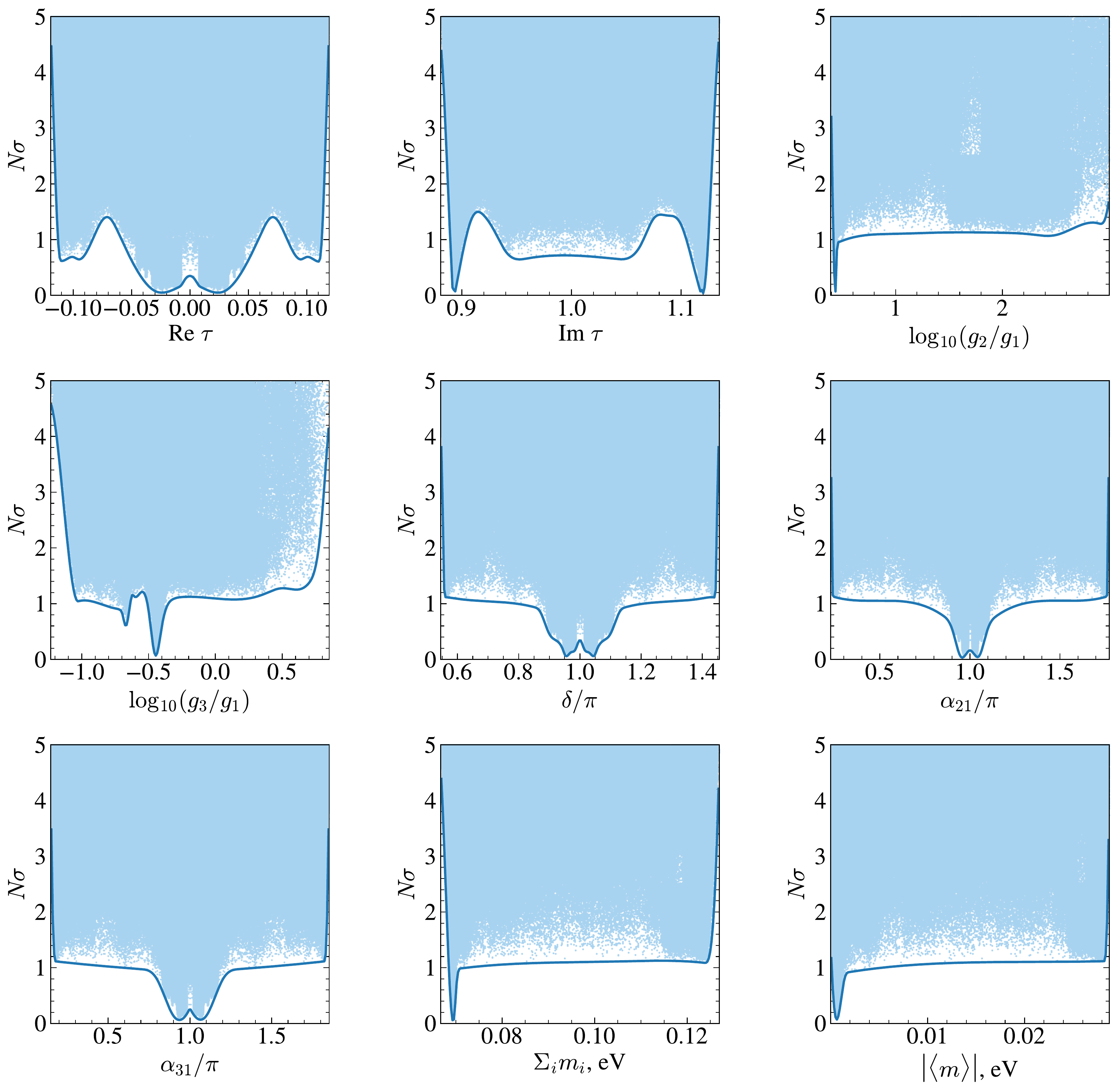}
  \caption{
    Projections of $N\sigma = \sqrt{\Delta \chi^2}$ across observables and model parameters
    in the NO fit of the $S_4'$ Weinberg operator model of subsection~\ref{sec:ModelWeinberg}.
    The lower boundary curves were obtained by fitting B-splines with generalised least squares~\cite{Cardiel:2009}.
    }
  \label{fig:WNOchi2}
\end{figure}

Finally, let us comment on the allowed values for 
the effective Majorana mass $|\vev{m}|$
entering the expression for the rate of $(\beta\beta)_{0\nu}$-decay.
At the $3\sigma$ level,
the IO fit one predicts $|\vev{m}| \simeq 22-29$ meV,
while for NO one has $|\vev{m}| \lesssim 29$ meV (see Tables~\ref{tab:WNO} and~\ref{tab:WIO}).
In the latter case, very small values of $|\vev{m}|$ are allowed,
contradicting a tendency of bottom-up modular-invariant models (see e.g.~\cite{Feruglio:2019ktm}).
This is also the case for the NO best fit point, for which a value of $|\vev{m}|$ slightly below the meV is preferred. 
However, $|\vev{m}|$ can be large, $|\vev{m}| > 20$ meV, already at the $1.2 \sigma$ level.
This can, for instance, be seen in Figure~\ref{fig:WNOchi2}, where we collect the $N\sigma$ projections for different model parameters and observables.

\begin{table}[t!]
  \centering
  \begin{tabular}{lccc}
    \toprule
    {} & Best fit value &            $2\sigma$ range &            $3\sigma$ range \\
    \midrule
    $\re \tau$                              &     $\pm 0.029725$ &       $-0.11437 - 0.11437$ &       $-0.11597 - 0.11597$ \\
    $\im \tau$                              &       $1.1181$ &         $0.88795 - 1.1262$ &         $0.88582 - 1.1289$ \\
    $|\tau - i|$                           &      $0.12174$ &        $0.10112 - 0.12848$ &       $0.099153 - 0.13229$ \\
    $\alpha_2 / \alpha_1$                  &       $1.7303$ &            $1.73 - 1.7307$ &          $1.7299 - 1.7309$ \\
    $\alpha_3 / \alpha_1$                  &       $-2.7706$ &          $-(2.7208 - 2.8229)$ &          $-(2.6926 - 2.8488)$ \\
    $g_2 / g_1$                            &        $2.716$ &           $2.5942 - 988.3$ &           $2.5493 - 998.8$ \\
    $g_3 / g_1$                            &     $-0.35786$ &      $-(0.080198 - 5.5555)$ &      $-(0.073447 - 6.3885)$ \\
    $v_d\,\alpha_1$, GeV                     &       $1.5958$ &          $0.9571 - 1.9425$ &         $0.89127 - 2.1572$ \\
    $v_u^2\, g_1 / \Lambda$, eV            &     $0.076533$ &   $0.00028142 - 0.12373$ &   $0.00027594 - 0.12636$ \\
    \midrule
    $m_e / m_\mu$                          &    $0.0048091$ &    $0.0044239 - 0.0051918$ &    $0.0042302 - 0.0053878$ \\
    $m_\mu / m_\tau$                       &     $0.056485$ &      $0.048273 - 0.065161$ &      $0.043297 - 0.069358$ \\
    $r$                                    &     $0.029554$ &      $0.028086 - 0.030967$ &      $0.027394 - 0.031744$ \\
    $\delta m^2$, $10^{-5} \text{ eV}^2$   &       $7.3431$ &          $7.0658 - 7.5968$ &          $6.9304 - 7.7309$ \\
    $|\Delta m^2|$, $10^{-3} \text{ eV}^2$ &       $2.4846$ &          $2.4532 - 2.5158$ &          $2.4354 - 2.5299$ \\
    $\sin^2 \theta_{12}$                   &        $0.305$ &        $0.28123 - 0.33069$ &        $0.26737 - 0.34545$ \\
    $\sin^2 \theta_{13}$                   &     $0.022247$ &      $0.020754 - 0.023378$ &      $0.020131 - 0.024027$ \\
    $\sin^2 \theta_{23}$                   &      $0.54509$ &        $0.48487 - 0.57838$ &        $0.48344 - 0.59162$ \\
    \midrule
    $m_1$, eV                              &     $0.007377$ &     $0.0067068 - 0.032237$ &     $0.0064624 - 0.032432$ \\
    $m_2$, eV                              &     $0.011307$ &      $0.010874 - 0.033328$ &      $0.010728 - 0.033518$ \\
    $m_3$, eV                              &     $0.050752$ &      $0.050412 - 0.059852$ &      $0.050227 - 0.060044$ \\
    $\Sigma_i m_i$, eV                     &     $0.069436$ &       $0.068192 - 0.12541$ &       $0.067748 - 0.12594$ \\
    $\left| \langle m \rangle \right|$, meV &   $0.63241$ &    $0.00018464 - 28.483$ &    $0.00012046 - 28.547$ \\
    $\delta / \pi$                         &       $\pm 1.0487$ &          $0.5572 - 1.4428$ &         $0.55293 - 1.4471$ \\
    $\alpha_{21} / \pi$                    &       $\pm 1.0395$ &           $0.24004 - 1.76$ &         $0.23476 - 1.7652$ \\
    $\alpha_{31} / \pi$                    &       $\pm 1.0718$ &         $0.16644 - 1.8336$ &         $0.15932 - 1.8407$ \\
    \midrule
    $N \sigma$                             &     $0.0695$ &                            &                            \\
    \bottomrule
  \end{tabular}
  \caption{
  Best fit values and $2\sigma$ and $3\sigma$ ranges for the parameters and observables in the fit of the $S_4'$ Weinberg operator model of subsection~\ref{sec:ModelWeinberg} with NO.
  }
\label{tab:WNO}
\end{table}

\vskip 2mm

For the seesaw model of subsection~\ref{sec:ModelSeesaw}, we find that
a fit of the data summarised in Table~\ref{tab:globalFit} is possible. As an example, the point in parameter space
described by $\tau = -0.14 + 1.43\,i$ and
\begin{align}
{g_2}/{g_1} = -1.778\,, \quad
{g_3}/{g_1} = -2.433\,, \quad
{\alpha_2}/{\alpha_1} = 2.128\,, \quad
{\alpha_3}/{\alpha_1} = -2.640\,, 
\end{align}
%
fits a neutrino mass spectrum with NO at the $N \sigma \simeq 1.79$ level, 
with the following values for the observables:
\begin{equation}
\begin{gathered}
m_e/m_\mu \simeq 0.004807,\quad 
m_\mu/m_\tau \simeq 0.06214,\\
r \simeq 0.02934,\quad
\delta m^2 \simeq 7.304 \cdot 10^{-5} \text{ eV}^2,\quad
|\Delta m^2| \simeq 2.489 \cdot 10^{-3} \text{ eV}^2,\\
m_2 \simeq 8.547 \cdot 10^{-3} \text{ eV},\quad
m_3 \simeq 5.026 \cdot 10^{-2} \text{ eV},\quad
\Sigma_i m_i  \simeq 5.880 \cdot 10^{-2} \text{ eV},\\
\sin^2 \theta_{12} \simeq 0.3140,\quad
\sin^2 \theta_{13} \simeq 0.02227,\quad
\sin^2 \theta_{23} \simeq 0.5619,\\
\delta / \pi \simeq 1.724,\quad
\alpha_{32} / \pi \simeq 0.8666,\quad
|\vev{m}| = 3.110 \cdot 10^{-3} \text{ eV},
\end{gathered}
\end{equation}
%
given the overall factors $v_u^2\, g_1^2 / \Lambda_1 \simeq 0.2347$ eV and $v_d\, \alpha_1 \simeq 1.778$ GeV. Note that in this scenario $m_1$ vanishes at tree level, such that only the difference $\alpha_{32} \equiv \alpha_{31}-\alpha_{21}$ of Majorana phases is physical.
The ratio of the masses $M_i$ of the two heavy Majorana neutrinos is additionally predicted to be $M_2/ M_1 \simeq 1.14$.
A full numerical exploration of this scenario is postponed to future work.

\begin{table}[t!]
  \centering
  \begin{tabular}{lccc}
    \toprule
    {} & Best fit value &            $2\sigma$ range &            $3\sigma$ range \\
    \midrule
    Re $\tau$                              &    $\mp 0.027941$ &    $\mp (0.019166 - 0.034317)$ &    $\mp (0.0091225 - 0.03702)$ \\
    Im $\tau$                              &       $1.5921$ &           $1.539 - 1.6365$ &          $1.5185 - 1.6634$ \\
    $\alpha_2 / \alpha_1$                  &       $1.7266$ &          $1.7253 - 1.7278$ &          $1.7244 - 1.7284$ \\
    $\alpha_3 / \alpha_1$                  &         $-2.17$ &          $-(2.1304 - 2.2089)$ &          $-(2.1107 - 2.2311)$ \\
    $g_2 / g_1$                            &       $0.4705$ &        $0.42608 - 0.53039$ &        $0.39954 - 0.56815$ \\
    $g_3 / g_1$                            &      $-1.2442$ &        $-(1.0788 - 1.5506)$ &       $-(0.98527 - 1.7423)$ \\
    $v_d\,\alpha_1$, GeV                     &       $2.4973$ &          $2.1623 - 2.9577$ &          $1.9847 - 3.2909$ \\
    $v_u^2\, g_1 / \Lambda$, eV            &      $0.23558$ &      $0.21555 - 0.24684$ &        $0.2085 - 0.2546$ \\
    \midrule
    $m_e / m_\mu$                          &    $0.0047923$ &    $0.0044167 - 0.0051648$ &    $0.0042241 - 0.0053747$ \\
    $m_\mu / m_\tau$                       &      $0.05649$ &      $0.048227 - 0.065009$ &      $0.043596 - 0.069331$ \\
    $r$                                    &     $0.029756$ &      $0.028387 - 0.031155$ &      $0.027599 - 0.031907$ \\
    $\delta m^2$, $10^{-5} \text{ eV}^2$   &        $7.336$ &           $7.073 - 7.5932$ &          $6.9165 - 7.7266$ \\
    $|\Delta m^2|$, $10^{-3} \text{ eV}^2$ &       $2.4654$ &          $2.4373 - 2.4916$ &          $2.4216 - 2.5061$ \\
    $\sin^2 \theta_{12}$                   &      $0.30312$ &        $0.27799 - 0.32846$ &        $0.26532 - 0.34422$ \\
    $\sin^2 \theta_{13}$                   &      $0.02225$ &      $0.021048 - 0.023615$ &        $0.0204 - 0.024279$ \\
    $\sin^2 \theta_{23}$                   &      $0.55029$ &         $0.46546 - 0.5795$ &        $0.44266 - 0.59413$ \\
    \midrule
    $m_1$, eV                              &     $0.052871$ &      $0.051464 - 0.053891$ &       $0.05099 - 0.054701$ \\
    $m_2$, eV                              &      $0.05356$ &      $0.052179 - 0.054563$ &      $0.051716 - 0.055363$ \\
    $m_3$, eV                              &     $0.019147$ &      $0.015034 - 0.021806$ &       $0.01339 - 0.023755$ \\
    $\Sigma_i m_i$, eV                     &      $0.12558$ &        $0.11873 - 0.13014$ &         $0.11614 - 0.1337$ \\
    $\left| \langle m \rangle \right|$, meV &     $25.024$ &      $22.877 - 28.031$ &      $21.624 - 29.217$ \\
    $\delta / \pi$                         &       $\pm 1.2172$ &           $\pm (1.138 - 1.2892)$ &          $\pm (1.0635 - 1.3166)$ \\
    $\alpha_{21} / \pi$                    &       $\pm 1.1906$ &          $\pm (1.1235 - 1.2685)$ &          $\pm (1.0569 - 1.2995)$ \\
    $\alpha_{31} / \pi$                    &      $\pm 0.31101$ &        $\pm (0.19899 - 0.41995)$ &       $\pm (0.091604 - 0.46069)$ \\
    \midrule
    $N \sigma$                             &     $0.0699$ &                            &                            \\
    \bottomrule
  \end{tabular}
  \caption{
  Best fit values and $2\sigma$ and $3\sigma$ ranges for the parameters and observables in the fit of the $S_4'$ Weinberg operator model of subsection~\ref{sec:ModelWeinberg} with IO.
  }
  \label{tab:WIO}
\end{table}

\clearpage
\vfill
\pagebreak
\section{Summary and Conclusions}
\label{sec:conclusions}

In the present article we have developed 
the  formalism of the finite modular group 
$\Gamma'_4$ 
-- the double cover group of $\Gamma_4$ --
that can be used, in particular, 
for theories of lepton and quark 
flavour. The finite modular group 
$\Gamma_4$, as is well known, 
is isomorphic to the permutation group $S_4$,
while $\Gamma'_4$ is isomorphic to the 
double cover of  $S_4$, $S'_4$. 
In comparison with $S_4$, the group $S'_4$ has twice 
as many elements and twice as many irreducible representations,
i.e.~it has 48 elements and admits 10 irreps: 
4 one-dimensional, 2 two-dimensional, and 4 three-dimensional. 
We have denoted them by:
\begin{align}
    \mathbf{1}\,,\,
    \mathbf{\hat{1}}\,,\,
    \mathbf{1'}\,,\,
    \mathbf{\hat{1}'}\,,\,
    \mathbf{2}\,,\,
    \mathbf{\hat{2}}\,,\,
    \mathbf{3}\,,\,
    \mathbf{\hat{3}}\,,\,
    \mathbf{3'}\,,\,
    \mathbf{\hat{3}'}\,.
\end{align}
%
Our notation has been chosen such that irreps without a hat have a direct 
correspondence with $S_4$ irreps, whereas hatted irreps are novel and 
specific to $S_4'$. Working in a symmetric basis for the generators of 
$S_4'$, we have derived the decompositions of tensor products of 
$S_4'$ irreps, as well as the corresponding Clebsch-Gordan coefficients
(see Appendix~\ref{app:S4p}).

Modular forms of level 4 transforming non-trivially 
under $S'_4$ can have even integer or odd integer weight $k>0$. 
In Section~\ref{sec:modforms}, we have explicitly constructed a basis 
for the 3-dimensional space of the modular 
forms of lowest weight $k = 1$, which furnishes 
a 3-dimensional representation 
$\mathbf{\hat{3}}$
of  $S'_4$, not present in $S_4$.
The three components of the weight 1 modular form $Y_{\mathbf{\hat{3}}}^{(1)}(\tau)$
transforming as a ${\bf \hat{3}}$ 
were shown to be quadratic polynomials of 
two ``weight 1/2'' Jacobi theta constants, 
denoted as $\varepsilon(\tau)$ and $\theta(\tau)$, 
$\tau$ being the modulus (cf.~eq.~\eqref{eq:k1triplet}). The functions 
$\varepsilon(\tau)$ and $\theta(\tau)$ are related to the 
Dedekind eta function, and their $q$-expansions are given in
eq.~\eqref{eq:theta_eps_qexp}.
We have further 
constructed $S'_4$ multiplets of modular forms 
of weights up to $k=10$.
The multiplets of weights $k\geq 2$ 
are expressed as homogeneous 
polynomials of even degree in the two 
functions $\varepsilon$ and $\theta$ -- see eqs.~\eqref{eq:k2}, \eqref{eq:k3} and \eqref{eq:k4}, and Appendix~\ref{app:multiplets}.

 We have also investigated the problem of combining modular and 
generalised CP (gCP) invariance in theories based on $S_4'$. 
We have shown, in particular, that in such theories the CP 
transformation can be defined in two possible ways, which we have 
denoted as CP$_1$ and CP$_2$ (see Section~\ref{sec:gCP}). They act in the same way on the 
(VEV of the) modulus $\tau$, but 
the corresponding automorphisms act differently
on the generators $S$ and $T$ of $S_4'$.
The CP$_1$ transformation coincides with 
the one that can be employed in $\Gamma_2 \simeq S_3$, $\Gamma_3 \simeq A_4$, $\Gamma_4 \simeq S_4$, and $\Gamma_5 \simeq A_5$
modular-invariant theories~\cite{Novichkov:2019sqv}. 
The second transformation, CP$_2$, may or may not differ from CP$_1$ in practice
and is incompatible with certain combinations of modular weights and irreps. Note that CP$_2$ may also be consistently combined with other finite modular groups, such as $\Gamma_2 \simeq S_3$ and $\Gamma_4 \simeq S_4$.

 We have analysed in detail, in Section~\ref{sec:residual},
 the possible residual symmetries in
theories with modular invariance, and with modular and gCP invariance.
Depending on the value of $\tau$,
some generators of the full symmetry group
may be preserved.
The possible residual symmetry groups can be non-trivial
and are summarised in Table~\ref{tab:residual}.

 Finally, we have provided examples of
application of our results in Section~\ref{sec:pheno}, constructing
phenomenologically viable lepton flavour models
based on the finite modular $S'_4$ symmetry 
in which neutrino masses are generated 
by the Weinberg operator and by the type I seesaw mechanism.
Part of the novelty of these models lies in using (hatted) modular
forms not present in the $\Gamma_4 \simeq S_4$ construction.

The approach developed by us in the present article
simplifies considerably the parameterisation of
modular forms of level 4 and given weight. 
In particular, the derivation of $k>1$ modular multiplets in terms of just
two independent functions $\varepsilon$ and $\theta$
automatically bypasses a typical need to search for non-linear
constraints, which would relate redundant multiplets coming from
tensor products.
This approach  
can be useful in other setups based on modular symmetry,
for both homogeneous (double cover) and inhomogeneous
finite modular groups.

\section*{Acknowledgements}

We would like to thank F.~Capozzi, E.~Di Valentino, E.~Lisi, A.~Marrone, A.~Melchiorri and A.~Palazzo for kindly sharing with us the data files for one-dimensional $\chi^2$ projections. 
  This project has received funding from the European Union's Horizon 2020
  research and innovation programme under the Marie
  Skłodowska-Curie grant agreements No 674896
  (ITN Elusives) and No 690575 (RISE InvisiblesPlus).
  This work was supported in part
  by the INFN program on Theoretical Astroparticle Physics (P.P.N. and S.T.P.)
  and by the  World Premier International Research Center
  Initiative (WPI Initiative, MEXT), Japan (S.T.P.).
  The work of J.T.P.~was supported by
  Fundação para a Ciência e a Tecnologia (FCT, Portugal) through the projects
  CFTP-FCT Unit 777 (UIDB/00777/2020 and UIDP/00777/2020)
  and PTDC/FIS-PAR/29436/2017
  which are partially funded through POCTI (FEDER), COMPETE, QREN and EU.

\vfill
\pagebreak

\appendix

\section{Dedekind Eta and Jacobi Theta}
\label{app:eta_theta}
The Dedekind eta function is a holomorphic function defined in the complex upper half-plane as
\begin{equation}
  \eta(\tau) \,\equiv\, q^{\frac{1}{24}} \, \prod_{n=1}^{\infty} \left( 1-q^n \right) \,,
  \label{eq:eta_def}
\end{equation}
where $q \equiv e^{2\pi i \tau}$ and $\im\tau > 0$. In this work, fractional powers $q^{1/n}$, $n$ being a non-zero integer, should be read as $ e^{2\pi i \tau/n}$.

The Jacobi theta functions $\Theta_i(z, \tau)$, $i = 1, \ldots, 4$, (see e.g.~\cite{Kharchev:2015tv}) are special functions of two complex variables.
We are primarily interested in the so-called theta constants $\Theta_i(\tau) \equiv \Theta_i(0, \tau)$ which are functions of one complex variable defined in the upper half-plane by%
\footnote{In the notation of Ref.~\cite{Kharchev:2015tv} $q \equiv e^{\pi i \tau}$, which corresponds to $q^{1/2}$ in our notation.}
\begin{equation}
  \begin{aligned}
    \Theta_2(\tau) &\,\equiv\, \sum_{k \in \mathbb{Z}} q^{\frac{1}{2} \left( k+\frac{1}{2} \right)^2} \,, \\
    \Theta_3(\tau) &\,\equiv\, \sum_{k \in \mathbb{Z}} q^{\frac{k^2}{2}} \,, \\
    \Theta_4(\tau) &\,\equiv\, \sum_{k \in \mathbb{Z}} (-1)^k q^{\frac{k^2}{2}} \\
  \end{aligned}
  \label{eq:theta_def}
\end{equation}
(the first theta constant, $\Theta_1(\tau)$, is identically zero).
The theta constants transform under the generators of the modular group as
\begin{equation}
  \begin{aligned}
    \Theta_2(\tau) &\,\xrightarrow{T}\, e^{\frac{\pi i}{4}} \Theta_2(\tau)\,,\quad &
    \Theta_2(\tau) \,\xrightarrow{S}\, \sqrt{-i\tau} \, \Theta_4(\tau)\,,
    \\
    \Theta_3(\tau) &\,\xrightarrow{T}\, \Theta_4(\tau)\,,\quad &
    \Theta_3(\tau) \,\xrightarrow{S}\, \sqrt{-i\tau} \, \Theta_3(\tau)\,,
    \\
    \Theta_4(\tau) &\,\xrightarrow{T}\, \Theta_3(\tau)\,,\quad &
    \Theta_4(\tau) \,\xrightarrow{S}\, \sqrt{-i\tau} \, \Theta_2(\tau)\,.
  \end{aligned}
  \label{eq:theta_mod_trans}
\end{equation}
Note that in the $S$ transformation the principal value of the square root is assumed.

Apart from the power series expansions~\eqref{eq:theta_def}, the theta constants admit the following infinite product representations:
\begin{equation}
  \begin{aligned}
    \Theta_2(\tau) &= 2 \, q^{\frac{1}{8}} \prod_{n=1}^{\infty} \left( 1-q^n \right) \left( 1+q^n \right)^2, \\
    \Theta_3(\tau) &= \prod_{n=1}^{\infty} \left( 1-q^n \right) \left( 1+q^{n-\frac{1}{2}} \right)^2, \\
    \Theta_4(\tau) &= \prod_{n=1}^{\infty} \left( 1-q^n \right) \left( 1-q^{n-\frac{1}{2}} \right)^2.
  \end{aligned}
  \label{eq:theta_prod_exp}
\end{equation}
By comparing the product expansions~\eqref{eq:theta_prod_exp} with the definition of the Dedekind eta function~\eqref{eq:eta_def}, one can relate the theta constants to the Dedekind eta as
\begin{equation}
  \Theta_2(\tau) = \frac{2 \eta^2(2\tau)}{\eta(\tau)} \,, \quad
  \Theta_3(\tau) = \frac{\eta^5(\tau)}{\eta^2 \left( \frac{\tau}{2} \right) \eta^2(2\tau)} \,.
  \label{eq:theta_eta_rel}
\end{equation}

Finally, using the power series expansions~\eqref{eq:theta_def} one can prove a useful identity:
\begin{equation}
  \Theta_3(2\tau) = \frac{1}{2} \left[ \Theta_3 \left( \frac{\tau}{2} \right) + \Theta_4 \left( \frac{\tau}{2} \right) \right] \,.
  \label{eq:theta_double_arg}
\end{equation}

\section{Modular Forms of Level 4 in Terms of Theta Constants}
\label{app:mod4_theta}
The correspondence between modular forms of level 4 and the theta constants is well-known.
The classical result is~\cite{Mumford1983}
\begin{equation}
  \mathcal{M} (\Gamma(4)) \simeq \mathbb{C}
  \left[ \Theta_2^2(\tau), \Theta_3^2(\tau), \Theta_4^2(\tau) \right]
  \big/
  \left\{ \Theta_3^4(\tau) - \Theta_2^4(\tau) - \Theta_4^4(\tau) = 0 \right\} \,,
  \label{eq:mod4_theta_sq}
\end{equation}
i.e.~the ring of modular forms of level 4 is generated by the three squares of the theta constants subject to one non-linear relation
\begin{equation}
  \Theta_3^4(\tau) - \Theta_2^4(\tau) - \Theta_4^4(\tau) = 0 \,.
  \label{eq:theta_nonlinear}
\end{equation}
The idea we employ to avoid the non-linear relation~\eqref{eq:theta_nonlinear} is to re-express $\Theta_i^2(\tau)$ in terms of $\Theta_j(2\tau)$ using bilinear identities on the theta functions~\cite{Kharchev:2015tv}:
\begin{equation}
  \begin{aligned}
    \Theta_2^2(\tau) &= 2 \, \Theta_2(2\tau) \, \Theta_3(2\tau) \,,\\
    \Theta_3^2(\tau) &= \Theta_3^2(2\tau) + \Theta_2^2(2\tau) \,,\\
    \Theta_4^2(\tau) &= \Theta_3^2(2\tau) - \Theta_2^2(2\tau) \,.
  \end{aligned}
  \label{eq:theta_sq_double}
\end{equation}
The relation~\eqref{eq:theta_nonlinear} is automatically satisfied for the right-hand sides of eq.~\eqref{eq:theta_sq_double}, therefore, comparing~\eqref{eq:theta_sq_double} with the original polynomial ring~\eqref{eq:mod4_theta_sq}, we conclude that
\begin{equation}
  \mathcal{M}(\Gamma(4)) \simeq \mathbb{C}
  \left[ \Theta_2^2(2\tau),\, \Theta_3^2(2\tau),\, \Theta_2(2\tau) \, \Theta_3(2\tau) \right] \,,
\end{equation}
which means that modular forms of level 4 are homogeneous even-degree polynomials in $\Theta_2(2\tau)$ and $\Theta_3(2\tau)$.

\section{Group Theory of \texorpdfstring{$S_4'$}{S4'}}
\label{app:S4p}
\subsection{Properties and Irreducible Representations}
\label{app:propirreps}
The homogeneous finite modular group $S_4' \equiv SL(2,\mathbb{Z}_4)$
can be defined by three generators $S$, $T$ and $R$ satisfying the relations:
\begin{align}
S^2 = R\,,\quad
T^4 = (ST)^3 = R^2 = \id\,, \quad
TR = RT
\,.
\end{align}
It is a group of 48 elements (twice as many as $S_4$), with group ID \texttt{[48,30]} in the computer algebra system GAP~\cite{GAP4,SmallGroups}.
It admits 10 irreducible representations: 4 one-dimensional, 2 two-dimensional, and 4 three-dimensional, which we denote by
\begin{align}
    \mathbf{1}\,,\,
    \mathbf{\hat{1}}\,,\,
    \mathbf{1'}\,,\,
    \mathbf{\hat{1}'}\,,\,
    \mathbf{2}\,,\,
    \mathbf{\hat{2}}\,,\,
    \mathbf{3}\,,\,
    \mathbf{\hat{3}}\,,\,
    \mathbf{3'}\,,\,
    \mathbf{\hat{3}'}\,.
\end{align}
The notation has been chosen such that irreps without a hat have a direct correspondence with $S_4$ irreps, whereas hatted irreps are novel and specific to $S_4'$. In fact, for the hatless irreps, the new generator $R$ is represented by the identity matrix and the construction effectively reduces to that of $S_4 \simeq S_4' \, \big/ \left\{ R = 1 \right\}$.
We also note that the hatless irreps are real, while the hatted irreps are complex except for $\mathbf{\hat{2}}$ which is pseudoreal.

The 48 elements of $S_4'$ are organised into 10 conjugacy classes.
The character table is given in Table~\ref{tab:characters} and shows at least one representative element for each class.
\begin{table}[ht]
  \centering
  \begin{tabular}{lccccccccccc}
    \toprule
    & Rep.~element(s)& $\mathbf{1}$ & $\mathbf{\hat{1}}$ & $\mathbf{1'}$ & $\mathbf{\hat{1}'}$ & $\mathbf{2}$ & $\mathbf{\hat{2}}$ & $\mathbf{3}$ & $\mathbf{\hat{3}}$ & $\mathbf{3'}$ & $\mathbf{\hat{3}'}$ \\
    \midrule
    $1C_1$ & $\id$ & 1 & 1 & 1 & 1 & 2 & 2 & 3 & 3 & 3 & 3 \\
    $1C_2$ & $R$ & 1 & $-1$ & 1 & $-1$ & 2 & $-2$ & 3 & $-3$ & 3 & $-3$ \\
    $3C_2$ & $T^2$ & 1 & $-1$ & 1 & $-1$ & 2 & $-2$ & $-1$ & 1 & $-1$ & 1 \\
    $3\hat{C}_2$ & $RT^2$ & 1 & 1 & 1 & 1 & 2 & 2 & $-1$ & $-1$ & $-1$ & $-1$ \\
    $6C_4$ & $S$ & 1 & $i$ & $-1$ & $-i$ & 0 & 0 & 1 & $i$ & $-1$ & $-i$ \\
    $6\hat{C}_4$ & $RS = S^{-1}$ & 1 & $-i$ & $-1$ & $i$ & 0 & 0 & 1 & $-i$ & $-1$ & $i$ \\
    $6C_4'$ & $T$ & 1 & $-i$ & $-1$ & $i$ & 0 & 0 & $-1$ & $i$ & 1 & $-i$ \\
    $6\hat{C}_4'$ & $RT$, $T^{-1}$ & 1 & $i$ & $-1$ & $-i$ & 0 & 0 & $-1$ & $-i$ & 1 & $i$\\
    $8C_3$ & $ST$ & 1 & 1 & 1 & 1 & $-1$ & $-1$ & 0 & 0 & 0 & 0 \\
    $8C_6$ & $RST$ & 1 & $-1$ & 1 & $-1$ & $-1$ & 1 & 0 & 0 & 0 & 0 \\
    \bottomrule
  \end{tabular}
  \caption{Character table for $S_4'$, obtained via the GAP \texttt{Irr()} function.
  $nC_k$ denotes a conjugacy class of $n$ elements of order $k$.}
  \label{tab:characters}
\end{table}

\subsection{Representation Basis}
\label{app:sym_basis}
In Table~\ref{tab:basis}, we summarise the working basis for the representation matrices of the group generators $S$, $T$ and $R$. In this basis, the group generators are represented by symmetric matrices, $\rho_\mathbf{r}(S,T,R) =\rho_\mathbf{r}(S,T,R)^T$, for all irreps $\mathbf{r}$ of $S_4'$. Such a basis is convenient for the study of modular symmetry extended by a gCP symmetry (see Section~\ref{sec:gCP} and Ref.~\cite{Novichkov:2019sqv}).

\begin{table}[ht]
\centering
\begin{tabular}{lccc}
    \toprule
    $\mathbf{r}$ & $\rho_{\mathbf{r}}(S)$ & $\rho_{\mathbf{r}}(T)$ & $\rho_{\mathbf{r}}(R)$ \\
    \midrule
    $\mathbf{1}$ & $1$ & $1$ & $1$ \\
    \addlinespace
    $\mathbf{\hat{1}}$ & $i$ & $-i$ & $-1$ \\
    \addlinespace
    $\mathbf{1'}$ & $-1$ & $-1$ & $1$ \\
    \addlinespace
    $\mathbf{\hat{1}'}$ & $-i$ & $i$ & $-1$ \\
    \addlinespace
    $\mathbf{2}$
        & $\dfrac{1}{2} \begin{pmatrix} -1 & \sqrt{3} \\ \sqrt{3} & 1 \end{pmatrix}$
        & $\begin{pmatrix} 1 & 0 \\ 0 & -1 \end{pmatrix}$
        & $\begin{pmatrix} 1 & 0 \\ 0 & 1 \end{pmatrix}$
    \\
    \addlinespace
    $\mathbf{\hat{2}}$
        & $\dfrac{i}{2} \begin{pmatrix} -1 & \sqrt{3} \\ \sqrt{3} & 1 \end{pmatrix}$
        & $\begin{pmatrix} -i & 0 \\ 0 & i \end{pmatrix}$
        & $-\begin{pmatrix} 1 & 0 \\ 0 & 1 \end{pmatrix}\phantom{-}$
    \\
    \addlinespace
    $\mathbf{3}$
        & $-\dfrac{1}{2} \begin{pmatrix} 0 & \sqrt{2} & \sqrt{2} \\ \sqrt{2} & -1 & 1 \\ \sqrt{2} & 1 & -1 \end{pmatrix}$
        & $\begin{pmatrix} -1 & 0 & 0 \\ 0 & -i & 0 \\ 0 & 0 & i \end{pmatrix}$
        & $\begin{pmatrix} 1 & 0 & 0 \\ 0 & 1 & 0 \\ 0 & 0 & 1 \end{pmatrix}$
    \\
    \addlinespace
    $\mathbf{\hat{3}}$
        & $-\dfrac{i}{2} \begin{pmatrix} 0 & \sqrt{2} & \sqrt{2} \\ \sqrt{2} & -1 & 1 \\ \sqrt{2} & 1 & -1 \end{pmatrix}$
        & $\begin{pmatrix} i & 0 & 0 \\ 0 & -1 & 0 \\ 0 & 0 & 1 \end{pmatrix}$
        & $-\begin{pmatrix} 1 & 0 & 0 \\ 0 & 1 & 0 \\ 0 & 0 & 1 \end{pmatrix}\phantom{-}$
    \\
    \addlinespace
    $\mathbf{3'}$
        & $\dfrac{1}{2} \begin{pmatrix} 0 & \sqrt{2} & \sqrt{2} \\ \sqrt{2} & -1 & 1 \\ \sqrt{2} & 1 & -1 \end{pmatrix}$
        & $\begin{pmatrix} 1 & 0 & 0 \\ 0 & i & 0 \\ 0 & 0 & -i \end{pmatrix}$
        & $\begin{pmatrix} 1 & 0 & 0 \\ 0 & 1 & 0 \\ 0 & 0 & 1 \end{pmatrix}$
    \\
    \addlinespace
    $\mathbf{\hat{3}'}$
        & $\dfrac{i}{2} \begin{pmatrix} 0 & \sqrt{2} & \sqrt{2} \\ \sqrt{2} & -1 & 1 \\ \sqrt{2} & 1 & -1 \end{pmatrix}$
        & $\begin{pmatrix} -i & 0 & 0 \\ 0 & 1 & 0 \\ 0 & 0 & -1 \end{pmatrix}$
        & $-\begin{pmatrix} 1 & 0 & 0 \\ 0 & 1 & 0 \\ 0 & 0 & 1 \end{pmatrix}\phantom{-}$
    \\
    \bottomrule
\end{tabular}
\caption{Representation matrices for the group generators in different $S_4'$ irreps $\mathbf{r}$.}
\label{tab:basis}
\end{table}

\subsection{Tensor Products and Clebsch-Gordan Coefficients}
\label{app:CGCs}

We present here the decompositions of tensor products of $S_4'$ irreps, as well as the corresponding Clebsch-Gordan coefficients,
given in the basis of Table~\ref{tab:basis}.
Entries of each multiplet entering the tensor product are denoted by $\alpha_i$ and $\beta_i$. Apart from the trivial products $\mathbf{1} \otimes \mathbf{r} = \mathbf{r}$, these results are collected in Tables~\ref{tab:1r}\,--\,\ref{tab:33}.

\begin{table}[ht]
\centering
\renewcommand{\arraystretch}{1.2}
\begin{tabular}{cc}
\toprule
\addlinespace
   ${ }$ \qquad Tensor product decomposition \qquad ${ }$
&  ${ }$ \qquad Clebsch-Gordan coefficients  \qquad ${ }$\\
\addlinespace
\midrule
\addlinespace
$\begin{array}{@{}c@{{}\,\otimes\,{}}c@{{}\,\,=\,\,{}}l@{}}
\mathbf{1'} & \mathbf{1'}       & \mathbf{1}        \\
\mathbf{1'} & \mathbf{\hat{1}}  & \mathbf{\hat{1}'} \\
\mathbf{1'} & \mathbf{\hat{1}'} & \mathbf{\hat{1}}  \\
\mathbf{\hat{1}}  & \mathbf{\hat{1}}  & \mathbf{1'} \\
\mathbf{\hat{1}}  & \mathbf{\hat{1}'} & \mathbf{1}  \\
\mathbf{\hat{1}'} & \mathbf{\hat{1}'} & \mathbf{1'}
\end{array}$
& $\alpha_1 \beta_1$ \\
\addlinespace
\midrule
\addlinespace
$\begin{array}{@{}c@{{}\,\otimes\,{}}c@{{}\,\,=\,\,{}}l@{}}
\mathbf{1'}       & \mathbf{2}       & \mathbf{2} \\
\mathbf{\hat{1}'} & \mathbf{2}       & \mathbf{\hat{2}} \\
\mathbf{1'}       & \mathbf{\hat{2}} & \mathbf{\hat{2}} \\
\mathbf{\hat{1}} & \mathbf{\hat{2}} & \mathbf{2}
\end{array}$
& $\alpha_1 \begin{pmatrix} \beta_2\\ -\beta_1 \end{pmatrix}$ \\
\addlinespace
\midrule
\addlinespace
$\begin{array}{@{}c@{{}\,\otimes\,{}}c@{{}\,\,=\,\,{}}l@{}}
\mathbf{\hat{1}}  & \mathbf{2}       & \mathbf{\hat{2}} \\
\mathbf{\hat{1}'} & \mathbf{\hat{2}} & \mathbf{2}
\end{array}$
&
$\alpha_1 \begin{pmatrix} \beta_1\\ \beta_2 \end{pmatrix}$ \\
\addlinespace
\midrule
\addlinespace
$\begin{array}{@{}c@{{}\,\otimes\,{}}c@{{}\,\,=\,\,{}}l@{}}
\mathbf{1'}       & \mathbf{3}        & \mathbf{3'}       \\
\mathbf{\hat{1}}  & \mathbf{3}        & \mathbf{\hat{3}}  \\
\mathbf{\hat{1}'} & \mathbf{3}        & \mathbf{\hat{3}'} \\
\mathbf{1'}       & \mathbf{3'}       & \mathbf{3}        \\
\mathbf{\hat{1}}  & \mathbf{3'}       & \mathbf{\hat{3}'} \\
\mathbf{\hat{1}'} & \mathbf{3'}       & \mathbf{\hat{3}}  \\
\mathbf{1'}       & \mathbf{\hat{3}}  & \mathbf{\hat{3}'} \\
\mathbf{\hat{1}}  & \mathbf{\hat{3}}  & \mathbf{3'}       \\
\mathbf{\hat{1}'} & \mathbf{\hat{3}}  & \mathbf{3}        \\
\mathbf{1'}       & \mathbf{\hat{3}'} & \mathbf{\hat{3}}  \\
\mathbf{\hat{1}}  & \mathbf{\hat{3}'} & \mathbf{3}        \\
\mathbf{\hat{1}'} & \mathbf{\hat{3}'} & \mathbf{3'}
\end{array}$
&
$\alpha_1 \begin{pmatrix} \beta_1\\ \beta_2 \\ \beta_3 \end{pmatrix}$ \\
\addlinespace
\bottomrule
\end{tabular}
\caption{Decomposition of all non-trivial tensor products involving 1-dimensional $S_4'$ irreps, and corresponding Clebsch-Gordan coefficients.}
\label{tab:1r}
\end{table}

\begin{table}[ht]
\centering
\renewcommand{\arraystretch}{1.2}
\begin{tabular}{cc}
\toprule
\addlinespace
   ${ }$ \qquad Tensor product decomposition \qquad ${ }$
&  ${ }$ \qquad Clebsch-Gordan coefficients  \qquad ${ }$\\
\addlinespace
\midrule
\addlinespace
$\begin{array}{@{}c@{{}\,\otimes\,{}}c@{{}\,\,=\,\,{}}l@{{}\,\oplus\,{}}l@{{}\,\oplus\,{}}l@{}}
\mathbf{2} & \mathbf{2} & \mathbf{1} & \mathbf{1'} & \mathbf{2} \\[1mm]
\mathbf{2} & \mathbf{\hat{2}} & \mathbf{\hat{1}} &\mathbf{\hat{1}'} & \mathbf{\hat{2}}
\end{array}$
&
$\begin{array}{rl}
&\dfrac{1}{\sqrt{2}}\left(\alpha_1 \beta_1 + \alpha_2 \beta_2 \right) \\[3mm]
\oplus& \dfrac{1}{\sqrt{2}}\left(\alpha_1 \beta_2 - \alpha_2 \beta_1 \right) \\[3mm]
\oplus& \dfrac{1}{\sqrt{2}} \begin{pmatrix}
\alpha_2 \beta_2 - \alpha_1 \beta_1 \\
\alpha_1 \beta_2 + \alpha_2 \beta_1
\end{pmatrix}
\end{array}$ \\
\addlinespace
\midrule
\addlinespace
$\begin{array}{@{}c@{{}\,\otimes\,{}}c@{{}\,\,=\,\,{}}l@{{}\,\oplus\,{}}l@{{}\,\oplus\,{}}l@{}}
\mathbf{\hat{2}} & \mathbf{\hat{2}} & \mathbf{1} & \mathbf{1'} & \mathbf{2}
\end{array}$
&
$\begin{array}{rl}
       &\dfrac{1}{\sqrt{2}}\left(\alpha_1 \beta_2 - \alpha_2 \beta_1 \right)\\
\oplus &\dfrac{1}{\sqrt{2}}\left(\alpha_1 \beta_1 + \alpha_2 \beta_2 \right)\\
\oplus &\dfrac{1}{\sqrt{2}} \begin{pmatrix}
\alpha_1 \beta_2 + \alpha_2 \beta_1 \\
\alpha_1 \beta_1 - \alpha_2 \beta_2
\end{pmatrix}
\end{array}$ \\
\addlinespace
\bottomrule
\end{tabular}
\caption{Decomposition of tensor products involving two 2-dimensional $S_4'$ irreps, and corresponding Clebsch-Gordan coefficients. Note that the order is important in order to match the left and right columns.}
\label{tab:22}
\end{table}

\begin{table}[ht]
\centering
\renewcommand{\arraystretch}{1.2}
\begin{tabular}{cc}
\toprule
\addlinespace
   ${ }$ \qquad Tensor product decomposition \qquad ${ }$
&  ${ }$ \qquad Clebsch-Gordan coefficients  \qquad ${ }$\\
\addlinespace
\midrule
\addlinespace
$\begin{array}{@{}c@{{}\,\otimes\,{}}c@{{}\,\,=\,\,{}}l@{{}\,\oplus\,{}}l@{}}
\mathbf{2}       & \mathbf{3}        & \mathbf{3}       & \mathbf{3'}       \\[1mm]
\mathbf{2}       & \mathbf{\hat{3}}  & \mathbf{\hat{3}} & \mathbf{\hat{3}'} \\[1mm]
\mathbf{\hat{2}} & \mathbf{3}        & \mathbf{\hat{3}} & \mathbf{\hat{3}'} \\[1mm]
\mathbf{\hat{2}} & \mathbf{\hat{3}'} & \mathbf{3}       & \mathbf{3'}
\end{array}$
&
$\begin{array}{rl}
& \begin{pmatrix}
\alpha_1\,\beta_1 \\
\left(\sqrt{3}/2\right)\alpha_2\,\beta_3 - \left(1/2\right)\alpha_1\,\beta_2 \\
\left(\sqrt{3}/2\right)\alpha_2\,\beta_2 - \left(1/2\right)\alpha_1\,\beta_3
\end{pmatrix}
\\
\oplus & \begin{pmatrix} - \alpha_2\,\beta_1 \\
\left(\sqrt{3}/2\right)\alpha_1\,\beta_3 + \left(1/2\right)\alpha_2\,\beta_2 \\
\left(\sqrt{3}/2\right)\alpha_1\,\beta_2 + \left(1/2\right)\alpha_2\,\beta_3
\end{pmatrix}
\end{array}$ \\
\addlinespace
\midrule
\addlinespace
$\begin{array}{@{}c@{{}\,\otimes\,{}}c@{{}\,\,=\,\,{}}l@{{}\,\oplus\,{}}l@{}}
\mathbf{2}       & \mathbf{3'}       & \mathbf{3}       & \mathbf{3'}       \\[1mm]
\mathbf{2}       & \mathbf{\hat{3}'} & \mathbf{\hat{3}} & \mathbf{\hat{3}'} \\[1mm]
\mathbf{\hat{2}} & \mathbf{3'}       & \mathbf{\hat{3}} & \mathbf{\hat{3}'} \\[1mm]
\mathbf{\hat{2}} & \mathbf{\hat{3}}  & \mathbf{3}       & \mathbf{3'}
\end{array}$
&
$\begin{array}{rl}
& \begin{pmatrix} - \alpha_2\,\beta_1 \\
\left(\sqrt{3}/2\right)\alpha_1\,\beta_3 + \left(1/2\right)\alpha_2\,\beta_2 \\
\left(\sqrt{3}/2\right)\alpha_1\,\beta_2 + \left(1/2\right)\alpha_2\,\beta_3
\end{pmatrix}
\\
\oplus & \begin{pmatrix}
\alpha_1\,\beta_1 \\
\left(\sqrt{3}/2\right)\alpha_2\,\beta_3 - \left(1/2\right)\alpha_1\,\beta_2 \\
\left(\sqrt{3}/2\right)\alpha_2\,\beta_2 - \left(1/2\right)\alpha_1\,\beta_3
\end{pmatrix}
\end{array}$ \\
\addlinespace
\bottomrule
\end{tabular}
\caption{The same as in Table~\ref{tab:22}, but for products involving a 2-dimensional and a 3-dimensional irrep.}
\label{tab:23}
\end{table}

\begin{table}[ht]
\centering
\renewcommand{\arraystretch}{1.2}
\begin{tabular}{cc}
\toprule
\addlinespace
   ${ }$ \qquad Tensor product decomposition \qquad ${ }$
&  ${ }$ \qquad Clebsch-Gordan coefficients  \qquad ${ }$\\
\addlinespace
\midrule
\addlinespace
$\begin{array}{@{}c@{{}\,\otimes\,{}}c@{{}\,\,=\,\,{}}l@{{}\,\oplus\,{}}l@{{}\,\oplus\,{}}l@{{}\,\oplus\,{}}l@{}}
\mathbf{3}       & \mathbf{3}        &
\mathbf{1}       & \mathbf{2}        & \mathbf{3}       & \mathbf{3'}       \\[1mm]
\mathbf{3}       & \mathbf{\hat{3}}  &
\mathbf{\hat{1}} & \mathbf{\hat{2}}  & \mathbf{\hat{3}} & \mathbf{\hat{3}'} \\[1mm]
\mathbf{3'}      & \mathbf{3'}       &
\mathbf{1}       & \mathbf{2}        & \mathbf{3}       & \mathbf{3'}       \\[1mm]
\mathbf{3'}      & \mathbf{\hat{3}'} &
\mathbf{\hat{1}} & \mathbf{\hat{2}}  & \mathbf{\hat{3}} & \mathbf{\hat{3}'} \\[1mm]
\mathbf{\hat{3}} & \mathbf{\hat{3}'} &
\mathbf{1}       & \mathbf{2}        & \mathbf{3}       & \mathbf{3'}
\end{array}$
&
$\begin{array}{rl}
& \dfrac{1}{\sqrt{3}}\left(\alpha_1\beta_1+\alpha_2\beta_3+\alpha_3\beta_2\right)
\\
\oplus & \dfrac{1}{\sqrt{2}}
\begin{pmatrix}
\left(2\alpha_1\beta_1- \alpha_2\beta_3- \alpha_3\beta_2\right)/\sqrt{3} \\
\alpha_2\beta_2+\alpha_3\beta_3
\end{pmatrix}
\\
\oplus & \dfrac{1}{\sqrt{2}}
\begin{pmatrix}
\alpha_3\beta_3-\alpha_2\beta_2 \\
\alpha_1\beta_3+\alpha_3\beta_1 \\
-\alpha_1\beta_2-\alpha_2\beta_1
\end{pmatrix}
\\
\oplus & \dfrac{1}{\sqrt{2}}
\begin{pmatrix}
\alpha_3\beta_2-\alpha_2\beta_3 \\
\alpha_2\beta_1-\alpha_1\beta_2\\
\alpha_1\beta_3-\alpha_3\beta_1
\end{pmatrix}
\end{array}$ \\
\addlinespace
\midrule
\addlinespace
$\begin{array}{@{}c@{{}\,\otimes\,{}}c@{{}\,\,=\,\,{}}l@{{}\,\oplus\,{}}l@{{}\,\oplus\,{}}l@{{}\,\oplus\,{}}l@{}}
\mathbf{3}        & \mathbf{3'}       &
\mathbf{1'}       & \mathbf{2}        & \mathbf{3}       & \mathbf{3'}       \\[1mm]
\mathbf{3}        & \mathbf{\hat{3}'} &
\mathbf{\hat{1}'} & \mathbf{\hat{2}}  & \mathbf{\hat{3}} & \mathbf{\hat{3}'} \\[1mm]
\mathbf{3'}       & \mathbf{\hat{3}}  &
\mathbf{\hat{1}'} & \mathbf{\hat{2}}  & \mathbf{\hat{3}} & \mathbf{\hat{3}'} \\[1mm]
\mathbf{\hat{3}}  & \mathbf{\hat{3}}  &
\mathbf{1'}       & \mathbf{2}        & \mathbf{3}       & \mathbf{3'}       \\[1mm]
\mathbf{\hat{3}'} & \mathbf{\hat{3}'} &
\mathbf{1'}       & \mathbf{2}        & \mathbf{3}       & \mathbf{3'}
\end{array}$
&
$\begin{array}{rl}
& \dfrac{1}{\sqrt{3}}\left(\alpha_1\beta_1+\alpha_2\beta_3+\alpha_3\beta_2\right)
\\
\oplus & \dfrac{1}{\sqrt{2}}
\begin{pmatrix}
\alpha_2\beta_2+\alpha_3\beta_3 \\
\left(-2\alpha_1\beta_1 + \alpha_2\beta_3+  \alpha_3\beta_2\right)/\sqrt{3}
\end{pmatrix}
\\
\oplus & \dfrac{1}{\sqrt{2}}
\begin{pmatrix}
\alpha_3\beta_2-\alpha_2\beta_3 \\
\alpha_2\beta_1-\alpha_1\beta_2\\
\alpha_1\beta_3-\alpha_3\beta_1
\end{pmatrix}
\\
\oplus & \dfrac{1}{\sqrt{2}}
\begin{pmatrix}
\alpha_3\beta_3-\alpha_2\beta_2 \\
\alpha_1\beta_3+\alpha_3\beta_1 \\
-\alpha_1\beta_2-\alpha_2\beta_1
\end{pmatrix}
\end{array}$ \\
\addlinespace
\bottomrule
\end{tabular}
\caption{The same as in Table~\ref{tab:22}, but for products involving two 3-dimensional irreps.}
\label{tab:33}
\end{table}

\vfill
\clearpage
\section{Higher Weight Modular Multiplets for \texorpdfstring{$S_4'$}{S4'}}
\label{app:multiplets}
Modular multiplets for the homogeneous finite modular group $S_4'$ can be written in terms of the functions $\theta(\tau)$ and $\varepsilon(\tau)$ of eqs.~\eqref{eq:theta_eps_def} and \eqref{eq:theta_eps_qexp}. For weights $k=1,\ldots,4$, they are given in eqs.~\eqref{eq:k1triplet}, \eqref{eq:k2}, \eqref{eq:k3}, and \eqref{eq:k4}, respectively.
In the present Appendix we collect higher weight multiplets, up to $k=10$.
All multiplets contained in this paper have been obtained from the $k=1$ triplet of eq.~\eqref{eq:k1triplet} using the Clebsch-Gordan coefficients of Appendix~\ref{app:CGCs} and respecting the corresponding normalisations, up to a sign.
For $k=5$, one has:
\begin{align*}
  Y_{\mathbf{\hat{2}}}^{(5)}(\tau) &=
  \begin{pmatrix}
  \frac{3}{2}\left( \varepsilon ^3 \,\theta ^7-\varepsilon ^7 \,\theta ^3 \right)\\[1mm]
  \frac{\sqrt{3}}{4}\left( \varepsilon  \,\theta ^9-\varepsilon ^9 \,\theta   \right)
  \end{pmatrix}\,,
  \\[2mm]
  Y_{\mathbf{\hat{3}},1}^{(5)}(\tau) &=
  \begin{pmatrix}
  \frac{6\sqrt{2}}{\sqrt{5}} \,
\varepsilon ^5\, \theta ^5
\\[1mm]
\:\:\:\frac{3}{8\sqrt{5}} \left(
5\, \varepsilon ^2\, \theta ^8
+10\, \varepsilon ^6\, \theta ^4
+\varepsilon ^{10}
\right) \\[1mm]
-\frac{3}{8\sqrt{5}} \left(
\theta ^{10}
+10\, \varepsilon ^4\, \theta ^6
+5\, \varepsilon ^8\, \theta ^2
\right)
  \end{pmatrix}\,,
  \\[2mm]
  Y_{\mathbf{\hat{3}},2}^{(5)}(\tau) &=
  \begin{pmatrix}
  \frac{3}{4}\left( \varepsilon  \, \theta ^9 -2 \,\varepsilon ^5 \, \theta ^5 + \varepsilon ^9 \, \theta \right)\\[1mm]
  \frac{3}{\sqrt{2}}\left(-\varepsilon ^2 \, \theta ^8 + \varepsilon ^6 \, \theta ^4\right)\\[1mm]
  \frac{3}{\sqrt{2}}\left(-\varepsilon ^4 \, \theta ^6 + \varepsilon ^8 \, \theta ^2\right)
  \end{pmatrix}\,,
  \\[2mm]
  Y_{\mathbf{\hat{3}'}}^{(5)}(\tau) &=
  \begin{pmatrix}
  2\left(\varepsilon ^3\, \theta ^7 +  \varepsilon ^7 \,\theta ^3\right)\\[1mm]
  \frac{1}{4\sqrt{2}}\left(\theta ^{10} -14 \,\varepsilon ^4\, \theta ^6  -3 \,\varepsilon ^8\, \theta ^2 \right) \\[1mm]
  \frac{1}{4\sqrt{2}}\left( 3\, \varepsilon ^2\, \theta ^8 + 14\, \varepsilon ^6\, \theta ^4 -\varepsilon ^{10 }\right)
  \end{pmatrix}\,.
\end{align*}

For $k=6$, one has:
\begin{align*}
  Y_{\mathbf{1}}^{(6)}(\tau) &=
  \frac{1}{4\sqrt{6}} \left(
\theta ^{12}
-33 \,\varepsilon ^4 \,\theta ^8
-33 \,\varepsilon ^8 \,\theta ^4
+\varepsilon ^{12}
\right)
  \,,\\[2mm]
  Y_{\mathbf{1'}}^{(6)}(\tau) &=
  \frac{3}{2}\sqrt{\frac{3}{2}} \left(
\varepsilon ^2\, \theta ^{10}
-2 \,\varepsilon ^6\, \theta ^6
+ \varepsilon ^{10} \,\theta ^2
\right)
  \,,\\[2mm]
  Y_{\mathbf{2}}^{(6)}(\tau) &=
  \begin{pmatrix}
  \frac{1}{8} \left(
\theta ^{12}
+15\, \varepsilon ^4\, \theta ^8
+15\, \varepsilon ^8\, \theta ^4
+\varepsilon ^{12}
\right) \\[1mm]
-\frac{\sqrt{3}}{4} \left(
\varepsilon ^2\, \theta ^{10}
+14\, \varepsilon ^6\, \theta ^6
+\varepsilon ^{10}\, \theta ^2
\right)
  \end{pmatrix}\,,
  \\[2mm]
  Y_{\mathbf{3}}^{(6)}(\tau) &=
  \begin{pmatrix}
\frac{3}{2}
\left(\varepsilon ^2 \,\theta ^{10}-\varepsilon ^{10}\, \theta ^2 \right)
		 \\[1mm]
\frac{3}{4\sqrt{2}} \left(
 5\, \varepsilon ^3\, \theta ^9
 -6\, \varepsilon ^7\, \theta ^5
 +\varepsilon ^{11} \,\theta
		\right) \\[1mm]
\frac{3}{4\sqrt{2}} \left(
 \varepsilon\,  \theta ^{11}
 -6 \,\varepsilon ^5\, \theta ^7
 +5 \,\varepsilon ^9\, \theta ^3
		\right)
  \end{pmatrix}\,,
  \\[2mm]
    Y_{\mathbf{3'},1}^{(6)}(\tau) &=
  \begin{pmatrix}
  -\frac{3}{8\sqrt{13}}\left(
\theta ^{12}
 -3\, \varepsilon ^4\, \theta ^8
 +3\, \varepsilon ^8\, \theta ^4
 -\varepsilon ^{12}
\right)
\\[1mm]
\frac{3 \sqrt{2}}{\sqrt{13}}\left(
 3 \,\varepsilon ^5\, \theta ^7
 + \varepsilon ^9\, \theta ^3
\right) \\[1mm]
\frac{3 \sqrt{2}}{\sqrt{13}} \left(
 \varepsilon ^3\, \theta ^9
 + 3\, \varepsilon ^7\, \theta ^5
\right)
  \end{pmatrix}\,,
  \\[2mm]
  Y_{\mathbf{3'},2}^{(6)}(\tau) &=
  \begin{pmatrix}
    3\left(
     \varepsilon ^4\, \theta ^8
 -\varepsilon ^8\, \theta ^4
     \right)\\[1mm]
  -\frac{3}{4\sqrt{2}}\left(
 \varepsilon\,  \theta ^{11}
 +2\, \varepsilon ^5\, \theta ^7
 -3\, \varepsilon ^9\, \theta ^3
  \right)\\[1mm]
 \:\:\:\, \frac{3}{4\sqrt{2}}\left(
 3\, \varepsilon ^3\, \theta ^9
 -2\, \varepsilon ^7\, \theta ^5
 -\varepsilon ^{11}\, \theta
  \right)
\end{pmatrix}\,.
\end{align*}

For $k=7$, one has:
\begin{align*}
  Y_{\mathbf{\hat{1}'}}^{(7)}(\tau) &=
  \frac{1}{4} \sqrt{\frac{3}{2}} \left(
-\varepsilon ^{13}\, \theta
-13\, \varepsilon ^9\, \theta ^5
+13\, \varepsilon ^5\, \theta ^9
+\varepsilon\,  \theta ^{13}
\right)
  \,,\\[2mm]
  Y_{\mathbf{\hat{2}}}^{(7)}(\tau) &=
  \begin{pmatrix}\frac{3}{2} \left(
\varepsilon ^3\, \theta ^{11}
-\varepsilon ^{11} \,\theta ^3
\right) \\[1mm]
-\frac{\sqrt{3}}{8} \left(
\varepsilon\,  \theta ^{13}
-11\, \varepsilon ^5\, \theta ^9
+11\, \varepsilon ^9\, \theta ^5
-\varepsilon ^{13} \,\theta
\right)
  \end{pmatrix}\,,
  \\[2mm]
  Y_{\mathbf{\hat{3}},1}^{(7)}(\tau) &=
  \begin{pmatrix}
  \frac{12}{\sqrt{13}} \left(
\varepsilon ^5\, \theta ^9
 +\varepsilon ^9\, \theta ^5
\right) \\[1mm]
\frac{3}{8\sqrt{26}} \left(
\varepsilon ^2\, \theta ^{12}
 +45\, \varepsilon ^6\, \theta ^8
 +19\, \varepsilon ^{10} \,\theta ^4
 -\varepsilon ^{14}
\right) \\[1mm]
\frac{3}{8\sqrt{26}} \left(
 \theta ^{14}
 -19\, \varepsilon ^4\, \theta ^{10}
 -45\, \varepsilon ^8\, \theta ^6
 -\varepsilon ^{12}\, \theta ^2
\right)
\end{pmatrix}\,,
  \\[2mm]
  Y_{\mathbf{\hat{3}},2}^{(7)}(\tau) &=
  \begin{pmatrix}
  \frac{3}{8} \left(
 \varepsilon\,  \theta ^{13}
 -\varepsilon ^5\, \theta ^9
 -\varepsilon ^9\, \theta ^5
 +\varepsilon ^{13}\, \theta
\right) \\[1mm]
\frac{3}{4\sqrt{2}} \left(
 \varepsilon ^2\, \theta ^{12}
 +6\, \varepsilon ^6\, \theta ^8
 -7\, \varepsilon ^{10} \,\theta ^4
\right) \\[1mm]
\frac{3}{4\sqrt{2}} \left(
 7\, \varepsilon ^4\, \theta ^{10}
 -6\, \varepsilon ^8\, \theta ^6
 -\varepsilon ^{12} \,\theta ^2
\right)
  \end{pmatrix}\,,
  \\[2mm]
  Y_{\mathbf{\hat{3}'},1}^{(7)}(\tau) &=
  \begin{pmatrix}
  \frac{3}{4\sqrt{37}}\left(
  7\, \varepsilon ^3 \,\theta ^{11}
 +50\, \varepsilon ^7\, \theta ^7
 +7\, \varepsilon ^{11}\, \theta ^3
\right) \\[1mm]
-\frac{3}{4\sqrt{74}} \left(
\theta ^{14}
 +14\, \varepsilon ^4\, \theta ^{10}
 +49\, \varepsilon ^8\, \theta ^6
\right) \\[1mm]
\:\:\:\frac{3}{4\sqrt{74}} \left(
49\, \varepsilon ^6\, \theta ^8
 +14\, \varepsilon ^{10} \,\theta ^4
 +\varepsilon ^{14}
\right)
\end{pmatrix}\,,
  \\[2mm]
  Y_{\mathbf{\hat{3}'},2}^{(7)}(\tau) &=
  \begin{pmatrix}
\frac{9}{4}
\left(  \varepsilon ^3\, \theta ^{11}
 -2\, \varepsilon ^7\, \theta ^7
 +\varepsilon ^{11} \,\theta ^3 \right)
\\[1mm]
\:\:\:\frac{9}{4\sqrt{2}} \left(
\varepsilon ^4\, \theta ^{10}
 -2\, \varepsilon ^8\, \theta ^6
 +\varepsilon ^{12}\, \theta ^2
\right) \\[1mm]
-\frac{9}{4\sqrt{2}} \left(
\varepsilon ^2\, \theta ^{12}
 -2\, \varepsilon ^6\, \theta ^8
 +\varepsilon ^{10} \,\theta ^4
\right)
  \end{pmatrix}\,.
\end{align*}

For $k=8$, one has:
\begin{align*}
  Y_{\mathbf{1}}^{(8)}(\tau) &=
\frac{1}{8\sqrt{6}}
\left(
 \theta^{16}
+28 \,\varepsilon^4 \,\theta^{12}
+198 \,\varepsilon^8 \,\theta^8
+28 \,\varepsilon^{12} \,\theta^4
+\varepsilon^{16}
\right)
  \,,\\[2mm]
  Y_{\mathbf{2},1}^{(8)}(\tau) &=
  \begin{pmatrix}
\frac{9}{16\sqrt{82}}
\left(
 \theta^{16}
-130 \,\varepsilon^8 \,\theta^8
+\varepsilon^{16}
\right)
\\[1mm]
\frac{3}{8}\sqrt{\frac{3}{82}}
\left(
 5 \,\varepsilon^2 \,\theta^{14}
+91 \,\varepsilon^6 \,\theta^{10}
+91 \,\varepsilon^{10} \,\theta^6
+5 \,\varepsilon^{14} \,\theta^2
\right)
  \end{pmatrix}\,,
  \\[2mm]
  Y_{\mathbf{2},2}^{(8)}(\tau) &=
  \begin{pmatrix}
\frac{9}{4}
\left(
 \varepsilon^4 \,\theta^{12}
-2 \,\varepsilon^8 \,\theta^8
+\varepsilon^{12} \,\theta^4
\right)
\\[1mm]
\frac{3\sqrt{3}}{8}
\left(
 \varepsilon^2 \,\theta^{14}
-\varepsilon^6 \,\theta^{10}
-\varepsilon^{10} \,\theta^6
+\varepsilon^{14} \,\theta^2
\right)
  \end{pmatrix}\,,
  \\[2mm]
  Y_{\mathbf{3},1}^{(8)}(\tau) &=
  \begin{pmatrix}
9\sqrt{\frac{2}{5}}
\left(
 \varepsilon^6 \,\theta^{10}
-\varepsilon^{10} \,\theta^6
\right)
\\[1mm]
\:\:\:\,\frac{9}{16\sqrt{5}}
\left(
 5 \,\varepsilon^3 \,\theta^{13}
+5 \,\varepsilon^7 \,\theta^9
-9 \,\varepsilon^{11} \,\theta^5
-\varepsilon^{15} \,\theta
\right)
\\[1mm]
-\frac{9}{16\sqrt{5}}
\left(
 \varepsilon \, \theta^{15}
+9 \,\varepsilon^5 \,\theta^{11}
-5 \,\varepsilon^9 \,\theta^7
-5 \,\varepsilon^{13} \,\theta^3
\right)
  \end{pmatrix}\,,
  \\[2mm]
  Y_{\mathbf{3},2}^{(8)}(\tau) &=
  \begin{pmatrix}
-\frac{9}{8}
\left(
 \varepsilon^2 \,\theta^{14}
-3 \,\varepsilon^6 \,\theta^{10}
+3 \,\varepsilon^{10} \,\theta^6
-\varepsilon^{14} \,\theta^2
\right)
\\[1mm]
\frac{9}{2\sqrt{2}}
\left(
\varepsilon^3 \,\theta^{13}
-2 \,\varepsilon^7 \,\theta^9
+\varepsilon^{11} \,\theta^5
\right)
\\[1mm]
\frac{9}{2\sqrt{2}}
\left(
 \varepsilon^5 \,\theta^{11}
-2 \,\varepsilon^9 \,\theta^7
+\varepsilon^{13} \,\theta^3
\right)
  \end{pmatrix}\,,
  \\[2mm]
  Y_{\mathbf{3'},1}^{(8)}(\tau) &=
  \begin{pmatrix}
\frac{3}{50}
\left(
 \theta^{16}
-\varepsilon^{16}
\right)
\\[1mm]
\frac{3}{200\sqrt{2}}
\left(
 \varepsilon \, \theta^{15}
+273 \,\varepsilon^5 \,\theta^{11}
+715 \,\varepsilon^9 \,\theta^7
+35 \,\varepsilon^{13} \,\theta^3
\right)
\\[1mm]
\frac{3}{200\sqrt{2}}
\left(
 35 \,\varepsilon^3 \,\theta^{13}
+715 \,\varepsilon^7 \,\theta^9
+273 \,\varepsilon^{11} \,\theta^5
+\varepsilon^{15} \,\theta
\right)
  \end{pmatrix}\,,
  \\[2mm]
  Y_{\mathbf{3'},2}^{(8)}(\tau) &=
  \begin{pmatrix}
3
\left(
 \varepsilon^4 \,\theta^{12}
-\varepsilon^{12} \,\theta^4
\right)
\\[1mm]
\frac{3}{8\sqrt{2}}
\left(
 \varepsilon \, \theta^{15}
-15 \,\varepsilon^5 \,\theta^{11}
+11 \,\varepsilon^9 \,\theta^7
+3 \,\varepsilon^{13} \,\theta^3
\right)
\\[1mm]
\frac{3}{8\sqrt{2}}
\left(
 3 \,\varepsilon^3 \,\theta^{13}
+11 \,\varepsilon^7 \,\theta^9
-15 \,\varepsilon^{11} \,\theta^5
+\varepsilon^{15} \,\theta
\right)
  \end{pmatrix}\,.
\end{align*}

For $k=9$, one has:
\begin{align*}
  Y_{\mathbf{\hat{1}}}^{(9)}(\tau) &=
\frac{9}{4}\sqrt{\frac{3}{2}}
\left(
 \varepsilon^3 \,\theta^{15}
-3 \,\varepsilon^7 \,\theta^{11}
+3 \,\varepsilon^{11} \,\theta^7
-\varepsilon^{15} \,\theta^3
\right)
  \,,\\[2mm]
  Y_{\mathbf{\hat{1}'}}^{(9)}(\tau) &=
\frac{1}{8}\sqrt{\frac{3}{2}}
\left(
 \varepsilon \, \theta^{17}
-34 \,\varepsilon^5 \,\theta^{13}
+34 \,\varepsilon^{13} \,\theta^5
-\varepsilon^{17} \,\theta
\right)
  \,,\\[2mm]
  Y_{\mathbf{\hat{2}}}^{(9)}(\tau) &=
  \begin{pmatrix}
\frac{3}{8}
\left(
 \varepsilon^3 \,\theta^{15}
+13 \,\varepsilon^7 \,\theta^{11}
-13 \,\varepsilon^{11} \,\theta^7
-\varepsilon^{15} \,\theta^3
\right)
\\[1mm]
\frac{\sqrt{3}}{16}
\left(
 \varepsilon \, \theta^{17}
+14 \,\varepsilon^5 \,\theta^{13}
-14 \,\varepsilon^{13} \,\theta^5
-\varepsilon^{17} \,\theta
\right)
  \end{pmatrix}\,,
  \\[2mm]
  Y_{\mathbf{\hat{3}},1}^{(9)}(\tau) &=
  \begin{pmatrix}
\frac{1}{2\sqrt{19}}
\left(
 \varepsilon \, \theta^{17}
-17 \,\varepsilon^5 \,\theta^{13}
-17 \,\varepsilon^{13} \,\theta^5
+\varepsilon^{17} \,\theta
\right)
\\[1mm]
\frac{1}{16\sqrt{38}}
\left(
 17 \,\varepsilon^2 \,\theta^{16}
-442 \,\varepsilon^6 \,\theta^{12}
+170 \,\varepsilon^{14} \,\theta^4
-\varepsilon^{18}
\right)
\\[1mm]
\frac{1}{16\sqrt{38}}
\left(
 \theta^{18}
-170 \,\varepsilon^4 \,\theta^{14}
+442 \,\varepsilon^{12} \,\theta^6
-17 \,\varepsilon^{16} \,\theta^2
\right)
  \end{pmatrix}\,,
  \\[2mm]
  Y_{\mathbf{\hat{3}},2}^{(9)}(\tau) &=
  \begin{pmatrix}
-\frac{3}{8\sqrt{5}}
\left(
 \varepsilon \, \theta^{17}
-2 \,\varepsilon^9 \,\theta^9
+\varepsilon^{17} \,\theta
\right)
\\[1mm]
\frac{3}{8\sqrt{10}}
\left(
 \varepsilon^2 \,\theta^{16}
+49 \,\varepsilon^6 \,\theta^{12}
-37 \,\varepsilon^{10} \,\theta^8
-13 \,\varepsilon^{14} \,\theta^4
\right)
\\[1mm]
\frac{3}{8\sqrt{10}}
\left(
 13 \,\varepsilon^4 \,\theta^{14}
+37 \,\varepsilon^8 \,\theta^{10}
-49 \,\varepsilon^{12} \,\theta^6
-\varepsilon^{16} \,\theta^2
\right)
  \end{pmatrix}\,,
  \\[2mm]
  Y_{\mathbf{\hat{3}},3}^{(9)}(\tau) &=
  \begin{pmatrix}
\frac{9}{2}
\left(
 \varepsilon^5 \,\theta^{13}
-2 \,\varepsilon^9 \,\theta^9
+\varepsilon^{13} \,\theta^5
\right)
\\[1mm]
-\frac{9}{8\sqrt{2}}
\left(
 \varepsilon^2 \,\theta^{16}
+\varepsilon^6 \,\theta^{12}
-5 \,\varepsilon^{10} \,\theta^8
+3 \,\varepsilon^{14} \,\theta^4
\right)
\\[1mm]
\:\:\,\,\frac{9}{8\sqrt{2}}
\left(
 3 \,\varepsilon^4 \,\theta^{14}
-5 \,\varepsilon^8 \,\theta^{10}
+\varepsilon^{12} \,\theta^6
+\varepsilon^{16} \,\theta^2
\right)
  \end{pmatrix}\,,
  \\[2mm]
  Y_{\mathbf{\hat{3}'},1}^{(9)}(\tau) &=
  \begin{pmatrix}
\frac{1}{4\sqrt{10}}
\left(
 7 \,\varepsilon^3 \,\theta^{15}
-39 \,\varepsilon^7 \,\theta^{11}
-39 \,\varepsilon^{11} \,\theta^7
+7 \,\varepsilon^{15} \,\theta^3
\right)
\\[1mm]
-\frac{1}{32\sqrt{5}}
\left(
 \theta^{18}
-90 \,\varepsilon^4 \,\theta^{14}
-182 \,\varepsilon^{12} \,\theta^6
+15 \,\varepsilon^{16} \,\theta^2
\right)
\\[1mm]
\:\:\:\frac{1}{32\sqrt{5}}
\left(
 15 \,\varepsilon^2 \,\theta^{16}
-182 \,\varepsilon^6 \,\theta^{12}
-90 \,\varepsilon^{14} \,\theta^4
+\varepsilon^{18}
\right)
  \end{pmatrix}\,,
  \\[2mm]
  Y_{\mathbf{\hat{3}'},2}^{(9)}(\tau) &=
  \begin{pmatrix}
\frac{3}{4}
\left(
 \varepsilon^3 \,\theta^{15}
-\varepsilon^7 \,\theta^{11}
-\varepsilon^{11} \,\theta^7
+\varepsilon^{15} \,\theta^3
\right)
\\[1mm]
\frac{3}{8\sqrt{2}}
\left(
 5 \,\varepsilon^4 \,\theta^{14}
-11 \,\varepsilon^8 \,\theta^{10}
+7 \,\varepsilon^{12} \,\theta^6
-\varepsilon^{16} \,\theta^2
\right)
\\[1mm]
\frac{3}{8\sqrt{2}}
\left(
 \varepsilon^2 \,\theta^{16}
-7 \,\varepsilon^6 \,\theta^{12}
+11 \,\varepsilon^{10} \,\theta^8
-5 \,\varepsilon^{14} \,\theta^4
\right)
  \end{pmatrix}\,.
\end{align*}

Finally, for $k=10$ one has:
\begin{align*}
  Y_{\mathbf{1}}^{(10)}(\tau) &=
\frac{1}{16\sqrt{6}}
\left(
 \theta^{20}
-19 \,\varepsilon^4 \,\theta^{16}
-494 \,\varepsilon^8 \,\theta^{12}
-494 \,\varepsilon^{12} \,\theta^8
-19 \,\varepsilon^{16} \,\theta^4
+\varepsilon^{20}
\right)
  \,,\\[2mm]
  Y_{\mathbf{1'}}^{(10)}(\tau) &=
\frac{3}{8}\sqrt{\frac{3}{2}}
\left(
 \varepsilon^2 \,\theta^{18}
+12 \,\varepsilon^6 \,\theta^{14}
-26 \,\varepsilon^{10} \,\theta^{10}
+12 \,\varepsilon^{14} \,\theta^6
+\varepsilon^{18} \,\theta^2
\right)
  \,,\\[2mm]
  Y_{\mathbf{2},1}^{(10)}(\tau) &=
  \begin{pmatrix}
\frac{3}{32\sqrt{10}}
\left(
 \theta^{20}
+5 \,\varepsilon^4 \,\theta^{16}
+250 \,\varepsilon^8 \,\theta^{12}
+250 \,\varepsilon^{12} \,\theta^8
+5 \,\varepsilon^{16} \,\theta^4
+\varepsilon^{20}
\right)
\\[1mm]
- \frac{3}{2} \sqrt{\frac{3}{10}}
\left(
 5 \,\varepsilon^6 \,\theta^{14}
+22 \,\varepsilon^{10} \,\theta^{10}
+5 \,\varepsilon^{14} \,\theta^6
\right)
  \end{pmatrix}\,,
  \\[2mm]
  Y_{\mathbf{2},2}^{(10)}(\tau) &=
  \begin{pmatrix}
\frac{9}{4}
\left(
 \varepsilon^4 \,\theta^{16}
-\varepsilon^8 \,\theta^{12}
-\varepsilon^{12} \,\theta^8
+\varepsilon^{16} \,\theta^4
\right)
\\[1mm]
-\frac{3\sqrt{3}}{16}
\left(
 \varepsilon^2 \,\theta^{18}
-12 \,\varepsilon^6 \,\theta^{14}
+22 \,\varepsilon^{10} \,\theta^{10}
-12 \,\varepsilon^{14} \,\theta^6
+\varepsilon^{18} \,\theta^2
\right)
  \end{pmatrix}\,,
  \\[2mm]
  Y_{\mathbf{3},1}^{(10)}(\tau) &=
  \begin{pmatrix}
\frac{3}{16\sqrt{2}}
\left(
 \varepsilon^2 \,\theta^{18}
-34 \,\varepsilon^6 \,\theta^{14}
+34 \,\varepsilon^{14} \,\theta^6
-\varepsilon^{18} \,\theta^2
\right)
\\[1mm]
\:\:\:\,\frac{3}{32}
\left(
 \varepsilon^3 \,\theta^{17}
-34 \,\varepsilon^7 \,\theta^{13}
+34 \,\varepsilon^{15} \,\theta^5
-\varepsilon^{19} \,\theta
\right)
\\[1mm]
-\frac{3}{32}
\left(
 \varepsilon \, \theta^{19}
-34 \,\varepsilon^5 \,\theta^{15}
+34 \,\varepsilon^{13} \,\theta^7
-\varepsilon^{17} \,\theta^3
\right)
  \end{pmatrix}\,,
  \\[2mm]
  Y_{\mathbf{3},2}^{(10)}(\tau) &=
  \begin{pmatrix}
\frac{9}{16}
\left(
 \varepsilon^2 \,\theta^{18}
-2 \,\varepsilon^6 \,\theta^{14}
+2 \,\varepsilon^{14} \,\theta^6
-\varepsilon^{18} \,\theta^2
\right)
\\[1mm]
\frac{9}{8\sqrt{2}}
\left(
 \varepsilon^3 \,\theta^{17}
+5 \,\varepsilon^7 \,\theta^{13}
-13 \,\varepsilon^{11} \,\theta^9
+7 \,\varepsilon^{15} \,\theta^5
\right)
\\[1mm]
\frac{9}{8\sqrt{2}}
\left(
 7 \,\varepsilon^5 \,\theta^{15}
-13 \,\varepsilon^9 \,\theta^{11}
+5 \,\varepsilon^{13} \,\theta^7
+\varepsilon^{17} \,\theta^3
\right)
  \end{pmatrix}\,,
  \\[2mm]
  Y_{\mathbf{3'},1}^{(10)}(\tau) &=
  \begin{pmatrix}
-\frac{3}{32\sqrt{29}}
\left(
 \theta^{20}
+59 \,\varepsilon^4 \,\theta^{16}
-182 \,\varepsilon^8 \,\theta^{12}
+182 \,\varepsilon^{12} \,\theta^8
-59 \,\varepsilon^{16} \,\theta^4
-\varepsilon^{20}
\right)
\\[1mm]
3 \sqrt{\frac{2}{29}}
\left(
 13 \,\varepsilon^9 \,\theta^{11}
+2 \,\varepsilon^{13} \,\theta^7
+\varepsilon^{17} \,\theta^3
\right)
\\[1mm]
3 \sqrt{\frac{2}{29}}
\left(
 \varepsilon^3 \,\theta^{17}
+2 \,\varepsilon^7 \,\theta^{13}
+13 \,\varepsilon^{11} \,\theta^9
\right)
  \end{pmatrix}\,,
  \\[2mm]
  Y_{\mathbf{3'},2}^{(10)}(\tau) &=
  \begin{pmatrix}
\frac{36}{\sqrt{13}}
\left(
 \varepsilon^8 \,\theta^{12}
-\varepsilon^{12} \,\theta^8
\right)
\\[1mm]
-\frac{9}{16\sqrt{26}}
\left(
 \varepsilon \, \theta^{19}
+20 \,\varepsilon^5 \,\theta^{15}
+14 \,\varepsilon^9 \,\theta^{11}
-28 \,\varepsilon^{13} \,\theta^7
-7 \,\varepsilon^{17} \,\theta^3
\right)
\\[1mm]
\:\:\:\:\frac{9}{16\sqrt{26}}
\left(
 7 \,\varepsilon^3 \,\theta^{17}
+28 \,\varepsilon^7 \,\theta^{13}
-14 \,\varepsilon^{11} \,\theta^9
-20 \,\varepsilon^{15} \,\theta^5
-\varepsilon^{19} \,\theta
\right)
  \end{pmatrix}\,,
  \\[2mm]
  Y_{\mathbf{3'},3}^{(10)}(\tau) &=
  \begin{pmatrix}
\frac{9}{8}
\left(
 \varepsilon^4 \,\theta^{16}
-3 \,\varepsilon^8 \,\theta^{12}
+3 \,\varepsilon^{12} \,\theta^8
-\varepsilon^{16} \,\theta^4
\right)
\\[1mm]
\:\:\:\frac{9}{8\sqrt{2}}
\left(
\varepsilon^5 \,\theta^{15}
-3 \,\varepsilon^9 \,\theta^{11}
+3 \,\varepsilon^{13} \,\theta^7
-\varepsilon^{17} \,\theta^3
\right)
\\[1mm]
-\frac{9}{8\sqrt{2}}
\left(
 \varepsilon^3 \,\theta^{17}
-3 \,\varepsilon^7 \,\theta^{13}
+3 \,\varepsilon^{11} \,\theta^9
-\varepsilon^{15} \,\theta^5
\right)
  \end{pmatrix}\,.
\end{align*}

\vfill
\clearpage
\newgeometry{left=2cm,right=2.5cm,top=1.5cm,bottom=1.5cm}
\begin{landscape}
\thispagestyle{empty}

\section{Explicit Expressions for Mass Matrices}
\label{app:full}

Making use of the expressions for modular form multiplets given
in eqs.~\eqref{eq:k3} and \eqref{eq:k4}, 
one can write 
the mass matrices of eqs.~\eqref{eq:MnuYs} and \eqref{eq:MeYs}
in terms of the $\theta(\tau)$ and $\varepsilon(\tau)$ functions.
These matrices explicitly read:
\begin{align}
M_\nu &\,=\,
\frac{2\,v_u^2\,g_1}{\sqrt{3} \,\Lambda}
\begin{pmatrix}
\dfrac{\theta^8}{2\sqrt{3}}\left[\left(1-\dfrac{\sqrt{3}}{2}\tilde{g}_2\right)\left(1+x^2\right) 
+ x\,\left(14 + 5\sqrt{3}\,\tilde{g}_2\right)\right] 
& -\dfrac{3}{4}\,\tilde{g}_3\,\varepsilon^3\,\theta^5\, (1-x)
&  -\dfrac{3}{4}\,\tilde{g}_3\,\varepsilon\,\theta^7\, (1-x)
\\[6mm]
*
& -\dfrac{3}{2}\,\varepsilon^2\,\theta^6\left[\tilde{g}_2\,(1+x) 
+ \dfrac{\tilde{g}_3}{\sqrt{2}}\,(1-x)\right] 
&  \dfrac{\theta^8}{2\sqrt{3}}\left[\left(1 + \dfrac{\sqrt{3}}{4}\tilde{g}_2\right)\left(1+x^2\right) 
+ x\,\left(14 - \dfrac{5\sqrt{3}}{2}\,\tilde{g}_2\right)\right] 
\\[6mm]
*
& *
& -\dfrac{3}{2}\,\varepsilon^2\,\theta^6\,\left[\tilde{g}_2\,(1+x) 
- \dfrac{\tilde{g}_3}{\sqrt{2}}\,(1-x)\right]
\end{pmatrix}\,,
\\
M_e^\dagger &\,=\,
v_d \, \alpha_1\,
\begin{pmatrix}
   \varepsilon\,\theta^5\,(1-x) 
& -\dfrac{\theta^6}{\sqrt{6}}
\left[\dfrac{\tilde\alpha_2}{2}\,(1+3x) - \dfrac{\tilde\alpha_3}{2\sqrt{2}}\,(1-5x)\right]  
& 
\varepsilon^2\,\theta^4\left[-\dfrac{\tilde\alpha_2}{2}\sqrt{\dfrac{3}{2}}\,\left(1+\dfrac{x}{3}\right) 
-\dfrac{5\,\tilde\alpha_3}{4\sqrt{3}}\,\left(1-\dfrac{x}{5}\right)\right] 
\\[6mm]
-\dfrac{\theta^6}{\sqrt{6}}
\left[\dfrac{\tilde\alpha_2}{2}\,(1+3x) + \dfrac{\tilde\alpha_3}{2\sqrt{2}}\,(1-5x)\right]  
& 
\dfrac{2}{\sqrt{3}}\,\tilde\alpha_2\, \varepsilon^3\,\theta^3
&  \varepsilon\,\theta^5 \left[(1-x) + \dfrac{\tilde\alpha_3}{\sqrt{6}}\,(1+x)\right] 
\\[6mm]
-\varepsilon^2\,\theta^4\left[\dfrac{\tilde\alpha_2}{2}\sqrt{\dfrac{3}{2}}\,\left(1+\dfrac{x}{3}\right) 
-\dfrac{5\,\tilde\alpha_3}{4\sqrt{3}}\,\left(1-\dfrac{x}{5}\right)\right] 
& 
\varepsilon\,\theta^5 \left[(1-x) - \dfrac{\tilde\alpha_3}{\sqrt{6}}\,(1+x)\right] 
&
-\dfrac{2}{\sqrt{3}}\,\tilde\alpha_2\, \varepsilon^3\,\theta^3
\end{pmatrix}\,,
\end{align}
where $x \equiv \varepsilon^4/\theta^4$, and
stars denote repeated elements of a symmetric matrix.
Here, $\tilde g_{2(3)} \equiv g_{2(3)} / g_1$
and $\tilde \alpha_{2(3)} \equiv \alpha_{2(3)} / \alpha_1$.
Exact expressions for the determinants of these matrices are
\begin{align}
\det M_\nu
&\,=\,
C\, \theta ^{24} 
\Bigg(
\frac{3}{32}\,\tilde g_2^2\bigg [
6\left(1+14\,x + x^2\right)
\left ( \left ( 1-10\,x+x^2 \right )^2+48\,x\,(1+x)^2 \right )
+
\sqrt{3}\,\tilde g_2\left ( 1-10\,x+x^2 \right )
\left ( \left ( 1-10\,x+x^2 \right )^2-144\,x\,(1+x)^2 \right )
  \bigg ]
\nonumber \\&\qquad\qquad
 -\left(1+14\,x + x^2\right)^3
 + \frac{81\sqrt{3}}{8}\,x\,(1-x)^4 \,\tilde g_3^2
\left [ 3\,\tilde g_2 +\sqrt{2}\,\tilde g_3 \right ]
 \Bigg)\,,
 \qquad
 \text{with } 
 C \equiv \left(\frac{g_1\,v_u^2}{3\,\Lambda}\right)^3
 \,,
\label{eq:detMnu}
\\
\det M_e^\dagger
&\,=\,
-\,v_d^3\,\alpha^3_{1}\, \varepsilon^3\,\theta^{15}(1 - x)^3
\left(1
- \frac{1}{12\sqrt{3}}\tilde\alpha^3_{2}
- \frac{3}{8}\tilde \alpha^2_{3}
+ \frac{\sqrt{3}}{8}\tilde\alpha_{2}\tilde\alpha^2_{3}
- \frac{1}{4}\tilde\alpha^2_{2}\right )\,.
\end{align}

\end{landscape}
\restoregeometry

\bibliographystyle{JHEPwithnote}
\bibliography{bibliography}

\end{document}